\documentclass[fleqn]{svmult}

\pdfoutput=1
\usepackage{svaips}      
\usepackage{makeidx}     
\usepackage{graphicx}    
\usepackage{multicol}    

\usepackage{cite}
\usepackage{amssymb}
\usepackage{multirow}

\def\lbullet{\mbox{\large $\bullet$}}

\def\sblacksquare{\mbox{\footnotesize $\blacksquare$}}

\def\3{\ss }           

\def\bsigma{\mbox{\boldmath $\sigma$}}

\def\bxi{\mbox{\boldmath $\xi$}}
\def\bsxi{\mbox{\boldmath \tiny $\xi$}}
\def\bPi{\mbox{\boldmath $\Pi$}}
\def\bB{\mbox{\boldmath $B$}}

\newcommand{\lla}{\left\langle}
\newcommand{\rra}{\right\rangle}

\begin{document}

\title*{
Multi-Particle Collision Dynamics --- a Particle-Based Mesoscale
Simulation Approach to the Hydrodynamics of Complex Fluids}
\titlerunning{Multi-Particle Collision Dynamics}
\author{G. Gompper\inst{1} \and T. Ihle \inst{2}
\and D.M. Kroll \inst{2} \and R.G. Winkler\inst{1}}
\institute{
Theoretical Soft Matter and Biophysics, Institut f\"ur Festk\"orperforschung,
Forschungszentrum J{\"u}lich, 52425 J{\"u}lich, Germany
\and
Department of Physics, North Dakota State University,
Fargo, North Dakota, 58105-5566 }

\maketitle
\begin{abstract}

In this review, we describe and analyze a mesoscale simulation method
for fluid flow, which was introduced by Malevanets and Kapral in 1999,
and is now called multi-particle collision dynamics (MPC) or
stochastic rotation dynamics (SRD).  The method consists of
alternating streaming and collision steps in an ensemble of point
particles. The multi-particle collisions are performed by grouping
particles in collision cells, and mass, momentum, and energy are
locally conserved. This simulation technique captures both full
hydrodynamic interactions and thermal fluctuations.
The first part of the review begins with a description of several widely
used MPC algorithms and then discusses important features of the original
SRD algorithm and frequently used variations.  Two complementary approaches
for deriving the hydrodynamic equations and evaluating the transport
coefficients are reviewed. It is then shown how MPC algorithms can
be generalized to model non-ideal fluids, and binary mixtures with a
consolute point. The importance of
angular-momentum conservation for systems like phase-separated liquids with
different viscosities is discussed.
The second part of the review describes a number of recent applications
of MPC algorithms to study colloid and polymer dynamics, the
behavior of vesicles and cells in hydrodynamic flows, and the dynamics
of viscoelastic fluids.

\end{abstract}

PACS number(s): 47.11.-j, 05.40.-a, 02.70.Ns


\section{Introduction}
\label{sec:intro}

``Soft Matter'' is a relatively new field of research that encompasses
traditional complex fluids such as amphiphilic mixtures, colloidal
suspensions, and polymer solutions, as well as a wide range of phenomena
including chemically reactive flows (combustion),
the fluid dynamics of self-propelled
objects, and the visco-elastic behavior of networks in cells. One
characteristic feature of all these systems is that phenomena of interest
typically occur on mesoscopic length-scales---ranging from nano- to
micrometers---and at energy scales comparable to the thermal energy $k_BT$.

Because of the complexity of these systems, simulations have played a
particularly important role in soft matter research. These systems are
challenging for conventional simulation techniques due to the presence
of disparate time, length, and energy scales.
Biological systems present additional challenges because they
are often far from equilibrium and are driven by strong spatially
and temporally varying forces. The modeling of these systems often
requires the use of ``coarse-grained'' or mesoscopic approaches that
mimic the behavior of atomistic systems on the
length scales of interest. The goal is to incorporate the essential
features of the microscopic physics in models which are
computationally efficient and are easily implemented in complex
geometries and on parallel computers, and can be used to predict
behavior, test physical theories, and
provide feedback for the design and analysis of experiments and
industrial applications.

In many situations, a simple continuum description based on the
Navier-Stokes equation is not sufficient, since molecular-level
details---including thermal fluctuations---play a central role in
determining the dynamic behavior. A key issue is to resolve the interplay
between thermal fluctuations, hydrodynamic interactions, and
spatio-temporally varying forces. One well-known example of such systems
are microemulsions---a dynamic bicontinuous network of intertwined
mesoscopic patches of oil and water---where
thermal fluctuations play a central role in {\em creating} this phase.
Other examples include flexible polymers in solution, where the coil state
and stretching elasticity are due to the large configurational
entropy. On the other hand, atomistic molecular
dynamics simulations retain too many microscopic degrees of freedom,
consequently requiring very small time steps in order to resolve the
high frequency modes. This makes it impossible to study long
timescale behavior such as self-assembly and other mesoscale phenomena.

In order to overcome these difficulties, considerable effort has been
devoted to the development of mesoscale simulation methods such
as Dissipative Particle Dynamics
\cite{hoog_92_smh,espa_95_hfd,espa_95_fk}, Lattice-Boltzmann
\cite{mcna_88_ube,shan_93_lbm,he_97_tlb}, and
Direct Simulation Monte Carlo \cite{bird_94_mgd,alex_97_dsm,garc_00_nmp}.
The common approach of all these methods is to ``average out'' irrelevant
microscopic details in order to achieve high computational efficiency
while keeping the essential features of the microscopic physics on the
length scales of interest. Applying these ideas to suspensions leads to
a simplified, coarse-grained description of the solvent degrees of freedom,
in which embedded macromolecules such as polymers are treated by
conventional molecular dynamics simulations.

All these approaches are essentially alternative ways of solving the
Navier-Stokes equation and its generalizations. This is because the
hydrodynamic equations are expressions for the local conservation laws of
mass, momentum, and energy, complemented by constitutive relations which
reflect some aspects of the microscopic details. Frisch {\it et al.}
\cite{fris_86_lan} demonstrated that discrete algorithms can be
constructed which recover the Navier-Stokes equation in the continuum limit
as long as these conservation laws are obeyed and space is discretized in
a sufficiently symmetric manner.

The first model of this type was a cellular automaton, called the
Lattice-Gas-Automaton (LG). The algorithm consists of particles
which jump between nodes of a regular lattice at discrete time intervals.
Collisions occur when more than one particle jumps to the same node,
and collision rules are chosen which impose mass and momentum conservation.
The Lattice-Boltzmann method (LB)---which follows the evolution of
the single-particle probability distribution at each node---was a natural
generalization of this approach. LB solves the Boltzmann
equation on a lattice with a small set of discrete velocities determined
by the lattice structure. The price for obtaining this efficiency
is numerical instabilities in certain parameter ranges. Furthermore,
as originally formulated, LB did not contain any thermal fluctuations.
It became clear only very recently (and only for simple liquids) how
to restore fluctuations by introducing additional noise terms
to the algorithm \cite{adhi_05_flb}.

Except for conservation laws and symmetry requirements, there are relatively
few constraints on the structure of mesoscale algorithms.
However, the constitutive relations and the transport coefficients depend
on the details of the algorithm, so that the temperature and
density dependencies of the transport coefficients can be quite different
from those of real gases or liquids. However,
this is not a problem as long as the {\em functional form} of the
resulting hydrodynamic equations is correct. The mapping to real systems
is achieved by tuning the relevant characteristic numbers, such as the
Reynolds and Peclet numbers \cite{dhon96,lars_99_src}, to those of a given
experiment. When it is not possible to match all characteristic numbers,
one concentrates on those which are of order unity, since this indicates that
there is a delicate balance between two effects which need to be reproduced
by the simulation. On occasion, this can be difficult, since changing one
internal parameter, such as the mean free path, usually affects all transport
coefficients in different ways, and it may happen that a given mesoscale
algorithm is not at all suited for a given application
\cite{ripo_04_lhc,ripo_05_drf,hecht_05_scc,padd_06_hib}.

In this review we focus on the development and application of a
particle-based mesoscopic simulation technique which was recently
introduced by Malevanets and Kapral \cite{male_99_mms,male_00_smd}.
The algorithm, which consists of discrete streaming and collision steps,
shares many features with Bird's Direct Simulation
Monte Carlo (DSMC) approach \cite{bird_94_mgd}. Collisions occur
at fixed discrete time intervals, and although space is discretized into cells
to define the multi-particle collision environment, both the spatial
coordinates and the velocities of the particles are continuous variables.
Because of this, the algorithm exhibits unconditional numerical stability
and has an $H$-theorem \cite{male_99_mms,ihle_03_srd_a}. In this review,
we will use the name multi-particle collision dynamics (MPC) to refer
to this class of algorithms. In the original and most widely used version
of MPC, collisions consist of a stochastic rotation of the
relative velocities of the particles in a collision cell. We will refer
to this algorithm as stochastic rotation dynamics (SRD) in the following.

One important feature of MPC algorithms is that the dynamics is well-defined
for an arbitrary time step, $\Delta t$. In contrast to methods such as
molecular dynamics simulations (MD) or dissipative particle dynamics (DPD),
which approximate
the continuous-time dynamics of a system, the time step does not have to be
small. MPC {\em defines} a discrete-time dynamics which has been shown to yield
the correct long-time hydrodynamics; one consequence of the discrete dynamics
is that the transport coefficients depend explicitly on $\Delta t$.
In fact, this freedom can be used to tune the Schmidt number, $Sc$
\cite{ripo_05_drf}; keeping
all other parameters fixed, decreasing $\Delta t$ leads to an increase in
$Sc$. For small time steps, $Sc$ is larger than unity (as in a dense fluid),
while for large time steps, $Sc$ is of order unity, as in a gas.


Because of its simplicity, SRD can be considered an ``Ising model'' for
hydrodynamics, since it is Galilean invariant (when a random grid shift
of the collision cells is performed before each collision step
\cite{ihle_01_srd}) and incorporates all the
essential dynamical properties in an algorithm which is remarkably easy
to analyze. In addition to the conservation of momentum and mass, SRD
also locally conserves energy, which enables simulations in the
microcanonical ensemble.  It also fully incorporates both
thermal fluctuations and hydrodynamic interactions. Other more established
methods, such as Brownian Dynamics (BD) can also be augmented to
include hydrodynamic interactions. However, the additional computational
costs are often prohibitive \cite{mohan_07_utp,kim_06_bdf}. In addition,
hydrodynamic interactions can be easily switched off in MPC algorithms,
making it easy to study the importance of hydrodynamic interactions
\cite{kiku_02_pcp,ripo_07_hss}.

It must, however, be emphasized that all local algorithms such as MPC, DPD,
and LB model {\em compressible} fluids, so that it takes time for the
hydrodynamic interactions to ``propagate'' over longer distances. As a
consequence, these methods become quite inefficient in the Stokes limit,
where the Reynolds number approaches zero. Algorithms which
incorporate an Oseen tensor do not share this shortcoming.

The simplicity of the SRD algorithm has made it possible to derive analytic
expressions for the transport coefficients which are valid for both large
and small mean free paths \cite{kiku_03_tcm,ihle_03_srd_b,ihle_05_ect}.
This is usually very difficult to do for other mesoscale particle-based
algorithms. Take DPD as an example: the viscosity measured in
Ref.~\cite{backer_05_pf} is about $50\%$ smaller than the value predicted
theoretically in the same paper. For SRD, the agreement is generally better
than $1\%$.

MPC is particularly well suited for (i) studying phenomena where both
thermal fluctuations and hydrodynamics are important, (ii) for systems
with  Reynolds and Peclet numbers of order $0.1$ to $10$, (iii) if exact
analytical expressions for the transport coefficients and consistent
thermodynamics are needed, and (iv) for modeling complex phenomena for
which the constitutive relations are not known. Examples include
chemically reacting flows, self-propelled objects, or solutions with
embedded macromolecules and aggregates.

If thermal fluctuations are not essential or undesirable, a more
traditional method such as a finite-element solver or a Lattice-Boltzmann
approach is recommended.
If, on the other hand, inertia and fully resolved hydrodynamics are not
crucial, but fluctuations are, one might be better served using
Langevin or Brownian Dynamics.

This review consists of two parts. The first part begins with a description
of several widely used MPC algorithms in Sec.~\ref{sec:algorithms}, and then
discusses important features of the original SRD algorithm and a frequently
used variation (MPC-AT), which effectively thermostats the system by
replacing the relative velocities of particles in a collision cell with newly
generated Gaussian random numbers in the collision step. After a qualitative
discussion of the static and dynamic properties of MPC fluids in
Sec.~\ref{sec:qualitative_discussion}, two
alternative approaches for deriving the hydrodynamic equations and evaluating
the transport coefficients are described. First, in Sec.~\ref{sec:eq_calc},
discrete-time projection
operator methods are discussed and the explicit form of the resulting
Green-Kubo relations for the transport coefficients are given and evaluated.
Subsequently, in Sec.~\ref{sec:NECTC}, an alternative non-equilibrium approach
is described. The
two approaches complement each other, and the predictions of both methods
are shown to be in complete agreement. It is then shown in Sec.~\ref{sec:GMPC}
how MPC algorithms can be generalized to model non-ideal fluids and binary
mixtures, Finally, various approaches
for implementing slip and no-slip boundary conditions---as well as the
coupling of embedded objects to a MPC solvent---are described in
Sec.~\ref{sec:BOUND_COND}.  In Sec.~\ref{sec:couette}, the importance of
angular-momentum conservation is discussed, in particular in systems of
phase-separated fluids with different viscosities under flow. An important
aspect of mesoscale simulations is the possibility to directly assert the
effect of hydrodynamic interactions by switching them off, while
retaining the same thermal fluctuations and similar friction coefficients;
in MPC, this can be done very efficiently by an algorithm described in
Sec.~\ref{sec:MPCwithout}.
The second part of the review describes a number of recent applications
of MPC algorithms to study colloid and polymer dynamics, and the
behavior of vesicles and cells in hydrodynamic flows.
Sec.~\ref{sec:applications} focuses on the non-equilibrium behavior of
colloidal suspensions, the dynamics of dilute solutions of linear polymers
both in equilibrium and under flow conditions, and the properties of star
polymers---also called ultra-soft colloids---in shear flow.
Sec.~\ref{sec:vesicles} is devoted to the review of recent simulation
results for membranes in flow. After a short introduction to the modeling
of membranes with different levels of coarse-graining, the behavior of
fluid vesicles and red blood cells, both in shear and capillary flow,
is discussed. Finally, a simple extension of MPC for viscoelastic solvents
is described in Sec.~\ref{sec:viscoelastic}, where the point particles of
MPC for Newtonian fluids are replaced by harmonic dumbbells.

A discussion of
several complementary applications---such as chemically reactive flows
and self-propelled objects---can be found in a recent review of MPC by
R.~Kapral \cite{kapr_08_mpc}.

\section{Algorithms}
\label{sec:algorithms}

In the following, we use the term multi-particle collision dynamics (MPC)
to describe the generic class of particle-based algorithms for fluid flow
which consist of successive free-streaming and multi-particle collision
steps. The name stochastic rotation dynamics (SRD) is reserved for the
most widely used algorithm which was introduced by Malevanets and Kapral
\cite{male_99_mms}. The name refers to the fact that the collisions consist
of a {\em random rotation} of the relative velocities $\delta {\bf v}_i =
{\bf v}_i - {\bf u}$ of the particles in a collision cell, where
${\bf u}$ is the mean velocity of all particles in a cell.
There are a number of other MPC algorithms with different collision
rules \cite{alla_02_mss,nogu_07_pmh,ihle_06_cpa}.
For example, one class of algorithms uses modified
collision rules which provide a nontrivial ``collisional'' contribution
to the equation of state \cite{ihle_06_cpa,tuzel_06_ctc}. As a result,
these models can be used to model non-ideal fluids or multi-component
mixtures with a consolute point.

\subsection{Stochastic Rotation Dynamics (SRD)}

In SRD, the solvent is modeled by a large number $N$ of point-like
particles of mass $m$ which move in continuous space with a continuous
distribution of velocities.
The algorithm consists of individual streaming and collision steps. In
the streaming step, the coordinates, ${\bf r}_i(t)$, of all solvent
particles at time $t$ are simultaneously updated according to
\begin{equation}\label{stream}
{\bf r}_i(t+\Delta t)=
{\bf r}_i(t) + \Delta t\, {\bf v}_i(t)\,,
\end{equation}
where ${\bf v}_i(t)$ is the velocity
of particle $i$ at time $t$ and $\Delta t$ is the value of the discretized
time step.

In order to define the collisions, particles are sorted into cells, and they
interact only with members of their own cell. Typically, the system is
coarse-grained into cells of a regular, typically cubic, grid with lattice
constant $a$. In practice, lengths are often measured in units of $a$,
which corresponds to setting $a=1$.
The average number of particles per cell, $M$, is typically chosen to be
between three and 20. The actual number of particles in cell
at a given time, which fluctuates, will be denoted by $N_c$.
The collision step consists of a random rotation {\bf R} of the relative
velocities $\delta{\bf v}_i={\bf v}_i-{\bf u}$ of all the particles in
the collision cell,
\begin{equation}\label{collide}
{\bf v}_i(t+\Delta t)={\bf u}(t)+{\bf R}\cdot \delta{\bf v}_i(t)\,.
\end{equation}
All particles in the cell are subject to the same rotation, but the rotations
in different cells and at different times are statistically independent.
There is a great deal of freedom in how the rotation step is implemented,
and any stochastic rotation matrix which satisfies semi-detailed balance
can be used. Here, we describe the most commonly used algorithm.
In two dimensions, ${\bf R}$ is a rotation by an angle $\pm\alpha$,
with probability $1/2$. In three dimensions, a rotation
by a fixed angle $\alpha$ about a randomly chosen axis is typically
used. Note that rotations by an angle $-\alpha$ need not be considered, since
this amounts to a rotation by an angle $\alpha$ about an axis with the
opposite orientation.
If we denote the randomly chosen rotation axis by
$\hat{\bf R}$, the explicit collision rule in three dimensions is
\begin{eqnarray}\label{3D}
{\bf v}_i(t+\Delta t) = {\bf u}(t) &+&
\delta{\bf v}_{i,\perp}(t)\cos(\alpha)\nonumber \\
&+&(\delta{\bf v}_{i,\perp}(t)\times\hat{\bf R})\sin(\alpha) +
\delta{\bf v}_{i,\Vert}(t)\, ,
\end{eqnarray}
where $\perp$ and $\Vert$ are the components of the vector which are
perpendicular and parallel to the random axis $\hat{\bf R}$, respectively.
Malevanets and Kapral \cite{male_99_mms} have shown that there is an
$H$-theorem for the algorithm, that the equilibrium distribution of
velocities is Maxwellian, and that it yields the correct hydrodynamic
equations with an ideal-gas equation of state.

In its original form \cite{male_99_mms,male_00_smd}, the SRD algorithm
was not
Galilean invariant. This is most pronounced at low temperatures or small
time steps, where the mean free path, $\lambda=\Delta t\sqrt{k_B T/m}$,
is smaller than the cell size $a$. If the particles travel a distance between
collisions which is small compared to the cell size, essentially the same
particles collide repeatedly before other particles enter the cell or
some of the participating particles leave the cell. For small $\lambda$,
large numbers of particles in a given cell remain correlated over
several time steps. This leads to a breakdown of the molecular chaos
assumption---i.e., particles become correlated and retain information of
previous encounters. Since these correlations are changed by a
homogeneous imposed flow field, ${\bf V}$, Galilean invariance is
destroyed, and the transport coefficients depend on both the magnitude
and direction of ${\bf V}$.

Ihle and Kroll \cite{ihle_01_srd,ihle_03_srd_a} showed that Galilean
invariance can be restored by performing a random shift of the entire
computational grid before every collision step. The grid shift constantly
groups particles into new collision neighborhoods;
the collision environment no longer depends on the magnitude of an imposed
homogeneous flow field, and the resulting hydrodynamic equations are Galilean
invariant for arbitrary temperatures and Mach number. This procedure is
implemented by shifting all particles by the {\it same} random vector with
components uniformly distributed
in the interval $[-a/2, a/2]$ before the collision step. Particles are
then shifted back to their original positions after the collision.

In addition to restoring Galilean invariance, this grid-shift procedure
accelerates momentum transfer between cells and leads to a collisional
contribution to the transport coefficients. If the mean free path
$\lambda$ is larger than $a/2$, the violation of Galilean invariance
without grid shift is negligible, and it is not necessary to use this
procedure.

\subsubsection{SRD with Angular Momentum Conservation}

As noted by Pooley and Yeomans \cite{pool_05_ktd} and confirmed in
Ref.~\cite{ihle_05_ect}, the macroscopic stress tensor of SRD is {\em not}
symmetric in $\partial_\alpha v_\beta$. The reason for this is that the
multi-particle collisions do not, in general, conserve angular momentum.
The problem is particularly pronounced for small mean free paths, where
asymmetric collisional contributions to the stress tensor dominate the
viscosity (see Sec.~\ref{sec:ee}). In contrast, for mean free paths
larger than the cell size, where kinetic contributions dominate, the effect
is negligible.

An anisotropic stress tensor means that there is non-zero dissipation if
the entire fluid undergoes a rigid-body rotation, which is clearly unphysical.
However, as emphasized in Ref.~\cite{ihle_05_ect}, this asymmetry
is not a problem for most applications in the incompressible (or
small Mach number) limit, since the form of the Navier-Stokes equation is
not changed. This is in accordance with results obtained in
SRD simulations of vortex shedding behind an obstacle \cite{lamu_01_mcd},
and vesicle \cite{nogu_04_fvv} and polymer dynamics \cite{ripo_04_lhc}.
In particular, it has been shown that the linearized hydrodynamic modes
are completely unaffected in two dimensions; in three
dimensions only the sound damping is slightly modified \cite{ihle_05_ect}.

However, very recently G\"otze {\it et al.} \cite{gotz_07_ram} identified
several situations involving rotating flow fields in which this asymmetry
leads to significant deviations from the behavior of a Newtonian fluid.
This includes (i) systems in which boundary conditions are defined by
torques rather than
prescribed velocities, (ii) mixtures of liquids with a viscosity
contrast, and (iii) polymers with a locally high monomer density
and a monomer-monomer distance on the order of or smaller than the
lattice constant, $a$, embedded in a MPC fluid. A more detailed
discussion will be presented in Sec.~\ref{sec:couette} below.

For the SRD algorithm, it is possible to restore angular momentum
conservation by having
the collision angle depend on the specific positions of the particles
within a collision cell. Such a modification was first suggested by
Ryder \cite{ryde_05_the} for SRD in two dimensions. She showed that
the angular momentum of the particles in a collision cell is conserved if
the collision angle $\alpha$ is chosen such that
\begin{equation}
\label{RYDER1}
{\rm sin}(\alpha)=-2AB/( A^2+B^2)\,\,\, {\rm and} \,\,\,
{\rm cos}(\alpha)=(A^2-B^2)/( A^2+B^2)
\end{equation}
where
\begin{equation} \label{RYDER2}
A=\sum_1^{N_c}\,[{\bf r}_i\times ( {\bf v}_i-{\bf u}) ] \vert_z
\ \ \ {\rm and} \ \ \
B=\sum_1^{N_c}\,{\bf r}_i\cdot ( {\bf v}_i-{\bf u}).
\end{equation}
When the collision angles are determined in this way,
the viscous stress tensor is symmetric. Note, however, that evaluating
Eq.~(\ref{RYDER1}) is time-consuming, since the collision
angle needs to be computed for every collision cell every time step.
This typically increases the CPU time by a factor close to two.

A general procedure for implementing angular-momentum conservation in
multi-particle collision algorithms was introduced
by Noguchi {\it et al.} \cite{nogu_07_pmh}; it is discussed in
the following section.

\subsection{Multi-Particle Collision Dynamics with Anderson
Thermostat (MPC-AT)}
\label{sec:mpcda}

A stochastic rotation of the particle velocities relative to the
center-of-mass velocity is not the only possibility for performing
multi-particle collisions.
In particular, MPC simulations can be performed directly in the
canonical ensemble by employing an Anderson thermostat (AT)
\cite{alla_02_mss,nogu_07_pmh}; the resulting algorithm will
be referred to as MPC-AT-a. In this algorithm, instead of performing
a rotation of the relative velocities,
$\{\delta{\bf v}_i\}$, in the collision step, new relative velocities
are generated. The components of $\{\delta{\bf v}^{ran}_i\}$
are Gaussian random numbers with variance $\sqrt{k_BT/m}$.
The collision rule is \cite{nogu_07_pmh,gotz_07_ram}
\begin{equation}
\label{mpc-at-a}
{\bf v}_i(t+\Delta t) = {\bf u}(t) + \delta{\bf v}_i^{ran} =
{\bf u}(t)+{\bf v}_i^{ran}-\sum _{j\in cell}{\bf v}_j^{ran}/N_c \,,
\end{equation}
where $N_c$ is the number of particles in the collision cell, and
the sum runs over all particles in the cell. It is important to
note that MPC-AT is both a collision procedure and a thermostat.
Simulations are performed in the canonical ensemble, and no
additional velocity rescaling is required in non-equilibrium
simulations, where there is viscous heating.

Just as SRD, this algorithm conserves momentum at the cell
level but not angular momentum.
Angular momentum conservation can be restored \cite{ryde_05_the,nogu_07_pmh}
by imposing constraints on the new relative velocities.
This leads to an angular-momentum
conserving modification of MPC-AT \cite{gotz_07_ram,nogu_07_pmh},
denoted MPC-AT$+a$. The collision rule in this case is
\begin{eqnarray}
\nonumber
{\bf v}_i(t+\Delta t)&=&{\bf u}(t)+{\bf v}_{i,ran}-\sum _{cell}
{\bf v}_{i,ran}/N_c \\
\label{mpc-at+a}
& &+\left\{m\bPi^{-1} \sum_{j\in cell}\left[{\bf r}_{j,c}
\times ({\bf v}_j-{\bf v}_j^{ran})\right] \times {\bf r}_{i,c}\right\},
\end{eqnarray}
where $\bPi$ is the moment of inertia tensor of the particles in the cell,
and ${\bf r}_{i,c}={\bf r}_i-{\bf R}_c$ is the relative position of
particle $i$ in the cell and ${\bf R}_c$ is the center of mass of all
particles in the cell.

When implementing this algorithm, an unbiased multi-particle collision
is first performed, which typically leads to a small
change of angular momentum, $\Delta\vec{L}$. By solving the linear equation
$-\Delta\vec{L}=\bf{\Pi}\cdot\vec{\omega}$, the angular velocity $\omega$
which is needed to cancel the initial change of angular momentum is then
determined. The last term in Eq.~(\ref{mpc-at+a}) restores this angular
momentum deficiency.
MPC-AT can be adapted for simulations in the micro-canonical
ensemble by imposing an additional constraint on the values of the new
random relative velocities \cite{nogu_07_pmh}.

\subsubsection{Comparison of SRD and MPC-AT}
\label{sec:cSRD_AT}

Because $d$ Gaussian random numbers per particle are required at every
iteration,
where $d$ is the spatial dimension, the speed of the random number generator
is the limiting factor for MPC-AT. In contrast, the efficiency of SRD is rather
insensitive to the speed of the random number generator since only $d-1$
uniformly distributed random numbers are needed in every box per iteration,
and even a low quality random number generator is sufficient, because the
dynamics is self-averaging.
A comparison for two-dimensional systems shows that MPC-AT-a is about a
factor 2 to 3 times slower than SRD, and that MPC-AT+a is
about a factor 1.3 to 1.5 slower than MPC-AT-a \cite{goetz_private}.

One important difference between SRD and MPC-AT is the fact that
relaxation times in MPC-AT generally {\em decrease} when the number of
particles per cell is increased, while they {\em increase} for SRD.
A longer relaxation time means that a larger number of time steps is
required for transport coefficients to reach their asymptotic value.
This could be of importance when fast oscillatory or transient processes
are investigated. As a consequence, when using SRD, the average number of
particles per cell should be in range of $3-20$; otherwise, the internal
relaxation times could be no longer negligible compared to physical time
scales. No such limitation exists for MPC-AT, where the relaxation times
scale as $(\ln M)^{-1}$, where $M$ is the average number of particles in
a collision cell.

\subsection{Computationally Efficient Cell-Level Thermostating for SRD}
\label{sec:THERMOSTAT}

The MPC-AT algorithm discussed in Sec.~\ref{sec:mpcda} provides a very
efficient particle-level thermostating of the system.
However, it is considerably slower than the original SRD algorithm, and
there are situations in which the additional freedom offered by the choice
of SRD collision angle can be useful.

Thermostating is required in any non-equilibrium MPC simulation, where there
is viscous heating. A basic requirement of any thermostat is that it does
not violate local momentum conservation, smear out local flow profiles, or
distort the velocity distribution too much. When there is homogeneous heating,
the simplest way to maintain a constant temperature is to just rescale
velocity components by a scale factor $S$, $v^{new}_{\alpha}=Sv_{\alpha}$,
which adjusts the total kinetic energy to the desired value.
This can be done with just a single global scale factor, or a local factor
which is different in every cell. For a known macroscopic flow profile,
${\bf u}$, like in shear flow, the relative velocities ${\bf v}-{\bf u}$
can be rescaled. This is known as a profile-unbiased thermostat; however,
it has been shown to have deficiencies in molecular dynamics simulations
\cite{erpe_84_svh}.

Here we describe an alternative thermostat which exactly conserves momentum
in every cell and is easily incorporated into the MPC collision step.
It was originally developed by Heyes for constant-temperature molecular
dynamics simulations; however, the original algorithm described in
Ref.~\cite{alle:87}
violates detailed balance. The thermostat consists of the following
procedure which is performed independently in every collision cell
as part of the collision step.
\begin{enumerate}
\item Randomly select a real number $\psi \in [1,1+c]$,
where $c$ is a small number between 0.05 and 0.3 which determines the
strength of the thermostat.
\item Accept this number as a scaling factor $S=\psi$ with probability $1/2$;
otherwise, take $S=1/\psi$.
\item Create another random number $\xi\in[0,1]$. Rescale the velocities
if $\xi$ is smaller than the acceptance probability $p_A=\min(1,A)$, where
\begin{equation}
\label{THERMO3}
A=S^{d\,(N_c-1)}{\rm exp}\left(-\frac{m}{2k_B T_0}\,
\sum_{i=1}^{N_c}\,({\bf v}_i-{\bf u})^2\{S^2-1\}\right).
\end{equation}
$d$ is the spatial dimension, and $N_c$ is the number of particles in the
cell. The prefactor in Eq.~(\ref{THERMO3}) is an entropic contribution
which accounts for the fact that the phase-space volume changes if the
velocities are rescaled.
\item If the attempt is accepted, perform a stochastic rotation with the
scaled rotation matrix $S\,{\bf R}$. Otherwise, use the rotation matrix
${\bf R}$.
\end{enumerate}

This thermostat reproduces the Maxwell velocity distribution and
does not change the viscosity of the fluid. It gives excellent equilibration,
and the deviation of the measured kinetic
temperature from $T_0$ is smaller than $0.01 \%$. The parameter $c$ controls
the rate at which the kinetic temperature relaxes to $T_0$, and in agreement
with experience from MC-simulations, an acceptance rate in the range of
$50\%$ to $65\%$ leads to the fastest relaxation. For these acceptance rates,
the relaxation time is of order $5-10$ time steps. The corresponding value
for $c$ depends on the particle number $N_c$; in two dimensions, it is
about $0.3$ for $N_c=7$ and decreases to $0.05$ for $N_c=100$.
This thermostat has been successfully applied to SRD simulations of
sedimenting charged colloids \cite{hecht_05_scc}.

\section{Qualitative Discussion of Static and Dynamic Properties}
\label{sec:qualitative_discussion}

The previous section outlines several multi-particle algorithms. A detailed
discussion of the link between the microscopic dynamics described
by Eqs.~(\ref{stream}) and (\ref{collide}) or (\ref{3D}) and the macroscopic
hydrodynamic equations, which describe the
behavior at large length and time scales, requires a more careful analysis
of the corresponding Liouville operator ${\cal L}$. Before describing
this approach in more detail, we provide a more heuristic discussion
of the equation of state and of one of the transport coefficients, the shear
viscosity, using more familiar approaches for analyzing the behavior of
dynamical systems.

\subsection{Equation of State}

In a homogeneous fluid, the pressure is the normal force
exerted by the fluid on one side of a unit area on the fluid on the
other side; expressed somewhat differently, it is
the momentum transfer per
unit area per unit time across an imaginary (flat) fixed surface.
There are both {\em kinetic} and {\em virial} contributions to
the pressure. The first arises from the momentum transported across
the surface by particles that cross the surface in the unit time interval;
it yields the ideal-gas contribution, $P_{id}=Nk_BT/V$, to the
pressure. For classical particles interacting via pair-additive,
central forces, the intermolecular ``potential'' contribution to the
pressure can be determined using the method introduced by
Irving and Kirkwood \cite{irvi_50_smt}. A clear discussion of this
approach is given by Davis in Ref.~\cite{davi_96_smp}, where it is shown
to lead to the virial equation of state of a homogeneous fluid,
\begin{equation}\label{Hvirial}
P = \frac{Nk_BT}{V} + \frac{1}{3V}\sum_i\langle{\bf r}_i\cdot{\bf F}_i
\rangle,
\end{equation}
in three dimensions, where ${\bf F}_i$ is the force on particle $i$ due
to all the other particles, and the sum runs over all particles of the system.

The kinetic contribution to the pressure, $P_{id}=Nk_BT/V$, is clearly
present in all MPC algorithms. For SRD, this is
the only contribution. The reason is that the stochastic
rotations, which define the collisions, transport (on average) no net
momentum across a fixed dividing surface. More general MPC algorithms
(such as those discussed
in Sec.~\ref{sec:GMPC}) have an additional contribution to the virial
equation of state. However, instead of an explicit force ${\bf F}_i$
as in Eq.~(\ref{Hvirial}), the contribution from the multi-particle collisions
is a force of the form $m\Delta v_i/\Delta t$, and the
role of the particle position, ${\bf r}_i$, is played by a variable which
denotes the cell-partners which participate in the collision
\cite{ihle_06_cpa,tuze_07_mmf}.

\subsection{Shear Viscosity}
\label{sec:QSF}

Just as for the pressure, there are both kinetic and collisional contributions
to the transport coefficients.
We present here a heuristic discussion of these contributions
to the shear viscosity, since it illustrates rather clearly the essential
physics and provides background for subsequent technical discussions.

Consider a reference plane (a line in two dimensions) with normal
in the $y$-direction embedded in a
homogeneous fluid in equilibrium. The fluid below the plane exerts
a mean force ${\bf p}_y$ per unit area on the fluid above the plane;
by Newton's third law, the fluid above the plane must exert a mean force
$-{\bf p}_y$ on the fluid below the plane. The normal force per unit area
is just the mean pressure, $P$, so that $p_{yy}=P$. In a homogeneous simple
fluid
in which there are no velocity gradients, there is no tangential force, so
that, for example, $p_{yx}=0$.
$p_{\alpha\beta}$ is called
the {\em pressure tensor}, and the last result is just a statement of the well
known fact that the pressure tensor in a homogeneous simple fluid at
equilibrium has no off-diagonal elements; the diagonal elements are all
equal to the mean pressure $P$.

Consider a shear flow with a shear rate $\dot{\gamma}=
\partial u_x(y)/\partial y$. In this case, there is
a tangential stress on the reference surface because of the velocity gradient
normal to the plane. In the small gradient limit, the {\em dynamic viscosity},
$\eta$, is defined as the coefficient of proportionality between the tangential
stress, $p_{yx}$, and the normal gradient of the imposed velocity gradient,
\begin{equation}\label{LRDEF}
p_{yx} = -\eta \dot\gamma.
\end{equation}
The {\em kinematic viscosity}, $\nu$, is related to $\eta$ by $\nu=\eta/\rho$,
where $\rho=nm$ is the mass density, with $n$ the number density of the
fluid and $m$ the particle mass.

\vspace{0.1cm}
\noindent {\em Kinetic contribution to the shear viscosity:}
The kinetic contribution to the shear viscosity comes from transverse
momentum transport by the flow of fluid particles. This is the dominant
contribution to the viscosity of gases. The following analogy may make
this origin of viscosity clearer. Consider two ships moving side by side
in parallel, but with different speeds. If the sailors on the two ships
constantly throw sand bags from their ship onto the other, there will be
a transfer of momentum between to two ships so that the slower ship accelerates
and the faster ship decelerates. This can be interpreted
as an effective friction, or kinetic viscosity, between the ships.
There are no direct forces between the ships, and the transverse momentum
transfer originates solely from throwing sandbags from one ship to the other.

A standard result from kinetic theory is that the kinetic contribution to
the shear viscosity in simple gases is \cite{reif_65}
\begin{equation}\label{qualvisc}
\eta^{kin} \sim nm\bar  v\lambda,
\end{equation}
where $\lambda$ is the mean free path and $\bar{v}$ is the thermal
velocity.
Using the fact that $\lambda\sim\bar v\Delta t$ for SRD and that
$\bar v\sim\sqrt{k_BT/m}$, relation (\ref{qualvisc}) implies that
\begin{equation}\label{qualvisc2}
\eta^{kin} \sim n k_BT \Delta t,\ \ \ \ {\rm or\ equivalently,}\ \ \ \
\nu^{kin} \sim k_BT\Delta t/m,
\end{equation}
which is, as more detailed calculations presented later will show,
the correct dependence on $n$, $k_BT$, and  $\Delta t$. In fact, the
general form for the kinetic contribution to the kinematic viscosity is
\begin{equation} \label{gen_kin_vis}
\nu^{kin} = \frac{k_BT \Delta t}{m}f_{kin}(d,M,\alpha),
\end{equation}
where $d$ is the spatial dimension, $M$ is the mean number of particles
per cell, and $\alpha$ is the SRD collision angle.
Another way of obtaining this result is to use the analogy
with a random walk: The kinematic viscosity is the diffusion coefficient
for momentum diffusion. At large mean free path, $\lambda/a\gg 1$, momentum
is primarily transported by particle translation (as in the ship analogy).
The mean distance a particle streams during one time step, $\Delta t$, is
$\lambda$. According to the theory of random walks, the corresponding
diffusion coefficient scales as
$\nu^{kin}\sim \lambda^2/\Delta t\sim k_B T\Delta t/m$.

Note that in contrast to a ``real'' gas, for which the viscosity
has a square root dependence on the temperature, $\nu^{kin}\sim T$
for SRD. This is because the mean free path of a particle in SRD does
not depend on density; SRD allows particles to stream right through
each other between collisions. Note, however, that SRD can be easily
modified to give whatever temperature dependence is desired. For
example, an additional temperature-dependent collision probability
can be introduced; this would be of interest, e.g., for a simulation
of realistic shock-wave profiles.

\vspace{0.1cm}
\noindent {\em Collisional contribution to the shear viscosity:}
At small mean free paths, $\lambda/a\ll1$, particles ``stream'' only a short
distance between collisions, and the multi-particle ``collisions'' are the
primary mechanism for momentum transport. These collisions
redistribute momenta within cells of linear size $a$. This means that
momentum ``hops'' an average distance $a$ in one time step,
leading to a momentum diffusion coefficient $\nu^{col}\sim a^2/\Delta t$.
The general form of the collisional contribution to the shear viscosity is
therefore
\begin{equation} \label{gen_col_vis}
\nu^{col} = \frac{a^2}{\Delta t}f_{col}(d,M,\alpha).
\end{equation}
This is indeed the scaling observed in numerical simulations at small mean
free path.

The kinetic contribution dominates for $\lambda\gg a$, while the collisional
contribution dominates in the opposite limit. Two other transport coefficients
of interest are the thermal diffusivity, $D_T$, and the single particle
diffusion coefficient, $D$. Both have the dimension m$^2$/sec. As dimensional
analysis would suggest, the kinetic and collisional contributions to $D_T$
exhibit the same characteristic dependencies on $\lambda$, $a$, and $\Delta t$
described by Eqs.~(\ref{gen_kin_vis}) and (\ref{gen_col_vis}). Since there
is no collisional contribution to the diffusion coefficient,
$D\sim \lambda^2/\Delta t$.

Two complementary approaches have been used to derive the transport
coefficients of the SRD fluid. The first is an equilibrium approach
which utilizes a discrete projection operator formalism to obtain Green-Kubo
(GK) relations which express the transport coefficients as sums over
the autocorrelation functions of reduced fluxes. This approach was
first utilized by Malevanets and Kapral
\cite{male_00_smd}, and later extended by Ihle, Kroll and
T{\"u}zel \cite{ihle_03_srd_a,ihle_03_srd_b,ihle_05_ect} to include
collisional contributions and arbitrary rotation angles. This approach is
described in Sec.~\ref{LHGK}.

The other approach uses kinetic theory to calculate the transport
coefficients in stationary non-equilibrium situation such as shear flow.
The first application of this approach to SRD was presented in
Ref.~\cite{ihle_01_srd}, where the collisional contribution
to the shear viscosity for large $M$, where particle number fluctuations
can be ignored, was calculated. This scheme was later extended by
Kikuchi {\it et al.} \cite{kiku_03_tcm} to include fluctuations in the
number of particles per cell, and then used to obtain expressions for
the kinetic contributions to shear viscosity and thermal conductivity
\cite{pool_05_ktd}. This non-equilibrium approach is described
in Sec.~\ref{sec:NECTC}.

\section{Equilibrium Calculation of Dynamic Properties}
\label{sec:eq_calc}

A projection operator formalism for deriving the linearized hydrodynamic
equations and Green-Kubo (GK) relations for the transport coefficients
of molecular fluids was originally
introduced by Zwanzig \cite{zwan_61_ltp,mori_65_tcm,mori_65_crt} and
later adapted for lattice gases by Dufty and Ernst \cite{duft_89_grl}.
With the help of this formalism, explicit expressions for both the
reversible (Euler) as well as dissipative terms of the long-time,
large-length-scale hydrodynamics equations for the coarse-grained
hydrodynamic variables were derived. In addition, the resulting
GK relations enable explicit calculations of the
transport coefficients of the fluid. This work is summarized in
Sec.~\ref{LHGK}. An analysis of the equilibrium fluctuations of the
hydrodynamic modes can then be used to directly measure the shear and
bulk viscosities as well as the thermal diffusivity. This approach is
described in Sec.~\ref{SF}, where SRD results for the dynamic structure
factor are discussed.

\subsection{Linearized Hydrodynamics and Green-Kubo Relations}
\label{LHGK}

The Green-Kubo (GK) relations for SRD differ from the
well-known continuous versions due to the discrete-time dynamics, the
underlying lattice structure, and the multi-particle interactions.
In the following, we briefly outline this approach for determining the
transport coefficients. More details can be found in
Refs.~\cite{ihle_03_srd_a,ihle_03_srd_b}.

The starting point of this theory are microscopic definitions of
local hydrodynamic densities $A_\beta$. These ``slow'' variables are the
local number, momentum, and energy density. At the cell level, they
are defined as
\begin{equation}
\label{cons}
A_{\beta}(\bxi) = \sum_{i=1}^{N}
a_{\beta,i} \prod_{\gamma=1}^d \Theta\left({a\over2}-\left\vert\xi_\gamma+{a\over 2}
-r_{i\gamma}\right\vert\right),
\end{equation}
with the discrete cell coordinates $\bxi=a{\bf m}$,
where $m_\beta=1,\dots,L$, for each spatial component.
$a_{1,i}=1$ is the particle density, $\{a_{\beta,i}\}=m\{v_{i(\beta-1)}\}$,
with $\beta=2,...,d+1$, are the components of the particle momenta, and
$a_{d+2,i}=mv_i^2/2$ is the kinetic energy of particle $i$.
$d$ is the spatial dimension, and ${\bf r}_i$ and ${\bf v}_i$ are position
and velocity of particle $i$, respectively.

$A_\beta(\bxi)$, for $\beta=2,\dots,d+2$, are cell level coarse-grained
densities. For example, $A_2(\bxi)$ is the $x$-component of the
total momentum of all the particles in cell $\bxi$ at the given time.
Note that the particle density, $A_1$, was not coarse-grained in
Ref.~\cite{ihle_03_srd_a}, i.e., the $\Theta$ functions in
Eq.~(\ref{cons}) were
replaced by a $\delta$-function. This was motivated by the fact that during
collisions the particle number is trivially conserved in areas of arbitrary
size, whereas energy and momentum are only conserved at the cell level.

The equilibrium correlation functions for the conserved variables are
defined by
$\langle\delta A_\beta({\bf r},t) \delta A_\gamma({\bf r}',t')\rangle$,
where $\langle\delta A\rangle = A - \langle A\rangle$, and the brackets denote
an average over the equilibrium distribution. In a stationary, translationally
invariant system, the correlation functions depend only on the differences
${\bf r}-{\bf r}'$ and $t-t'$, and the Fourier transform of the matrix
of correlation functions is
\begin{equation}
G_{\alpha\beta}({\bf k},t) = \frac{1}{V}\langle\delta A_\beta^*({\bf k},0)
\delta A_\gamma({\bf k},t)\rangle,
\end{equation}
where the asterisk denotes the complex conjugate, and the
spatial Fourier transforms of the densities are given by
\begin{equation}
\label{ftc}
A_\beta({\bf k}) =
\sum_j a_{\beta,j} {\rm e}^{i{\bf k}\cdot\bsxi_j},
\end{equation}
where $\bxi_j$ is the coordinate of the cell occupied by particle $j$.
${\bf k} = 2\pi{\bf n}/(aL)$ is the wave vector, where $n_\beta=0,\pm1,\dots,
\pm(L-1),L$ for the spatial components. To simplify notation, we omit the
wave-vector dependence of $G_{\alpha\beta}$ in this section.

The collision invariants for the conserved densities are
\begin{equation}
\label{conserv}
\sum_j {\rm e}^{i{\bf k}\cdot{\bsxi}^s_j(t+\Delta t)}
[a_{\beta,j}(t+\Delta t)-
                 a_{\beta,j}(t)]=0,
\end{equation}
where ${\bxi}^s_j$ is the coordinate of the cell occupied by particle $j$
in the {\it shifted} system. Starting from these conservation laws, a
projection operator can be constructed that projects the full SRD dynamics
onto the conserved fields \cite{ihle_03_srd_a}. The central result is that
the discrete Laplace transform of the linearized hydrodynamic equations
can be written as
\begin{equation}\label{HDE}
[s+ik\Omega+k^2\Lambda]G({\bf k},s) = \frac{1}{\Delta t}G(0)R(k),
\end{equation}
where $R(k) = [1+\Delta t(ik\Omega+k^2\Lambda]^{-1}$ is the residue of the
hydrodynamic pole \cite{ihle_03_srd_a}. The linearized hydrodynamic equations
describe the long-time large-length-scale dynamics of the system, and are
valid in the limits of small $k$ and $s$.
The frequency matrix $\Omega$ contains the reversible (Euler)
terms of the hydrodynamic equations. $\Lambda$ is the matrix of transport
coefficients. The discrete Green-Kubo relation for the matrix of viscous
transport coefficients is \cite{ihle_03_srd_a}
\begin{equation}
\label{VTCGK}
\Lambda_{\alpha\beta}(\hat{\bf k}) \equiv
\frac{\Delta t}{Nk_BT}\left.\sum_{t=0}^\infty\right.'
\langle\hat k_\lambda \sigma_{\alpha\lambda}(0)\vert
\hat k_{\lambda'} \sigma_{\beta\lambda'}(t)\rangle,
\end{equation}
where the prime on the sum indicates that the $t=0$ term has the relative
weight 1/2.
$\sigma_{\alpha\beta}=P\delta_{\alpha\beta}-p_{\alpha\beta}$ is the viscous
stress tensor.
The reduced fluxes in Eq.~(\ref{VTCGK}) are given by
\begin{equation}
\label{VIS2}
\hat k_\lambda\sigma_{\alpha\lambda}(t) =
\frac{m}{\Delta t}\sum_j\left(-v_{j\alpha}(t){\bf \hat k}\cdot
[\Delta\bxi_j(t) + \Delta v_{j\alpha}(t) \Delta\bxi_j^s(t)] +
\frac{\Delta t}{d} \hat k_\alpha v_j^2(t)\right)\nonumber
\end{equation}
for $\alpha=1,\dots,d$,
with $\Delta\bxi_j(t) = \bxi_j(t+\Delta t) - \bxi_j(t)$,
$\Delta\bxi^s_j(t+\Delta t) = \bxi_j(t+\Delta t) - \bxi^s_j(t+\Delta t)$,
and
$\Delta v_{xj}(t)=v_{xj}(t+\Delta t) -v_{xj}(t)$.
$\bxi_j(t)$ is the cell coordinate of particle $j$ at time $t$,
while $\bxi_j^s$ is it's cell coordinate in the (stochastically) shifted
frame. The corresponding expressions for the thermal diffusivity and
self-diffusion coefficient can be found in Ref.~\cite{ihle_03_srd_a}.

The straightforward evaluation of the GK relations for the viscous (\ref{VIS2})
and thermal transport coefficients leads to three---kinetic,
collisional, and mixed---contributions. In addition, it was found that
for mean free paths $\lambda$ smaller than the cell size $a$, there are finite
cell-size corrections which could not be summed in a controlled fashion.
The origin of the problem was the explicit appearance of $\Delta\bxi$ in the
stress correlations. However, it was subsequently shown
\cite{ihle_04_rgr,ihle_05_ect}
that the Green-Kubo relations can be re-summed by introducing a
stochastic variable, $\bB_i$, which is the difference between change in the
shifted cell coordinates of particle $i$ during one streaming step and the
actual distance traveled, $\Delta t\,{\bf v}_i$. The resulting microscopic
stress tensor for the viscous modes is
\begin{equation}
\label{BCOR1}
\bar\sigma_{\alpha\beta} = \sum_i\,\left[mv_{i\alpha} v_{i\beta}
+\frac{m}{\Delta t} v_{i\alpha} B_{i\beta}\right]
\end{equation}
where $B_{j\beta}(t) = \xi^s_{j\beta}(t+\Delta t)-\xi^s_{j\beta}(t) -
\Delta t\, v_{j\beta}(t)$.
It is interesting to compare this result to the corresponding expression
\begin{equation}
\sigma_{\alpha\beta} = \sum_i\delta({\bf r}-{\bf r}_i)
\left[mv_{i\alpha}v_{i\beta} +
\frac{1}{2}\sum_{j\ne i} r_{ij\alpha}{\cal F}_{ij\beta}({\bf r}_{ij})
\right]
\end{equation}
for molecular fluids. The first term in both expressions, the ideal-gas
contribution, is the same in both cases. The collisional contributions,
however, are quit different. The primary reason is that in SRD, the
collisional contribution corresponds to a nonlocal (on the scale of the
cell size) force which acts only at discrete time intervals.

$\bB_i$ has a number of important properties which simplify the calculation
of the transport coefficients. In particular, it is shown in
Refs.~\cite{ihle_04_rgr,ihle_05_ect} that stress-stress correlation functions
involving one $\bB_i$ in the GK relations for the transport coefficients
are zero, so that, for example, $\Lambda_{\alpha\beta}(\hat{\bf k}) =
\Lambda_{\alpha\beta}^{kin}(\hat{\bf k}) +
\Lambda_{\alpha\beta}^{col}(\hat{\bf k})$, with
\begin{equation}\label{VIS_KIN}
\Lambda_{\alpha\beta}^{kin}(\hat{\bf k}) = \frac{\Delta t}{Nmk_BT}
\left.\sum_{n=0}^\infty\right.'
\langle {\hat k}_\lambda\sigma^{kin}_{\alpha\lambda}(0)\vert{\hat k}_{\lambda'}
\sigma^{kin}_{\beta\lambda}(n\Delta t)\rangle
\end{equation}
and
\begin{equation}\label{VIS_COL}
\Lambda_{\alpha\beta}^{col}(\hat{\bf k}) = \frac{\Delta t}{Nmk_BT}
\left.\sum_{n=0}^\infty\right.'
\langle {\hat k}_\lambda\sigma^{col}_{\alpha\lambda}(0)\vert{\hat k}_{\lambda'}
\sigma^{col}_{\beta\lambda}(n\Delta t)\rangle], \nonumber
\end{equation}
with
\begin{equation}\label{SIG_KIN}
\sigma^{kin}_{\alpha\beta}(n\Delta t) =
\sum_j mv_{j\alpha}(n\Delta t)v_{j\beta}(n\Delta t)
\end{equation}
and
\begin{equation}\label{SIG_COL}
\sigma^{col}_{\alpha\beta}(n\Delta t) =
\frac{1}{\Delta t} \sum_j mv_{j\alpha}(n\Delta t) B_{j\beta}(n\Delta t),
\end{equation}
where $B_{j\beta}(n\Delta t) = \xi^s_{j\beta}([n+1]\Delta t)-
\xi^s_{j\beta}(n\Delta t) - \Delta t v_{j\beta}(n\Delta t)$.
Similar relations were obtained for the thermal diffusivity in
Ref.~\cite{ihle_05_ect}.

\subsubsection{Explicit Expressions for the Transport Coefficients}
\label{sec:ee}

Analytical calculations of the SRD transport coefficients are greatly
simplified by the fact that collisional and kinetic contributions to the
stress-stress autocorrelation functions decouple. Both the kinetic and
collisional contributions have been calculated explicitly in two and three
dimension, and numerous numerical tests have shown that the resulting
expressions for all the transport coefficients are in excellent agreement
with simulation data. Before summarizing the results of this work, it is
important to emphasize that because of the cell structure introduced to
define coarse-grained collisions, angular momentum is not conserved in
a collision \cite{pool_05_ktd,ihle_05_ect,ryde_05_the}.
As a consequence, the macroscopic viscous
stress tensor is not, in general, a symmetric function of the derivatives
$\partial_\alpha v_\beta$. Although the kinetic contributions to the transport
coefficients lead to a symmetric stress tensor, the collisional do not.
Before evaluating the transport coefficients, we discuss the general
form of the macroscopic viscous stress tensor.

Assuming only cubic symmetry and allowing for a non-symmetric stress tensor,
the most general form of the linearized Navier-Stokes equation is
\begin{equation}
\partial_t v_\alpha({\bf k}) = -\partial_\alpha p +
\Lambda_{\alpha\beta}(\hat{\bf k})
v_\beta({\bf k}),
\end{equation}
where
\begin{eqnarray}
\label{ST}
\Lambda_{\alpha\beta}({\bf \hat k}) &\equiv&
\nu_1\left(\delta_{\alpha,\beta}+
              \frac{d-2}{d}\hat k_\alpha \hat k_\beta\right)\\
&+&
\nu_2\left(\delta_{\alpha,\beta} - \hat k_\alpha \hat k_\beta \right)
+ \gamma \hat k_\alpha \hat k_\beta
+ \kappa\ \hat k_\alpha^2 \delta_{\alpha,\beta}.\nonumber
\end{eqnarray}
In a normal simple liquid, $\kappa=0$ (because of invariance with respect
to infinitesimal rotations) and $\nu_2=0$ (because the stress tensor is
symmetric in $\partial_\alpha v_\beta$), so that the kinematic shear
viscosity is $\nu=\nu_1$.  In this case, Eq.~(\ref{ST}) reduces to the
well-known form \cite{ihle_03_srd_a}
\begin{equation}
\Lambda_{\alpha\beta}({\bf \hat k})
=\nu\left(\delta_{\alpha,\beta}+\frac{d-2}{d}
\hat k_\alpha \hat k_\beta\right) + \gamma \hat k_\alpha \hat k_\beta\,,
\end{equation}
where $\gamma$ is the bulk viscosity.

\begin{figure}[t]
\vspace{0.2cm}
 \hfill
  \begin{minipage}[t]{.48\textwidth}
    \begin{center}
      \includegraphics*[scale=0.23]{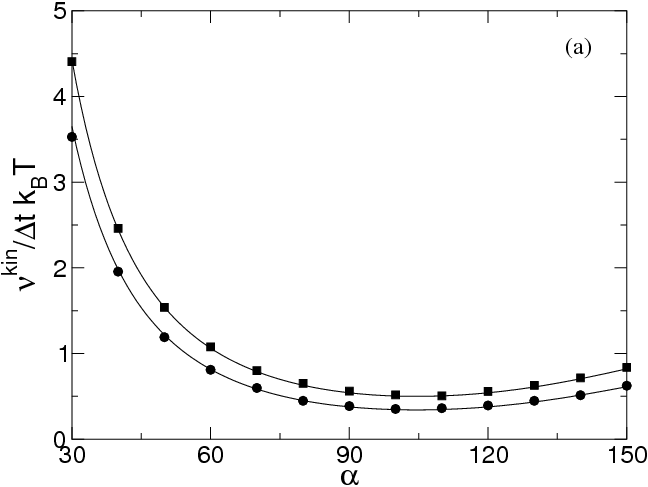}
    \end{center}
  \end{minipage}
  \hfill
  \begin{minipage}[t]{.48\textwidth}
   \begin{center}
      \includegraphics*[scale=0.23]{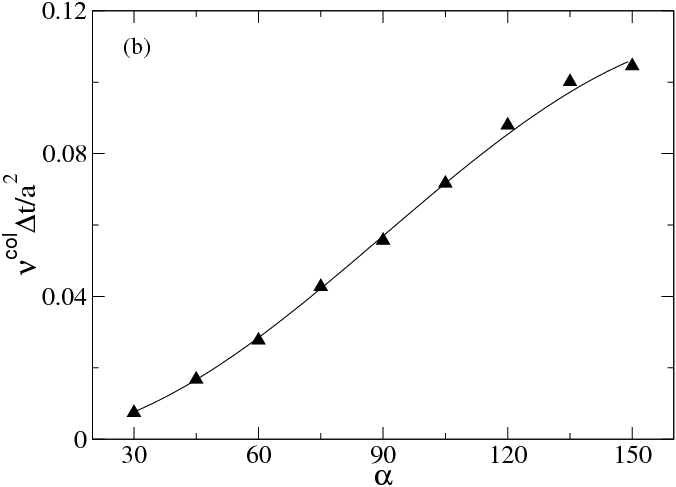}
    \end{center}
  \end{minipage}
   \hfill
\caption{(a) Normalized kinetic contribution to the viscosity,
$\nu^{kin}/(\Delta t k_BT)$, in three dimensions as a function of the
collision angle $\alpha$. Data were obtained by time averaging the
Green-Kubo relation over 75,000 iterations using $\lambda/a=2.309$ for $M=5$
($\sblacksquare$) and $M=20$ ($\lbullet$). The lines are the theoretical
prediction, Eq.~(\ref{3dkin}). Parameters: $L/a=32$, $\Delta t=1$.
From Ref.~\cite{tuzel_03_tcs}.
\hfill\break
(b) Normalized collisional contribution to the viscosity, $\nu^{col}\Delta t/
a^2$, in three dimensions as a function of the collision angle $\alpha$.
The solid line is the theoretical prediction, Eq.~(\ref{nu_rot_2df}).
Data were obtained by time averaging the Green-Kubo relation over
300,000 iterations. Parameters: $L/a=16$, $\lambda/a=0.1$, $M=3$,
and $\Delta t=1$. From Ref.~\cite{tuzel_thesis}.
}
\label{fig:viscosities}
\end{figure}

\noindent {\em Kinetic contributions:}
Kinetic contributions to the transport coefficients dominate when the
mean free path is larger than the cell size, {\em i.e.}, $\lambda>a$.
As can be seen from Eqs.~(\ref{VIS_KIN}) and (\ref{SIG_KIN}), an analytic
calculations of these contributions requires the evaluation of
time correlation functions of products of the particle velocities.
This is straightforward if one makes the basic assumption of {\em molecular
chaos} that successive collisions between particles are not correlated.
In this case, the resulting time-series in Eq.~(\ref{VIS_KIN}) is geometrical,
and can be summed analytically. The resulting expression for the shear
viscosity in two dimensions is
\begin{equation}\label{kin_vis}
\nu^{kin} = \frac{k_BT\Delta t}{2m}\left(\frac{M}{(M-1+e^{-M})\sin^2(\alpha)}
-1\right).
\end{equation}
Fluctuations in the number of particles per cell are included in
(\ref{kin_vis}). This result agrees with the non-equilibrium calculations
of Pooley {\em et al.} \cite{pool_03_the,pool_05_ktd}, measurements in
shear flow \cite{kiku_03_tcm}, and the numerical evaluation of the GK
relation in equilibrium simulations (see Fig.~\ref{fig:viscosities}).

The corresponding result in three dimensions for collision rule (\ref{3D})
is
\begin{equation}\label{3dkin}
\nu^{kin} = \frac{k_BT\Delta t}{2m}\left(\frac{5M}{(M-1+e^{-M})
[2-\cos(\alpha)-\cos(2\alpha)]}-1\right).
\end{equation}
The kinetic contribution to the stress tensor is symmetric, so that
$\nu_2^{kin}=0$ and the kinetic contribution to the shear viscosity
is $\nu^{kin}\equiv\nu_1^{kin}$.

\noindent{\em Collisional contributions:}
Explicit expressions for the collisional contributions to the
viscous transport coefficients can be obtained by considering various
choices for ${\bf \hat k}$ and $\alpha$ and $\beta$ in Eqs.~(\ref{VIS_COL}),
(\ref{SIG_COL}) and (\ref{ST}). Taking
${\bf \hat k}$ in the $y$-direction and $\alpha=\beta=1$ yields
\begin{equation}
\label{NU_COL}
\nu_1^{col}+\nu_2^{col} =
\frac{1}{\Delta t Nk_BT}\left.\sum_{t=0}^\infty\right.'
\sum_{i,j} \langle v_{ix}(0)B_{iy}(0)v_{ix}(t)B_{iy}(t)\rangle.
\end{equation}
Other choices lead to relations between the collisional
contributions to the viscous transport coefficients, namely
\begin{equation}\label{r1}
[1+(d-2)/d]\nu_1^{col}+\gamma^{col}+\kappa^{col} =
\nu_1^{col}+\nu_2^{col} .
\end{equation}
and
\begin{equation}
\label{r2}
[(d-2)/d]\nu_1^{col}-\nu_2^{col}+\gamma^{col}=0.
\end{equation}
These results imply that $\kappa^{col}=0$, and $\gamma^{col}-2\nu_1^{col}/d
=\nu_2^{col}-\nu_1^{col}$. It follows that the collision
contribution to the macroscopic viscous stress tensor is
\begin{eqnarray}
\label{ST1}
\hat\sigma^{col}_{\alpha\beta}/\rho &=& \nu_1^{col}(\partial_\beta v_\alpha +
\partial_\alpha v_\beta) + \nu_2^{col}(\partial_\beta v_\alpha -
\partial_\alpha v_\beta) +  (\nu_2^{col}-\nu_1^{col})
\delta_{\alpha\beta} \partial_\lambda v_\lambda \nonumber  \\
&=& (\nu_1^{col}+\nu_2^{col}) \partial_\beta v_\alpha +
(\nu_2^{col}-\nu_1^{col})Q_{\alpha\beta},
\end{eqnarray}
where $Q_{\alpha\beta}\equiv \delta_{\alpha\beta} \partial_\lambda v_\lambda
- \partial_\alpha v_\beta$. Since $Q_{\alpha\beta}$ has zero divergence,
$\partial_\beta Q_{\alpha\beta}=0$, the term containing $Q$ in
Eq.~(\ref{ST1}) will not appear in the linearized hydrodynamic equation
for the momentum density, so that
\begin{equation}\label{LH}
\rho\frac{\partial{\bf v}}{\partial t}
= -\nabla p + \rho(\nu^{kin}+\nu^{col})\Delta{\bf v}
+ \frac{d-2}{d}\nu^{kin}\nabla (\nabla\cdot{\bf v}),
\end{equation}
where $\nu^{col}=\nu_1^{col}+\nu_2^{col}$. In writing Eq.~(\ref{LH})
we have used the fact that the kinetic contribution to the microscopic
stress tensor, $\bar\sigma^{kin}$, is symmetric, and
$\gamma^{kin}=0$ \cite{ihle_03_srd_b}. The viscous contribution to the
sound attenuation coefficient is $\nu^{col}+2(d-1)\nu^{kin}/d$ instead
of the standard result, $2(d-1)\nu/d+\gamma$, for simple isotropic fluids.
The collisional contribution to the effective shear viscosity is
$\nu^{col}\equiv\nu_1^{col}+\nu_2^{col}$.
It is interesting to note that the kinetic theory approach discussed
in Ref.~\cite{pool_05_ktd} is able to show explicitly that
$\nu_1^{col}=\nu_2^{col}$, so that $\nu^{col}=2\nu_1^{col}$.

It is straightforward to evaluate the various contributions
to the right hand side of (\ref{NU_COL}). In particular, note that since
velocity correlation functions are only required at equal times and for a
time lag of one time step, molecular chaos can be assumed \cite{ihle_04_rgr}.
Using the relation \cite{ihle_05_ect}
\begin{equation}
\langle B_{i \alpha}(n\Delta t) B_{j \beta}(m\Delta t) \rangle={a^2 \over 12}
\, \delta_{\alpha\beta} (1+\delta_{ij})
\left[2 \delta_{n,m} - \delta_{n,m+1}-\delta_{n,m-1}  \right]\,,
\end{equation}
and averaging over the number of particles in a cell assuming that the
number of particles in any cell is Poisson distributed at each time step,
with an average number $M$ of particles per cell, one then finds
\begin{equation}
\label{VIS11}
\nu^{col} = \nu_1^{col}+\nu_2^{col}=
\frac{a^2}{6 d \Delta t} \left(\frac{M-1+e^{-M}}{M}\right)
[1-\cos(\alpha)] \label{nu_rot_2df} \;,
\end{equation}
for the SRD collision rules in both two and three dimensions.
Eq.~(\ref{VIS11}) agrees with the result of Refs.~\cite{kiku_03_tcm} and
\cite{pool_05_ktd} obtained using a completely different non-equilibrium
approach in shear flow. Simulation results for the collisional contribution
to the viscosity are in excellent agreement with this result (see
Fig.~\ref{fig:viscosities}).

\noindent {\em Thermal diffusivity and self-diffusion coefficient:} As
with the viscosity, there are both kinetic and collisional contributions
to the thermal diffusivity, $D_T$. A detailed analysis of both contributions
is given in Ref.~\cite{ihle_05_ect}, and the results are summarized in
Table 1. The self-diffusion coefficient, $D$, of particle $i$ is
defined by
\begin{equation}
D = \lim_{t\to\infty}\frac{1}{2dt}\langle[{\bf r}_i(t)-{\bf r}_i(0)]^2\rangle
= \frac{\Delta t}{d}\left.\sum_{n=0}^\infty\right.'
\langle {\bf v}_i(n \Delta t)\cdot{\bf v}_i(0)\rangle,
\end{equation}
where the second expressions is the corresponding discrete GK relation.
The self-diffusion coefficient is unique in that the collisions do not
explicitly contribute to $D$. With the assumption of molecular
chaos, the kinetic contributions are easily summed \cite{ihle_03_srd_b}
to obtain the result given in Table 1.

\subsubsection{Beyond Molecular Chaos}

The kinetic contributions to the transport coefficients presented
in Table 1 have all been derived under the assumption
of molecular chaos, i.e., that particle velocities are not correlated.
Simulation results for the shear viscosity and thermal diffusivity have
generally been found to be in good agreement with these results. However,
it is known that there are correlation effects for $\lambda/a$ smaller
than unity \cite{ripo_05_drf,ihle_06_sdp}. They arise from correlated
collisions between particles that are in the same collision cell for
more than one time step.

For the viscosity and thermal conductivity, these corrections are
generally negligible, since they are only significant in the small
$\lambda/a$ regime, where the collisional contribution to the
transport coefficients dominates. In this regard, it is important to
note that there are no correlation corrections to $\nu^{col}$ and $D_T^{col}$
\cite{ihle_05_ect}. For the self-diffusion coefficient---for which there
is no collisional contribution---correlation corrections dramatically
increase the value of this transport coefficient for $\lambda\ll a$,
see Refs.~\cite{ripo_05_drf,ihle_06_sdp}. These correlation corrections, which
arise from particles which collide with the same particles in consecutive
time steps, are distinct from the correlations effects which are
responsible for the long-time tails. This distinction is important,
since long-time tails are also visible at large mean free paths, where
these corrections are negligible.

\begin{table}
\def\arraystretch{2}
\begin{center}
\begin{tabular}{|c|c|c|c|}\hline
 {}  & {\ $d$\ } & Kinetic$~~(\times k_BT \Delta t/2m)$
& Collisional$~~(\times a^2/\Delta t)$ \\ \hline \hline
\multirow{2}{*} \ \   {$\nu$} \ \    & 2 & $\frac{M}{(M-1+e^{-M})
\sin^2(\alpha)}-1$ & \multirow{2}{*}{\ \ $\frac{\left(M-1+e^{-M}\right)}{6 d M}
                                      [1-\cos(\alpha)]$\ \ } \\
 &3 & $ \frac{5M}{(M-1+e^{-M})[2-\cos(\alpha)-\cos(2
\alpha)]}-1 $ & \\ \hline

\multirow{2}{*}    {\ \ $D_T$\ \ }         & 2 &
\multirow{2}{*} {\ \ $ \frac{d}{1-\cos(\alpha)} -1
    + \frac{2d}{M} \left[\frac{7-d}{5} - \frac{1}{4} \,{\csc^2
(\alpha/2)} \right]$\ \ }
 &  \multirow{2}{*}{$\frac{(1-1/M)}{3 (d+2)M }
 [1-\cos(\alpha)]
$}    \\
& 3 &  & \\ \hline
\multirow{2}{*}{$D$} & 2 &
\multirow{2}{*}{$\frac{dM}{(1-\cos(\alpha))(M-1+e^{-M})}
 -1$} & \multirow{2}{*}{-} \\
 {} & 3 &  & \\
\hline
\end{tabular}
\end{center}
\label{transport_coef}
\caption{Theoretical expressions for the kinematic shear viscosity $\nu$,
the thermal diffusivity, $D_T$, and the self-diffusion coefficient, $D$,
in both two ($d=2$) and three ($d=3$) dimensions. $M$ is the average
number of particles per cell, $\alpha$ is the collision angle, $k_B$
is Boltzmann's constant, $T$ is the temperature, $\Delta t$ is the time
step, $m$ is the particle mass, and $a $ is the cell size. Except for
self-diffusion constant, for which there is no collisional contribution,
both the kinetic and collisional contributions are listed. The
expressions for shear viscosity and self-diffusion coefficient
include the effect of fluctuations in the number of particles per
cell; however, for brevity, the relations for thermal diffusivity
are correct only up to $O(1/M)$ and $O(1/M^2)$ for the kinetic and
collisional contributions, respectively. For the complete expressions,
see Refs.~\cite{ihle_05_ect,tuzel_03_tcs,tuzel_thesis}.}
\end{table}

\subsection{Dynamic Structure Factor}
\label{SF}

Spontaneous thermal fluctuations of the density, $\rho({\bf r},t)$, the
momentum density, ${\bf g}({\bf r},t)$, and the energy density,
$\epsilon({\bf r},t)$,  are dynamically coupled, and an analysis of their
dynamic correlations in the limit of small wave numbers and frequencies
can be used to measure a fluid's transport coefficients. In particular,
because it is easily measured in dynamic light scattering, x-ray, and
neutron scattering experiments, the Fourier transform of the
density-density correlation function---the dynamics structure
factor---is one of the most widely used
vehicles for probing the dynamic and transport properties of
liquids \cite{berne_00_dls}.

A detailed analysis of equilibrium dynamic correlation functions---the
dynamic structure factor as well as the vorticity and entropy-density
correlation functions---using the SRD algorithm is presented in
Ref.~\cite{tuzel_06_dcs}. The results---which are in good agreement
with earlier
numerical measurements and theoretical predictions---provided further
evidence that the analytic expressions or the transport coefficients  are
accurate and that we have an excellent understanding of the SRD
algorithm at the kinetic level.

Here, we briefly summarize the results for the dynamic structure factor.
The dynamic structure exhibits three peaks, a central ``Rayleigh'' peak
caused by the thermal diffusion, and two symmetrically placed
``Brillouin peaks'' caused by sound. The width of the central peak
is determined by the thermal diffusivity, $D_T$, while that of the
two Brillouin peaks is related to the sound attenuation coefficient,
$\Gamma$. For the SRD algorithm \cite{tuzel_06_dcs},
\begin{equation}
\Gamma = D_T\left(\frac{c_p}{c_v}-1\right)+2\left(\frac{d-1}{d}\right)
\nu^{kin} + \nu^{col}.
\end{equation}
Note that in two-dimensions, the sound attenuation coefficient for a
SRD fluid has the same functional dependence on $D_T$ and
$\nu=\nu^{kin}+\nu^{col}$ as an isotropic fluid with an ideal-gas
equation of state (for which $\gamma=0$).

\begin{figure}[h]
    \begin{center}
      \includegraphics*[scale=0.22]{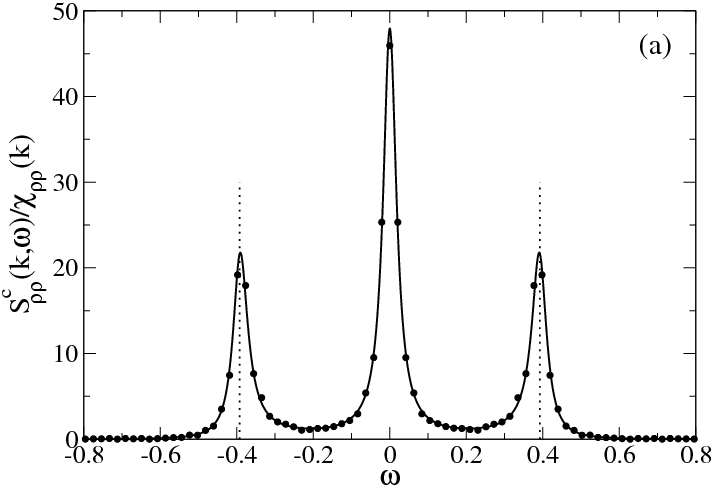}
      \includegraphics*[scale=0.22]{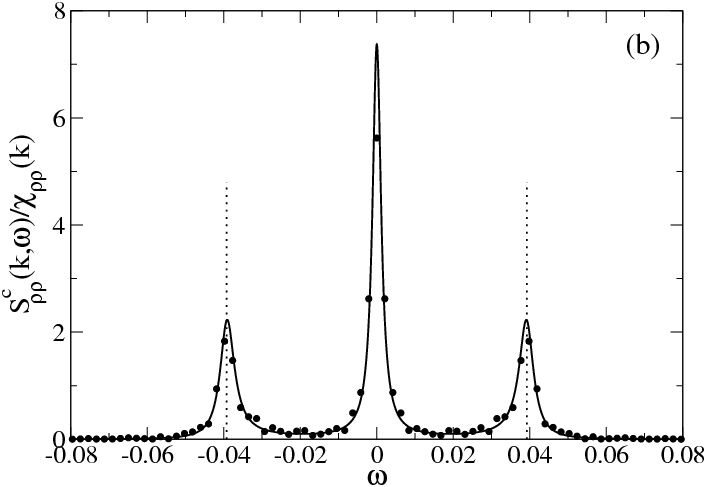}
    \end{center}
\caption{Normalized dynamic structure, $S^c_{\rho\rho}(k\omega)/
\chi_{\rho\rho}(k)$, for ${\bf k} = 2\pi(1,1)/L$ and (a)
$\lambda/a=1.0$ with $\alpha=120^\circ$, and (b) $\lambda/a=0.1$ with
$\alpha=60^\circ$. The solid lines are the theoretical prediction for
the dynamic structure factor (see Eq.~(36) of Ref.~\cite{tuzel_06_dcs})
using values for the transport coefficients obtained using the expressions
in Table 1. The dotted lines show the predicted
positions of the Brillouin peaks, $\omega=\pm ck$, with
$c=\sqrt{2k_BT/m}$. Parameters: $L/a=32$, $M=15$, and $\Delta t=1.0$.
From Ref.~\cite{tuzel_06_dcs}.
}
\end{figure}

Simulation results for the structure factor in two-dimensions with
$\lambda/a=1.0$ and collision angle $\alpha=120^\circ$, and
$\lambda/a=0.1$ with collision angle $\alpha=60^\circ$ are
shown in Figs.~2a and 2b, respectively. The solid lines are the
theoretical prediction for the dynamic structure factor (see
Eq.~(36) of Ref.~\cite{tuzel_06_dcs}) using $c=\sqrt{2k_BT/m}$ and values
for the transport coefficients obtained using the expressions in
Table 1, assuming that the bulk viscosity $\gamma=0$.
As can be seen, the agreement is excellent.

\section{Non-Equilibrium Calculations of Transport Coefficients}
\label{sec:NECTC}

MPC transport coefficients have also be evaluated by calculating the linear
response of the system to imposed gradients. This approach was introduced
by Kikuchi {\it et al.} \cite{kiku_03_tcm} for the shear viscosity and
then extended and refined in Ref.~\cite{pool_05_ktd} to determine the
thermal diffusivity and bulk viscosity. Here, we summarize the derivation of
the shear viscosity.

\subsection{Shear Viscosity of SRD: Kinetic Contribution}
\label{sec:shear_SRD_kin}

Linear response theory provides an alternative, and complementary,
approach for evaluating the shear viscosity. This non-equilibrium approach
is related to equilibrium calculations described in the previous section
through the fluctuation-dissipation theorem. Both methods yield identical
results. For the more complicated analysis of the hydrodynamic equations,
the stress tensor, and the longitudinal transport coefficients such as
the thermal conductivity, the reader is referred to Ref.~\cite{pool_05_ktd}.

Following Kikuchi {\it et al.} \cite{kiku_03_tcm},
we consider a two-dimensional liquid with an imposed shear
$\dot{\gamma}=\partial u_x(y)/\partial y$. On average, the velocity
profile is given by ${\bf v}=(\dot{\gamma}y,0)$.
The dynamic shear viscosity $\eta$ is the proportionality
constant between the velocity gradient $\dot{\gamma}$
and the frictional force acting on a plane perpendicular to $y$; i.e.
\begin{equation}
\label{VISC_DEF1}
\sigma_{xy}=\eta \dot{\gamma} ,
\end{equation}
where $\sigma_{xy}$ is the off-diagonal element of the viscous stress
tensor. During the streaming step, particles
will cross this plane only if $\vert v_y\,\Delta t\vert$ is greater
than the distance to the plane. Assuming that the fluid particles
are homogeneously distributed, the momentum flux is obtained by integrating
over the coordinates and velocities of all particles that cross the plane
from above and below during the time step $\Delta t$.
The result is \cite{kiku_03_tcm}
\begin{equation}
\label{STRESS_KIN1}
\sigma_{xy}= \rho \left({\dot{\gamma} \Delta t \over 2}
\,\langle v_y^2\rangle- \langle v_x v_y\rangle\right)\,,
\end{equation}
where the mass density $\rho=mM/a^d$, and the averages are taken over the
steady-state distribution $P(v_x-\dot{\gamma}y,v_y)$. It is important to
note that this is {\em not} the Maxwell-Boltzmann distribution, since
we are in a non-equilibrium steady state where the shear has induced
correlations between $v_x$ and $v_y$. As a consequence,
$\langle v_x v_y\rangle$ is nonzero.
To determine the behavior of $\langle v_x v_y\rangle$, the
effect of streaming and collisions are calculated separately.
During streaming, particles which arrive at $y_0$ with positive velocity
$v_y$ have started from $y_0-v_y\, \Delta t$; these particles bring a velocity
component $v_x$ which is smaller than that of particles originally located
at $y_0$. On the other hand, particles starting out at $y>y_0$ with negative
$v_y$ bring a larger $v_x$. The velocity distribution is therefore sheared
by the streaming, so that
$P^{after}(v_x,v_y)=P^{before}(v_x+\dot{\gamma}v_y \Delta t,v_y)$.
Averaging $v_xv_y$ over this distribution gives \cite{kiku_03_tcm}
\begin{equation}
\label{CORREL_KIN1}
\langle v_x v_y\rangle^{after}=\langle v_x v_y\rangle-\dot{\gamma}\Delta t
\langle v_y^2\rangle\, ,
\end{equation}
where the superscript denotes the quantity {\it after} streaming.
The streaming step therefore reduces correlations by
$-\dot{\gamma}\Delta t\langle v_y^2\rangle$, making $v_x$ and $v_y$
increasingly anti-correlated.

The collision step redistributes momentum between particles and tends
to reduce correlations. Making the assumption of molecular chaos, i.e.,
that is that the velocities of different particles are uncorrelated,
and averaging over the two possible rotation directions, one finds,
\begin{equation}
\label{CORREL_KIN2}
\langle v_x v_y\rangle^{after}=\left[1-{N_c-1\over N_c}
[1-\cos(2\alpha)]\right]\,\langle v_x v_y\rangle^{before}
\end{equation}
The number of particles in a cell, $N_c$ is not constant, and density
fluctuations have to be included. The probability to find $n$ uncorrelated
particles in a given cell is given by the Poisson distribution,
$w(n)=\exp(-M) M^n/n!$; the
probability of a given particle being in a cell together with $n-1$
others is $nw(n)/M$. Taking an average over this distribution gives
\begin{equation}\label{CORREL_KIN3}
\langle v_x v_y\rangle^{after} = f\, \langle v_x v_y\rangle^{before},
\end{equation}
with
\begin{equation}\label{CORREL_KIN5}
f= \left[1-\frac{M-1+\exp(-M)}{M}[1-\cos(2\alpha)]\right] .
\end{equation}
The difference between this result and just replacing $N_c$ by $M$ in
Eq.~(\ref{CORREL_KIN2}) is small, and only important for
$M\leq 3$. One sees that
$\langle v_x v_y\rangle$ is first modified by streaming
and then multiplied by a factor $f$ in the subsequent collision step.
In the steady state, it therefore oscillates between two values.
Using Eqs.~(\ref{CORREL_KIN1}), (\ref{CORREL_KIN3}), and (\ref{CORREL_KIN5}),
we obtain the self-consistency condition
$(\langle v_x v_y\rangle-\dot{\gamma}\Delta t\langle v_y^2\rangle)f=
\langle v_x v_y\rangle$. Solving for $\langle v_x v_y\rangle$,
assuming equipartition of energy, $\langle v_y^2 \rangle=k_B T/m$, and
substituting into (\ref{STRESS_KIN1}), we have
\begin{equation}
\label{CORREL_KIN4}
\sigma_{xy}={\dot{\gamma}\,M\Delta t k_B T\over m}
\left({1\over 2}+{f\over 1-f} \right)\, ,
\end{equation}
Inserting this result into the definition of the viscosity, (\ref{VISC_DEF1}),
yields the same expression for the kinetic viscosity in two-dimensional
as obtained by the equilibrium Green-Kubo approach discussed in
Sec.~\ref{sec:ee}.

\subsection{Shear Viscosity of SRD: Collisional Contribution}
\label{sec:shear_SRD_coll}

The collisional contribution to the shear viscosity is proportional to
$a^2/\Delta t$; as discussed in Sec.~\ref{sec:QSF}, it results from the
momentum transfer between particles in a cell of size $a$ during the
collision step. Consider again a collision cell of linear dimension
$a$ with a shear flow $u_x(y)=\dot\gamma y$.
Since the collisions occur in a shifted grid, they cause a transfer of
momentum between neighboring cells of the original unshifted reference frame
\cite{ihle_01_srd,ihle_03_srd_b}. Consider now the
momentum transfer due to collisions across the line
$y=h$, the coordinate of a cell boundary in the unshifted frame.
If we assume a homogeneous distribution of particles in the collision cell,
the mean velocities in the upper ($y>h$) and lower partitions are
\begin{equation}
{\bf u}_1=\frac{1}{M_1}\,\sum_{i=1}^{M_1}\,{\bf v}_i\ \ \ \ {\rm and}\ \ \ \
{\bf u}_2=\frac{1}{M_2}\,\sum_{i=M_1+1}^M\,{\bf v}_i\,,
\end{equation}
respectively, where $M_1=M(a-h)/a$ and $M_2=Mh/a$. Collisions transfer
momentum between the two parts of the cell. The $x$-component of the momentum
transfer is
\begin{equation}
\Delta p_x(h)\equiv \sum_{i=1}^{M_1}\,[v_{ix}^{after}-v_{ix}^{before}] \;\;.
\end{equation}
The use of the rotation rule (\ref{collide}) together with an average over
the sign of the stochastic rotation angle yields
\begin{equation}
\Delta p_x(h)=[\cos(\alpha)-1] M_1 (u_{1x}-u_x) \;\;.
\end{equation}
Since $M{\bf u} = M_1{\bf u}_1 + M_2{\bf u}_2$,
\begin{equation}
\Delta p_x(h)=[1-\cos(\alpha)]\,M\,(u_{2x}-u_{1x})\,\,
\frac{h}{a}\,\left(1-\frac{h}{a}\right) \;\;.
\end{equation}
Averaging over the position $h$ of the dividing line, which corresponds to
averaging over the random shift, we find
\begin{equation}\label{chap4:appendix:adpx}
\langle\Delta p_x\rangle= \frac{1}{a}\int_{0}^{a} \Delta p_{x}(h) dh
= \frac{1}{6}[1-\cos(\alpha)]M(u_{2x}-u_{1x}).
\end{equation}
Since the dynamic viscosity $\eta$ is defined as the ratio of the tangential
stress, $P_{yx}$, to $\partial u_x/\partial y$, we have
\begin{equation}\label{etaap}
\eta = \frac{\langle\Delta p_x\rangle/(a^2\Delta t)}
             {\partial u_x/\partial y} =
       \frac{\langle\Delta p_x\rangle/(a^2\Delta t)} {(u_{2x}-u_{1x})/(a/2)},
\end{equation}
so that the kinematic viscosity, $\nu=\eta/\rho$, in two-dimensions for
SRD is
\begin{equation}
\label{rotviscosity}
\nu^{col}=\frac{a^2}{12 \Delta t}[1-\cos(\alpha)]
\end{equation}
in the limit of small mean free path.  Since we have neglected the
fluctuations in the particle number, this expression corresponds to the
limit $M\rightarrow \infty$. Even though this derivation is somewhat
heuristic, it gives a remarkably accurate expression; in particular, it
contains the correct dependence on the cell size,
$a$, and the time step, $\Delta t$, in the limit of small free path,
\begin{equation}
\nu^{col} = \frac{a^2}{\Delta t} f_{col}(d,M,\alpha),
\end{equation}
as expected from simple random walk arguments.
Kikuchi {\em et al.} \cite{kiku_03_tcm} included particle number fluctuations
and obtained identical results for the collisional contribution to the viscosity
as was obtained in the Green-Kubo approach (see Table 1).

\subsection{Shear Viscosity of MPC-AT}

For MPC-AT, the viscosities have been calculated in Ref.~\cite{nogu_07_pmh}
using the methods described
in Secs.~\ref{sec:shear_SRD_kin} and \ref{sec:shear_SRD_coll}. The total
viscosity of MPC-AT is given by the sum of two terms, the collisional and
kinetic contributions. For MPC-AT-a, it was found
for both two and three dimensions that \cite{nogu_07_pmh}
\begin{eqnarray}
\nonumber
\nu^{kin}&=&{k_B T\Delta t\over m }\left( {M\over M-1+{\rm e}^{-M}}-{1\over 2}
\right) \ \ \ \ \ \ {\rm and} \\
\label{mpc_at-a1}
\nu^{col}&=&{a^2\over 12 \Delta t}\left( {M-1+{\rm e}^{-M}\over M}\right).
\end{eqnarray}
The exponential terms ${\rm e}^{-M}$ are due to the fluctuation of
the particle number per cell and become important for $M\le 3$.
As was the case for SRD, the kinetic viscosity has no anti-symmetric
component; the collisional contribution, however, does. Again, as discussed
in Sec.~\ref{sec:ee} for SRD, one finds $\nu_1^{col}=\nu_2^{col}=\nu^{col}/2$.
This relation is true for all $-a$ versions of MPC discussed in
Refs.~\cite{gg:gomp07f,nogu_07_pmh,nogu_08_ang}. Simulation results were found
to be in good agreement with theory.

For MPC-AT+a it was found for sufficiently large $M$
that \cite{gotz_07_ram,nogu_08_ang}
\begin{eqnarray}
\nonumber
\nu^{kin}&=&{k_B T\Delta t\over m }
\left( {M\over M-(d+2)/4}-{1\over 2}\right), \\
\label{mpc_at+a1}
\nu^{col}&=&{a^2\over 24 \Delta t}
\left( {M-7/5\over M}\right).
\end{eqnarray}
MPC-AT-a and MPC-AT+a both have the same kinetic contribution to the
viscosity in two
dimensions; however, imposing angular-momentum conservation makes the
collisional contribution to the stress tensor symmetric, so that the
asymmetric contribution, $\nu_2$, discussed in Sec.~\ref{sec:ee} vanishes.
The resulting collisional contribution to the viscosity is then reduced by
a factor close to two.

\section{Generalized MPC Algorithms for Dense Liquids and Binary Mixtures}
\label{sec:GMPC}

The original SRD algorithm models a single-component fluid with an
ideal-gas equation of state.
The fluid is therefore very compressible, and the speed of sound, $c_s$,
is low. In order to have negligible compressibility effects, as in real
liquids, the Mach number has to be kept small, which means that there are
limits on the flow velocity in the simulation. The SRD algorithm can be
modified to model both excluded volume effects, allowing for a more
realistic modeling of dense gases and liquids, as well as repulsive
hard-core interactions between components in mixtures, which
allow for a thermodynamically consistent modeling of phase separating
mixtures.

\subsection{Non-Ideal Model}

\begin{figure}[h]
  \begin{minipage}[t]{.30\textheight}
    \begin{center}
      \includegraphics*[scale=0.28]{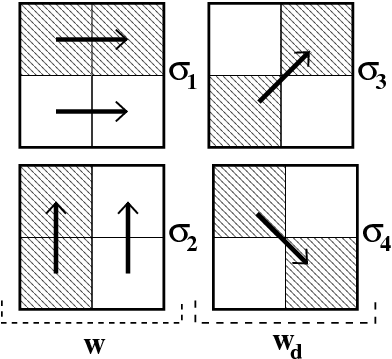}
    \end{center}
  \end{minipage}
  \hfill
  \begin{minipage}[t]{.30\textheight}
    \begin{center}
      \includegraphics*[scale=0.22]{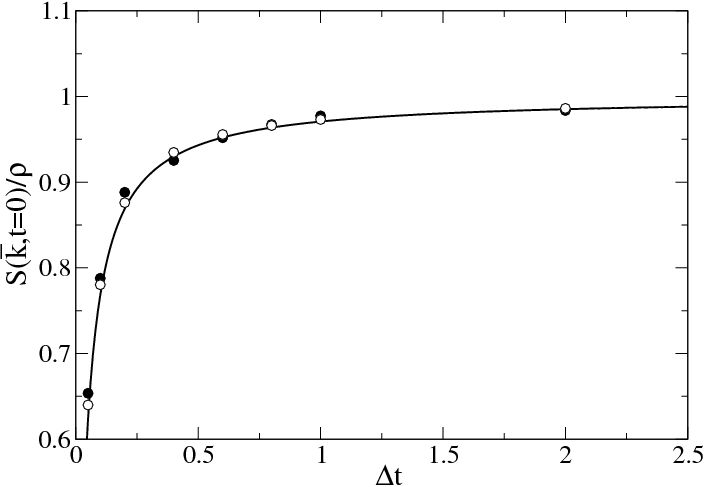}
    \end{center}
  \end{minipage}
   \hfill
\caption{
(left panel) Schematic of collision rules. Momentum is exchanged in three
ways: a) horizontally along $\bsigma_1$, b) vertically along $\bsigma_2$,
and c) diagonally along $\bsigma_3$ and $\bsigma_4$.
$w$ and $w_d$ denote the probabilities of choosing collisions a), b), and
c) respectively. \hfill\break
(right panel) Static structure factor $S(\bar k,t=0)$ as a function of
$\Delta t$ for $M=3$. The open circles ($\circ$) show results obtained by
taking the numerical derivative of the pressure. The bullets ($\bullet$)
are data obtained from direct measurements of the density fluctuations.
The solid line is the theoretical prediction obtained using the first
term in Eq.~(\ref{PRESS_MAV}) and Eq.~(\ref{S_THERMO}). $\bar k$ is the
smallest wave vector, $\bar k=(2\pi/L)(1,0)$. Parameters: $L/a=32$,
$A=1/60$, and $k_BT=1.0$.
From Ref.~\cite{ihle_06_cpa}.
}
\label{Nonideal}
\end{figure}

As in SRD, the algorithm consists of individual streaming and
collision steps. In order to define the collisions, a second grid
with sides of length $2a$ is introduced, which (in $d=2$) groups four
adjacent cells into one ``supercell''. The cell structure is sketched in
Fig.~\ref{Nonideal} (left panel). To initiate a collision, pairs of
cells in every supercell are chosen at random. Three different choices are
possible: a) horizontal (with $\bsigma_1=\hat x$), b) vertical ($\bsigma_2
=\hat y$), and c) diagonal collisions (with $\bsigma_3=
(\hat x+\hat y)/\sqrt{2}$ and $\bsigma_4=(\hat x-\hat y)/\sqrt{2}$).
%
%

For a mean particle velocity
${\bf u}_n=(1/M_n)\,\sum_{i=1}^{M_n}\,{\bf v}_i$, of cell $n$, the
projection of the difference of the mean velocities of the selected cell
pairs on
${\bf \sigma}_j$, $\Delta u={\bf \sigma}_j\cdot ({\bf u}_1-{\bf u}_2)$,
is then used to determine the probability of collision.
If $\Delta u<0$, no collision will be performed. For positive $\Delta u$, a
collision will occur with an acceptance probability, $p_A$,  which depends on
$\Delta u$ and the number of particles in the two cells, $N_1$ and $N_2$.
The choice of $p_A$ determines both the equation of state and the values
of the transport coefficients. While there is considerable freedom in choosing
$p_A$, the requirement of thermodynamic consistency imposes certain
restrictions \cite{ihle_06_cpa,ihle_06_sdp,tuzel_06_ctc}. One possible choice
is
\begin{equation}
\label{NONID0}
p_A(M_1,M_2,\Delta u)=\Theta(\Delta u)\,\,{\rm tanh}(\Lambda)
 \ \ \ \ {\rm with}\ \ \ \ \Lambda = A\,\Delta u\, N_1N_2 ,
\end{equation}
where $\Theta$ is the unit step function and $A$ is a parameter which
is used to tune the equation of state. The choice $\Lambda\sim N_1N_2$
leads to a non-ideal contribution to the pressure which is quadratic
in the particle density.

The collision rule chosen in Ref.~\cite{ihle_06_cpa} maximizes the momentum
transfer parallel to the connecting vector $\bsigma_j$ and does not change
the transverse momentum. It exchanges the parallel component of the mean
velocities of the two cells, which is equivalent to a ``reflection'' of
the relative velocities,
$v_i^{\Vert}(t+\Delta t)-u^{\Vert}=-(v_i^{\Vert}(t)-u^{\Vert})$,
where  $u^{\Vert}$ is the parallel component of the mean velocity of the
particles of {\it both} cells. This rule conserves momentum and energy in
the cell pairs.

Because of $x-y$ symmetry, the probabilities for choosing cell pairs in the
$x$- and $y$- directions (with unit vectors $\bsigma_1$ and $\bsigma_2$
in Fig.~\ref{Nonideal}) are equal, and will be denoted by $w$.
The probability for choosing diagonal pairs ($\bsigma_3$ and $\bsigma_4$ in
Fig.~\ref{Nonideal}) is given by $w_d=1-2w$. $w$ and $w_d$ must be chosen
so that the hydrodynamic equations are isotropic and do not depend on
the orientation of the underlying grid. An equivalent criterion is to
guarantee that the relaxation of the velocity distribution is isotropic.
These conditions require $w=1/4$ and $w_d=1/2$.
This particular choice also ensures that the kinetic part
of the viscous stress tensor is isotropic \cite{tuze_07_mmf}.

\subsubsection{Transport Coefficients}

The transport coefficients can be determined using the same Green-Kubo
formalism as was used for the original SRD algorithm
\cite{ihle_01_srd,ihle_04_rgr}. Alternatively, the non-equilibrium
approach describe in Sec.~\ref{sec:NECTC} can be used.
Assuming molecular chaos and ignoring fluctuations in the number of
particles per cell, the kinetic contribution to the viscosity is found to
be
\begin{equation}
\label{EXACT_NON_VIS}
\nu^{kin}={k_B T\over m}\,\Delta t\left( {1\over p_{col}}
-{1 \over 2} \right)\, \ \ \ {\rm with}\ \ \
p_{col} = A\,\sqrt{k_B T\over m \pi}\;M^{3/2},
\end{equation}
which is in good agreement with simulation data. $p_{col}$ is essentially
the collision rate,
and can be obtained by averaging the acceptance probability,
Eq.~(\ref{NONID0}). The collisional contribution to the viscosity is
$\nu^{col} = p_{col} (a^2/3\Delta t)$ \cite{ihle_08}.
The self-diffusion constant, $D$, is evaluated by summing over the
velocity-autocorrelation function (see, e.g. Ref.~\cite{ihle_01_srd});
which yields $D=\nu_{kin}$.

\subsubsection{Equation of State}

The collision rules conserve the kinetic energy, so that the internal energy
should be the same as that of an ideal gas. Thermodynamic consistency
therefore requires that the non-ideal contribution to the pressure is linear
in $T$. This is possible if the coefficient $A$ in Eq.~(\ref{NONID0})
is sufficiently small.

The mechanical definition of pressure---the average longitudinal momentum
transfer across a fixed interface per unit time and unit surface
area---can be used to determine the equation of state. Only the momentum
transfer due to collisions needs to be considered, since that coming from
streaming
constitutes the ideal part of the pressure. Performing this calculation
for a fixed interface and averaging over the position of the interface,
one finds the non-ideal part of the pressure,
\begin{equation}
\label{PRESS_MAV}
P_n=\left({1 \over 2 \sqrt{2}}+{1 \over 4} \right) {A\,M^2 \over 2}
{k_B T \over a \Delta t} + O(A^3T^2) .
\end{equation}
$P_n$ is quadratic in the particle density, $\rho=M/a^2$,
as it would be expected from a virial expansion. The prefactor $A$
must be chosen small enough that higher order terms in this expansion are
negligible. Prefactors $A$ leading to acceptance rates of
about $15\%$ are sufficiently small to guarantee that the pressure is
linear in $T$.

The total pressure is the average of the diagonal part of the
microscopic stress tensor,
\begin{equation}
\label{VIRIAL}
P=P_{id}+P_n = \frac{1}{\Delta t L_x\,L_y}\left\langle \sum_j \{
\Delta t v_{jx}^2 - \Delta v_{jx}\, z^s_{jlx}/2 \} \right\rangle .
\end{equation}
The first term gives the ideal part of the pressure, $P_{id}$, as discussed
in Ref.~\cite{ihle_01_srd}.
The average of the second term is the non-ideal part of the pressure, $P_n$.
${\bf z}^s_{jl}$ is a vector which indexes collision partners. The first
subscript denotes the particle number and the second, $l$,
is the index of the collision vectors $\bsigma_l$ in Fig.~\ref{Nonideal}
(left panel). The components of ${\bf z}^s_{jl}$ are either $0$, $1$, or $-1$
\cite{ihle_06_sdp}.
Simulation results for $P_n$ obtained using Eq.~(\ref{VIRIAL}) are in good
agreement with the analytical expression, Eq.~(\ref{PRESS_MAV}).
In addition, measurements of the static structure factor $S(k\to0,t=0)$ agree
with the thermodynamic prediction
\begin{equation}\label{S_THERMO}
S(k\to0,t=0) = \rho k_BT \partial\rho/ \partial P\vert_T
\end{equation}
when result (\ref{PRESS_MAV}) is used [see Fig.~\ref{Nonideal} (right panel)].
The adiabatic speed of sound
obtained from simulations of the dynamic structure factor is also
in good agreement with the predictions following from Eq.~(\ref{PRESS_MAV}).
These results provide strong evidence for the thermodynamic consistency of the
model. Consistency checks are particularly important because the
non-ideal algorithm does not conserve phase-space volume. This is because the
collision probability depends on the difference of collision-cell velocities,
so that two different states can be mapped onto the same state by a collision.
While the dynamics presumably still obeys detailed---or at least
semi-detailed---balance, this is very hard to prove, since it would
require knowledge not only of the transition probabilities, but also of
the probabilities of the
individual equilibrium states. Nonetheless, no inconsistencies due to the
absence of time-reversal invariance or a possible violation of detailed
balance have been observed.

The structure of $S(k)$ for this model is also very similar to that of a
simple dense fluid. In particular, for fixed $M$, both the depth of the
minimum at small $k$ and the height of the first peak increase with decreasing
$\Delta t$, until there is an order-disorder transition. The four-fold
symmetry of the resulting ordered state---in which clusters of particles are
concentrated at sites with the periodicity close, but not necessarily
equal, to that of the underlying grid---is
clearly dictated by the structure of the collision cells. Nevertheless,
these ordered structures are similar to the low-temperature phase of
particles with a strong repulsion at intermediate distances, but a soft
repulsion at short distances.
The scaling behavior of both the self-diffusion constant and
the pressure persists until the order/disorder transition.

\subsection{Phase-Separating Multi-Component Mixtures}

In a binary mixture of A and B particles, phase separation can occur
when there is an effective repulsion between A-B pairs. In the current
model, this is achieved by introducing velocity-dependent multi-particle
collisions between A and B particles. There are $N_A$ and $N_B$ particles
of type A and B, respectively. In two dimensions, the system is
coarse-grained into $(L/a)^2$ cells of a square lattice of linear dimension
$L$ and lattice constant $a$. The generalization to three dimensions
is straightforward.

Collisions are defined in the same way as in the non-ideal model discussed
in the previous section. Now, however, two types of collisions
are possible for each pair of cells: particles of
type A in the first cell can undergo a collision with particles of type B
in the second cell; vice versa, particles of type B in the first cell can
undergo a collision with particles of type A in the second cell. There are
no A-A or B-B collisions, so that there is an effective repulsion between
A-B pairs. The rules and probabilities for these collisions are chosen in
the same way as in the non-ideal single-component fluid described in
Refs.~\cite{ihle_06_cpa,ihle_06_sdp}. For example, consider the collision
of A particles in the first cell with the B particles in the second.
The mean particle velocity of A particles in the first cell is
${\bf u}_A=(1/N_{c,A})\,\sum_{i=1}^{N_{c,A}}\,{\bf v}_i$,
where the sum runs over all A particles, $N_{c,A}$, in the first cell.
Similarly, ${\bf u}_B=(1/N_{c,B})\,\sum_{i=1}^{N_{c,B}}\,{\bf v}_i$ is the mean
velocity of B particles in the second cell. The projection of the
difference of the mean velocities of the selected cell-pairs on
${\bf \sigma}_j$, $\Delta u_{AB}={\bf \sigma}_j\cdot ({\bf u}_A-{\bf u}_B)$,
is then used to determine the probability of collision.
If $\Delta u_{AB}<0$, no collision will be performed. For positive
$\Delta u_{AB}$, a collision will occur with an acceptance probability
\begin{equation}
\label{NONID1}
p_A(N_{c,A},N_{c,B},\Delta u_{AB})=A\,\Delta u_{AB}\,\Theta(\Delta u_{AB})\, N_{c,A} N_{c,B}\, ,
\end{equation}
where $\Theta$ is the unit step function and $A$ is a parameter which
allows us to tune the equation of state; in order to ensure thermodynamic
consistency, it must be sufficiently small that
that $p_A<1$ for essentially all collisions. When a collision occurs,
the parallel component of the mean
velocities of colliding particles in the two cells,
$v_i^{\Vert}(t+\Delta t)-u_{AB}^{\Vert}=-(v_i^{\Vert}(t)-u_{AB}^{\Vert}) $,
is exchanged,
where  $u_{AB}^{\Vert}=(N_{c,A}u^{\Vert}_A+N_{c,B}u^{\Vert}_B)/(N_{c,A}+N_{c,B})$ is
the parallel component of the mean velocity of the colliding
particles. The perpendicular component remains
unchanged. It is easy to verify that these rules conserve momentum and
energy in the cell pairs. The collision of B particles in the
first cell with A particles in the second is handled in a similar
fashion.

Because there are no A-A and B-B collisions, additional SRD collisions
at the cell level are incorporated in order to mix particle momenta.
The order of A-B and SRD collision is random,
i.e., the SRD collision is performed first with a probability $1/2$.
If necessary, the viscosity can be tuned by not performing
SRD collisions every time step. The results presented here
were obtained using a SRD collision angle of $\alpha=90^\circ$.

The transport coefficients can be calculated in the same way as for the
one-component non-ideal system. The resulting kinetic contribution to the
viscosity is
\begin{equation}
\nu^{kin} = \frac{\Delta t k_BT}{2}\left(
\frac{1}{A}\sqrt{\frac{2\pi}{k_BT}}[M_AM_B(M_A+M_B)]^{-1/2}-1\right)\,,
\end{equation}
where $M_A=\langle N_{c,A}\rangle$, $M_B=\langle N_{c,B} \rangle$.
In deep quenches, the concentration of the minority component is very
small, and the non-ideal contribution to the viscosity approaches zero.
In this case, the SRD collisions provide the dominant contribution to
the viscosity.

\subsubsection{Free Energy}

An analytic expression for the equation of state of this model can be derived
by calculating the momentum transfer across a fixed surface, in much the
same way as was done for the non-ideal model in Ref.~\cite{ihle_06_cpa}.
Since there are only non-ideal collisions between A-B particles, the resulting
contribution to the pressure is
\begin{equation}
\label{PTH}
P_n = \left(w+\frac{w_d}{\sqrt{2}}\right)AM_AM_B \frac{k_BT}{a\Delta t} =
\Gamma\rho_A\rho_B,
\end{equation}
where $\rho_A$ and $\rho_B$ are the densities of A and B and
$\Gamma\equiv(w+w_d/\sqrt{2})a^3A/\Delta t$.
In simulations, the total pressure can be measured by taking the ensemble
average of the diagonal components of the microscopic stress tensor. In
this way, the pressure can be measured locally, at the cell level. In
particular, the pressure in a region consisting of $N_{cell}$ cells is
\begin{equation}
\label{VSP}
P_n = \frac{1}{\Delta t a^2 N_{cell}}\left\langle\sum_{c=1}^{N_c}\sum_{j\in c}
[{\Delta t v_{jx}^2 - \Delta v_{jx}z^s_{jlx}/2}]\right\rangle,
\end{equation}
where the second sum runs over the particles in cell $c$.
The first term in Eq.~(\ref{VSP}) is the ideal-gas contribution to the
pressure; the second comes from the momentum transfer between cells
involved in the collision indexed by ${\bf z}^s_{jl}$ \cite{tuze_07_mmf}.
\begin{figure}[h]
  \begin{minipage}[t]{.47\linewidth}
    \begin{center}
      \includegraphics*[scale=0.193]{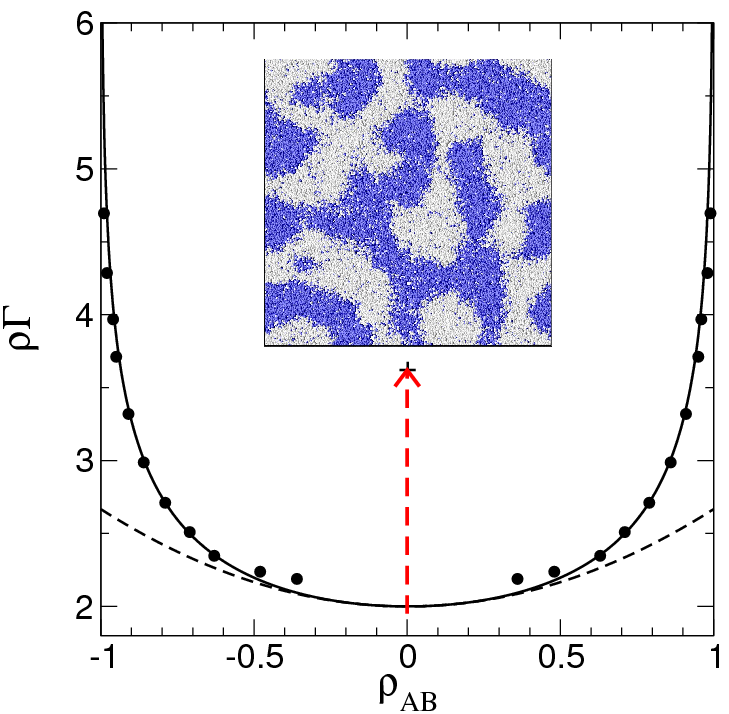}
    \end{center}
  \end{minipage}
  \hfill
  \begin{minipage}[t]{.47\linewidth}
    \begin{center}
      \includegraphics*[scale=0.21]{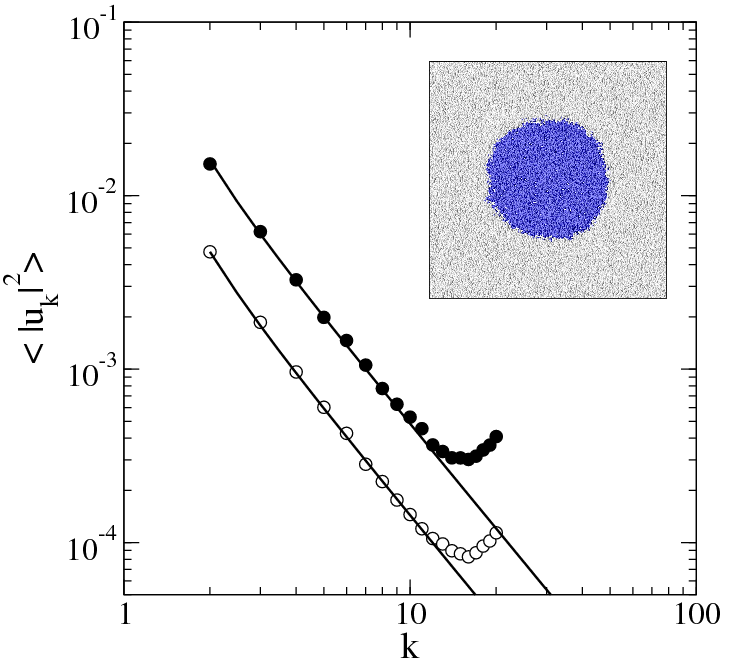}
    \end{center}
  \end{minipage}
  \hfill
\caption{(left panel) Binary phase diagram. There is
phase separation for $\rho\Gamma>2$. Simulation results
for $\rho_{AB}$ obtained from concentration histograms
are shown as bullets. The dashed line is a plot of the
leading singular behavior, $\rho_{AB}=\sqrt{3(\rho\Gamma-2)/2}$,
of the order parameter at the critical point.
The inset shows
a configuration 50,000 time steps after a quench along
$\rho_{AB}=0$ to $\rho\Gamma=3.62$ (arrow). The dark
(blue) and light (white) spheres are A and B particles,
respectively. Parameters: $L/a=64$, $M_A=M_B=5$,
$k_BT=0.0004$, $\Delta t=1$, and $a=1$. From Ref.~\cite{tuze_07_mmf}.
\hfill\break
(right panel) Dimensionless radial fluctuations,
$\langle \vert u_k^2 \vert\rangle$, as a function of
the mode number $k$ for $A=0.45$ ($\bullet$) and
$A=0.60$ ($\circ$) with $k_BT=0.0004$. The average
droplet radii are $r_0=11.95\,a$ and $r_0=15.21\,a$,
respectively. The solid lines are fits to Eq.~(\ref{SURFT}).
The inset shows a typical droplet configuration for
$\rho_{AB}=-0.6$, $\rho\Gamma=3.62$ ($A=0.60$ and
$k_BT=0.0004$). Parameters: $L/a=64$,
$M_A=2$, $M_B=8$, $\Delta t=1$, and $a=1$. From Ref.~\cite{tuze_07_mmf}.}
\end{figure}

Expression (\ref{PTH}) can be used to determine the entropy density, $s$.
The ideal-gas contribution to $s$ has the form \cite{callen_60_t}
\begin{equation}
\label{SI}
s_{ideal} = \rho\,\varphi(T) - k_B[\rho_A\ln\rho_A + \rho_B\ln\rho_B],
\end{equation}
where $\rho=\rho_A+\rho_B$. Since $\varphi(T)$ is independent
of $\rho_A$ and $\rho_B$, this term does not play a role in the current
discussion. The non-ideal contribution to the entropy density, $s_n$,
can be obtained from Eq.~(\ref{PTH}) using the thermodynamic relation
\begin{equation}
P_n/T = - s_n + \rho_A\partial s_n/\partial\rho_A +
\rho_B\partial s_n/\partial\rho_B.
\end{equation}
The result is $s_n = \Gamma \rho_A\rho_B$, so that
the total configurational contribution to the entropy density is
\begin{equation}
\label{BIN6}
s = -k_B\left\{\rho_A\ln\rho_A + \rho_B\ln\rho_B +
\Gamma \rho_A \rho_B\right\}.
\end{equation}

Since there is no configurational contribution to the internal energy
in this model, the mean-field phase diagram can be determined
by maximizing the entropy at fixed density $\rho$. The resulting
demixing phase diagram as a function of $\rho_{AB}=(\rho_A-\rho_B)/\rho$
is given by the solid line in Fig.~4 (left panel). The critical point is
located at
$\rho_{AB}=0$, $\rho\Gamma^*=2$. For $\rho\Gamma<2$, the order parameter
$\rho_{AB}=0$; for $\rho\Gamma>2$, there is phase separation into
coexisting A- and B-rich phases. As can be seen, the agreement
between the mean-field predictions and simulation results is very good
except close to the critical point, where the histogram method of determining
the coexisting densities is unreliable and critical fluctuations
influence the shape of the coexistence curve.

\subsubsection{Surface Tension}

A typical configuration for $\rho_{AB}=0$, $\rho\Gamma=3.62$ is shown
in the inset to Fig.~4 (left panel), and a snapshot of a fluctuating
droplet at $\rho_{AB}=-0.6$, $\rho\Gamma=3.62$ is shown in the inset
to Fig.~4 (right panel). The amplitude of the capillary wave fluctuations
of a droplet is determined by the surface tension, $\sigma$. Using the
parameterization $r(\phi) = r_0\left[1+\sum_{k=-\infty}^{\infty}u_k
\exp(ik\phi)\right]$ and choosing $u_0$ to fix the area of the droplet,
it can be shown that \cite{tuzel_thesis}
\begin{equation}
\label{SURFT}
\langle \vert u_k\vert^2\rangle = \frac{k_BT}{2\pi r_0\sigma}
\left(\frac{1}{k^2-1}\right).
\end{equation}
Figure~4 (right panel) contains a plot of $\langle \vert u_k\vert^2\rangle$
as a function of mode number $k$ for $\rho\Gamma=3.62$ and $\rho\Gamma=2.72$.
Fits to the data yield $\sigma\simeq2.9\,k_BT$ for
$\rho\Gamma=3.62$ and $\sigma\simeq1.1\,k_BT$ for $\rho\Gamma=2.72$.
Mechanical equilibrium requires that the pressure difference across
the interface of a droplet satisfies the Laplace equation
\begin{equation}
\label{LE}
\Delta p = p_{in}-p_{out} = (d-1) \sigma/r_0
\end{equation}
in $d$ spatial dimensions.
Measurements of $\Delta p$ [using Eq.~(\ref{VSP})] as a function of the
droplet radius for $A=0.60$ at $k_BT=0.0005$ yield results in
excellent agreement with the Laplace equation for the correct value
of the surface tension \cite{tuze_07_mmf}.

\begin{figure}[h]
  \begin{minipage}[t]{.47\linewidth}
    \begin{center}
      \includegraphics*[scale=0.57]{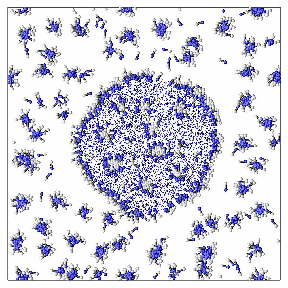}
     \label{droplet}
    \end{center}
  \end{minipage}
  \hfill
  \begin{minipage}[t]{.47\linewidth}
    \begin{center}
     \includegraphics*[scale=0.57]{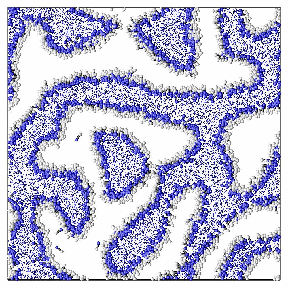}
     \label{emulsion}
    \end{center}
  \end{minipage}
  \hfill
\caption{(left panel) Droplet configuration in a mixture
with $N_A=8192$, $N_B=32768$, and $N_d=1500$ dimers after
$10^5$ time steps. The initial configuration is a droplet
with a homogeneous distribution of dimers. The dark (blue)
and light (white) colored spheres indicate A and B particles,
respectively. For clarity, A particles in the bulk are smaller
and B particles in the bulk are not shown. Parameters:
$L/a=64$, $M_A=2$, $M_B=8$, $A=1.8$, $k_BT=0.0001$,
$\Delta t=1$, and $a=1$. \hfill\break
(right panel) Typical configuration showing the bicontinuous
phase for $N_A=N_B=20480$ and $N_d=3000$. Parameters:
$L/a=64$, $M_A=5$, $M_B=5$, $A=1.8$, $k_BT=0.0001$, $\Delta t=1$,
and $a=1$. From Ref.~\cite{tuze_07_mmf}. }
\label{dimer}
\end{figure}

The model therefore displays the correct thermodynamic
behavior and interfacial fluctuations. It can also
be extended to model amphiphilic mixtures by introducing dimers
consisting of tethered A and B particles. If the A and B components of
the dimers participate in the same collisions as the solvent, they
behave like amphiphilic molecules in
binary oil-water mixtures. The resulting model displays a rich phase
behavior as a function of $\rho\Gamma$ and the number of dimers, $N_d$.
Both the formation of droplets and micelles, as
shown in Fig.~\ref{dimer} (left panel), and a bicontinuous phase,
as illustrated in Fig.~\ref{dimer} (right panel), have been observed
\cite{tuze_07_mmf}.  The coarse-grained nature
of the algorithm therefore enables the study of large time scales with
a feasible computational effort.

\subsubsection{Color Models for Immiscible Fluids}

There have been other generalizations of SRD to model binary mixtures
by Hashimoto {\em et al.} \cite{hash_00_irl} and Inoue {\em et al.}
\cite{inou_04_msm}, in which a color charge, $c_i=\pm 1$ is assigned to
two different species of particles. The rotation angle $\alpha$ in the
SRD rotation step is then chosen such that the color-weighted momentum
in a cell, ${\bf m}=\sum_{i=1}^{N_c} c_i({\bf v}_i-{\bf u})$, is
rotated to point in the direction of the gradient of the color field
$\bar c=\sum_{i=1}^{N_c}c_i$. This rule also leads to phase separation.
Several tests of the model have been performed; Laplace's equation
was verified numerically, and simulation studies of spinodal decomposition
and the deformation of a falling droplet were performed \cite{hash_00_irl}.
Later applications include a study of the transport of slightly deformed
immiscible droplets in a bifurcating channel \cite{inou_06_ams}.
Subsequently, the model was generalized through the addition of
dumbbell-shaped surfactants to model micellization \cite{sakai_00_fmr}
and the behavior of ternary amphiphilic mixtures in both two and three
dimensions \cite{saka_02_rlm,sakai_02_taf}. Note that since the color
current after the collision is always parallel to the color gradient,
thermal fluctuations of the order parameter are neglected in
this approach.

\section{Boundary Conditions and Embedded Objects}
\label{sec:BOUND_COND}

\subsection{Collisional Coupling to Embedded Particles}
\label{sec:COLL_COUPL}

A very simple procedure for coupling embedded objects such as colloids or
polymers to a MPC solvent has been proposed in Ref.~\cite{male_00_dsp}.
In this approach, every colloid particle or monomer in the polymer chain is
taken to be a point-particle which participates in the SRD collision. If
monomer $i$ has mass $m_m$ and velocity ${\bf w}_i$, the center of mass
velocity of the particles in the collision cell is
\begin{equation}
{\bf u}=\frac{m\sum_{i=1}^{N_c}{\bf v}_i+
m_m\sum_{i=1}^{N_m}{\bf w}_i}{N_cm+N_mm_m},
\end{equation}
where $N_m$ is the number of monomers in the collision cell. A stochastic
collision of the relative velocities of both the solvent particles and
embedded monomers is then performed in the collision step. This results in
an exchange of momentum between the solvent and embedded monomers.
The same procedure can of course be employed for other MPC algorithms,
such as MPC-AT.
The new monomer momenta are then used as initial conditions for a
molecular-dynamics update of the polymer degrees of freedom during the
subsequent streaming time step, $\Delta t$. Alternatively, the momentum
exchange, $\Delta p$, can be included as an additional force
$\Delta p/\Delta t$ in the molecular-dynamics integration.
If there are no other interactions
between monomers---as might be the case for embedded colloids---these
degrees of freedom stream freely during this time interval.

When using this approach, the average mass of solvent particles per cell,
$mN_c$, should be of the order of the monomer or colloid mass $m_m$
(assuming one embedded particle per cell) \cite{ripo_05_drf}.
This corresponds to a neutrally buoyant
object which responds quickly to the fluid flow but is not kicked around
too violently. It is also important to note that the average number of
monomers per cell, $\langle N_m\rangle$, should be smaller than unity
in order to properly resolve hydrodynamic interactions between the monomers.
On the other hand, the average bond length in a semi-flexible polymer
or rod-like colloid should also
not be much larger than the cell size $a$, in order to capture the
anisotropic friction of rod-like molecules due to hydrodynamic
interactions \cite{gg:gomp08b} (which leads to a twice as large
perpendicular than parallel friction coefficient for long stiff rods
\cite{doi:86}), and to
avoid an unnecessarily large ratio of the number of solvent to solute
particles. For a polymer, the average bond length should therefore be
of the order of $a$.

In order to use SRD to model suspended colloids with a radius of order
$1\mu$m in water, this approach would require approximately 60 solvent
particles per cell in order to match the Peclet number \cite{hecht_05_scc}.
This is much larger than the
optimum number (see discussion in Sec.~\ref{sec:cSRD_AT}), and the
relaxation to the Boltzmann distribution is very slow.
Because of its simplicity and efficiency, this monomer-solvent coupling
has been used in many polymer
\cite{falc_03_dst,ripo_04_lhc,wink:04,muss:05,webs_05_mtp}
and colloid simulations
\cite{falc_04_ihm,hecht_05_scc,ripo_05_drf,hecht_07_sta}.

\subsection{Thermal Boundaries}
\label{sec:THERMAL_WALL}

In order to accurately resolve the local flow field around a colloid,
more accurate methods have been proposed which exclude
fluid-particles from the interior of the colloid and mimic slip
\cite{male_00_smd,lee_04_fdb} or no-slip \cite{inou_02_dsm} boundary
conditions. The latter procedure is similar to what is known in molecular
dynamics as a ``thermal wall'' boundary condition: fluid particles which
hit the colloid particle are given a new, random velocity drawn from
the following probability distributions for the normal velocity component,
$v_N$, and the tangential component, $v_T$,
\begin{eqnarray}
\nonumber
p_N(v_N)&=& (m v_N/k_B T)\, \exp\left( -{m v_N^2/2k_B T}\right)\,,
\ \ {\rm with} \ \  v_N>0 \,,\\
\label{THERMAL1}
p_T(v_T)&=& \sqrt{m / 2 \pi k_B T}\, \exp\left( -{m v_T^2/2k_B T}\right) \,.
\end{eqnarray}
These probability distributions are constructed so that the probability
distribution for particles near the wall remains Maxwellian.
The probability distribution, $p_T$, for the
tangential components of the velocity is Maxwellian, and both positive and
negative values are permitted. The normal component must be positive, since
after scattering at the surface, the particle must move away from the wall.
The form of $p_N$ is a reflection of the fact that more particles with large
$\vert v_N\vert$ hit the wall per unit time than with small $\vert v_N\vert$
\cite{inou_02_dsm}.

This procedure models a no-slip boundary condition at the surface of the
colloid, and also thermostats the fluid at the boundaries. For many
non-equilibrium flow conditions, this may not be sufficient, and it may
also be necessary to thermostat the bulk fluid also (compare
Sec.~\ref{sec:THERMOSTAT}). It should also be noted that
Eqs.~(\ref{THERMAL1}) will be a good approximation only if the radius of
the embedded objects is much larger than the mean free path $\lambda$.
For smaller particles, corrections are needed.

If a particle hits the surface at time $t_0$ in the interval
between $n\Delta t$ and $(n+1)\Delta t$, the correct way to proceed would
be to give the particle its new velocity and then have it stream the
remaining time $(n+1)\Delta t-t_0$. However, such detailed resolution
is not necessary. It has been found \cite{hecht_05_scc} that good results
are also obtained using the following simple stochastic procedure.
If a particle is found to have penetrated the colloid during the streaming
step, one simply moves it to the boundary and then stream a distance
${\bf v}_{new}\,\Delta t\,\epsilon$, where $\epsilon$ is a uniformly
distributed random number in the interval [0,1].

Another subtlety is worth mentioning. If two colloid particles
are very close, it can happen that a solvent particle could hit the second
colloid after scattering off the first, all in the interval $\Delta t$.
Naively, one might be tempted to simply forbid this from happening or ignore
it. However, this would lead to a strong depletion-like attractive force
between the colloids
\cite{hecht_05_scc}. This effect can be greatly reduced by allowing multiple
collisions in which one solvent particle is repeatedly scattered off the
two colloids. In every collision, momentum is transferred to one of the
colloids, which pushes the colloids further apart. In practice, even allowing
for up to ten multiple collisions cannot completely cancel the depletion
interaction---one needs an additional repulsive force to eliminate
this unphysical attraction.
The same effect can occur when a colloid particle is near a wall.

Careful tests of this thermal coupling have been performed by Padding
{\it et al.} \cite{padd_05_stick,padd_06_hib}, who were able to reproduce
the correct rotational diffusion of a colloid. It should be noted that
because the coupling between the solvent particles and the surface occurs
only through the movement of the fluid particles, the coupling is quite
weak for small mean free paths.

\subsection{Coupling using Additional Forces}
\label{sec:ADDIT_FORCE}

Another procedure for coupling an embedded object to the solvent has been
pursued by Kapral {\it et al.} \cite{male_00_smd,lee_01_csd,kapr_08_mpc}.
They introduce a central repulsive force
between the solvent particles and the colloid. This force has to
be quite strong in order to prohibit a large number of solvent particles
from penetrating the colloid. When implementing this procedure, a
small time step $\delta t$ is therefore required in order to resolve these
forces correctly, and a large number of molecular dynamics time steps are
needed during the SRD streaming step.
In its original form, central forces were used, so that only slip
boundary conditions could be modeled. In principle, non-central forces
could be used to impose no-slip conditions.

This approach is quite natural and very easy to implement; it does, however,
require the use of small time steps and therefore may not be the
optimal procedure for many applications.

\subsection{``Ghost'' or ``Wall'' Particles}
\label{sec:walls}

One of the first approaches employed to impose a non-slip boundary
condition at an external wall or at a moving object in a MPC solvent
was to use ``ghost'' or ``wall'' particles
\cite{lamu_01_mcd,lamu_02_nsf}. In other mesoscale methods such as
Lattice-Boltzmann, no-slip conditions are modeled using the bounce-back
rule: the velocity of particle is inverted from ${\bf v}$ to $-{\bf v}$
when it intersects a wall. For planar walls which coincide with the
boundaries of the collision cells, the same procedure can be used in
MPC. However, the walls will generally not coincide with, or even be
parallel to, the cell walls. Furthermore, for small mean free paths, where a
shift of the cell lattice is required to guarantee Galilean invariance,
partially occupied boundary cells are unavoidable, even in the simplest
flow geometries.

\begin{figure}[ht]
\begin{center}
\includegraphics[width=10.0cm]{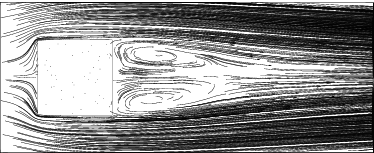}
\end{center}
\caption{\label{fig:flow_obstacle}
Velocity field of a fluid near a square cylinder in a Poiseuille
flow at Reynolds number $Re=v_{max}L/\nu=30$. The channel width is eight
times larger than the cylinder size $L$. A pair of stationary vortices
are seen behind the obstacle, as expected for $Re\le 60$.
From Ref.~\cite{lamu_02_nsf}.
}
\end{figure}

The simple bounce-back rule fails to guarantee no-slip boundary conditions
in the case of partially filled cells. The following generalization of
the bounce-back rule has therefore been suggested.
For all cells that are cut by walls, fill the ``wall'' part of the cell
with a sufficient number of virtual particles in order to make the
total number of particles equal to $M$, the average number of particles
per cell. The velocities of the wall particles are drawn from a
Maxwell-Boltzmann distribution with zero mean velocity and the same
temperature as the fluid. The collision step is then carried out using the
mean velocity of all particles in the cell. Note that since Gaussian
random numbers are used, and the sum of Gaussian random numbers is also
Gaussian-distributed, the velocities of the individual wall particles
need not to be determined explicitly. Instead, the average velocity
${\bf u}$ can be written as ${\bf u}=(\sum_{i=1}^n{\bf v}_i+{\bf a})/M$,
where ${\bf a}$ is vector whose components are Gaussian random numbers
with zero mean and variance $(M-n)k_BT$. Results for Poiseuille flow
obtained using this procedure, both with and without cell shifting, were
found to be in excellent agreement with the correct parabolic profile
\cite{lamu_01_mcd}. Similarly, numerical results for the recirculation
length, the drag coefficient, and the Strouhal number for flows around
a circular and square cylinder in two dimensions were shown to be in good
agreement with experimental results and computational fluid dynamics data
for a range of Reynolds numbers between $Re=10$ and $Re=130$
(see Fig.~\ref{fig:flow_obstacle}) \cite{lamu_01_mcd,lamu_02_nsf}.

\section{Importance of Angular-Momentum Conservation: Couette Flow}
\label{sec:couette}

\begin{figure}
\begin{center}
\includegraphics[width=8.0cm]{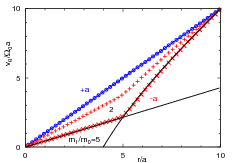}
\end{center}
\caption{\label{fig:Couette_2phase}
Azimuthal velocity of binary fluids in a rotating cylinder with
$\Omega_0=0.01 (k_BT/m_0a^2)^{1/2}$.
The viscous fluids with particle mass $m_1$ and $m_0$ are located at
$r<R_1$ and $R_1<r<R_2$, respectively, with $R_1=5a$ and $R_2=10a$.
Symbols represent the simulation results of MPC-AT$-a$ with
$m_1/m_0=2$ ($+$) or $m_1/m_0=5$ ($\times$), and MPC-AT$+a$ for
$m_1/m_0=5$ ($\circ$). Solid lines represent the analytical results
for MPC-AT$-a$ at $m_1/m_0=5$. Error bars are smaller than the size
of the symbols. From Ref.~\cite{gotz_07_ram}.
}
\end{figure}

As an example of a situation in which it is important to use an
algorithm which conserves angular momentum,
consider a drop of a highly viscous fluid inside a lower-viscosity
fluid in circular Couette flow. In order to avoid the complications
of phase-separating two-component fluids, the high viscosity fluid
is confined to a radius $r<R_1$ by an impenetrable boundary with
reflecting boundary conditions (i.e., the momentum parallel to the
boundary is conserved in collisions). No-slip boundary conditions
between the inner and outer fluids are guaranteed because
collision cells reach across the boundary. When a torque is applied
to the outer circular wall (with no-slip, bounce-back boundary conditions)
of radius $R_2>R_1$, a solid-body rotation of {\em both fluids} is
expected. The results of simulations with both MPC-AT$-a$ and MPC-AT$+a$
are shown in Fig.~\ref{fig:Couette_2phase}. While MPC-AT$+a$
reproduces the expected behavior, MPC-AT$-a$ produces different
angular velocities in the two fluids, with a low (high) angular
velocity in the fluid of high (low) viscosity \cite{gotz_07_ram}.

The origin of this behavior is that the viscous stress tensor in general has
symmetric and antisymmetric contributions (see Sec.~\ref{sec:ee}),
\begin{eqnarray}
\label{eq:stress}
\sigma_{\alpha\beta} = \lambda(\partial_\gamma v_\gamma)\delta_{\alpha\beta}
+ \bar{\eta} \left(\partial_\beta v_\alpha
           +\partial_\alpha  v_\beta \right)
+ \check{\eta} \left(\partial_\beta v_\alpha
           -\partial_\alpha  v_\beta \right),
\end{eqnarray}
where $\lambda$ is the second viscosity coefficient and
$\bar{\eta}\equiv\rho\nu_1$ and $\check{\eta}\equiv\rho\nu_2$ are the
symmetric and anti-symmetric components of the
viscosity, respectively. The last term in Eq.~(\ref{eq:stress}) is linear
in the vorticity $\nabla\times{\bf v}$, and does not conserve angular
momentum. This term therefore vanishes ({\em i.e.}, $\check{\eta}=0$) when
angular momentum is conserved.


The anti-symmetric part of the stress tensor implies an additional
torque, which becomes relevant when the boundary condition is given by
forces.  In cylindrical coordinates ($r,\theta,z$), the azimuthal
stress is given by \cite{gotz_07_ram}
\begin{equation}
\label{eq:strs_rq}
\sigma_{r\theta}=
(\bar{\eta}+\check{\eta}) \frac{r\partial (v_{\theta}/r)}{\partial r}
           + 2\check{\eta}\frac{v_{\theta}}{r}.
\end{equation}
The first term is the stress of the  angular-momentum-conserving fluid,
which depends on the derivative of the angular velocity
$\Omega=v_{\theta}/r$. The second term is an additional stress caused
by the lack of angular momentum conservation; it is proportional to $\Omega$.

In the case of the phase-separated fluids in circular Couette flow, this
implies that if both fluids rotate at the same angular velocity,
the inner and outer stresses do not coincide.
Thus, the angular velocity of the inner fluid $\Omega_1$ is smaller than
the outer one, with $v_\theta(r)=\Omega_1 r$ for $r<R_1$ and
\begin{equation} \label{eq:v_cf}
v_\theta(r) = Ar+ B/r, {\rm \ \ with\ \ }
A = \frac{\Omega_2 R_2^2 - \Omega_1 R_1^2}{R_2^2 - R_1^2}, \ \
B = \frac{(\Omega_1 - \Omega_2) R_1^2R_2^2}{R_2^2 - R_1^2}
\end{equation}
for $R_1<r<R_2$.
$\Omega_1$ is then obtained from the stress balance at $r=R_1$, {\em i.e.},
$2\check{\eta}_1\Omega_1 = (8/3)\eta_2 (\Omega_0-\Omega_1) +
2\check{\eta}_2\Omega_1$.
This calculation reproduces the numerical results very well, see
Fig.~\ref{fig:Couette_2phase}. Thus, it is essential to employ an $+a$
version of MPC in simulations of multi-phase flows of binary fluids with
different viscosities.

There are other situations in which the lack angular momentum conservation
can cause significant deviations. In Ref.~\cite{gotz_07_ram}, a star
polymer with small monomer spacing was placed in the middle of a rotating
Couette cell. As in the previous case, it was observed that
the polymer fluid rotated with a smaller angular velocity than the outer fluid.
When the angular momentum conservation was switched on, everything rotated
at the same angular velocity, as expected.

\section{MPC without Hydrodynamics}
\label{sec:MPCwithout}

The importance of hydrodynamic interactions (HI) in complex fluids is
generally accepted. A standard procedure for determining the influence of
HI is to investigate the same system with and without HI. In order to
compare results, however, the two simulations must differ as little as
possible---apart from the inclusion of HI. A well-known example of this
approach is Stokesian dynamics simulations (SD), where the original Brownian
dynamics (BD) method can be extended by including hydrodynamic
interactions in the mobility matrix by employing the Oseen
tensor \cite{doi:86,dhon96}.

A method for switching off HI in MPC has been proposed in
Refs.~\cite{kiku_02_pcp,kiku_03_tcm}.
The basic idea is to randomly interchange velocities of all solvent particles
after each collision step, so that momentum (and energy) are {\em not}
conserved {\em locally}. Hydrodynamic correlations are therefore destroyed,
while leaving friction coefficients and fluid self-diffusion coefficients
largely unaffected. Since this approach requires the same numerical
effort as the original MPC algorithm, a more efficient method has been
suggested recently in Ref.~\cite{ripo_07_hss}. If the
velocities of the solvent particles are not correlated, it is no longer
necessary to follow their trajectories. In a random solvent, the
solvent-solute interaction in the collision step can thus be replaced by the
interaction with a heat bath. This strategy is related to that
proposed in Ref.~\cite{lamu_01_mcd} to model no-slip boundary
conditions of solvent particles at a planar wall, compare
Sec.~\ref{sec:walls}. Since the positions
of the solvent particles within a cell are not required in the
collision step, no explicit particles have to be considered.  Instead,
each monomer is coupled with an effective solvent momentum ${\bf P}$
which is directly chosen from a Maxwell-Boltzmann distribution of
variance $m M k_B T$ and a mean given by the average momentum of
the fluid field---which is zero at rest, or $(m M\dot{\gamma}
r^i_y,0,0)$ in the case of an imposed shear flow. The total
center-of-mass velocity, which is used in the collision step,
is then given by \cite{ripo_07_hss}
\begin{equation}
{\bf v}_{cm,i} = \frac{m_m {\bf v}_i + {\bf P}}{m M+m_m},
\end{equation}
where $m_m$ is the mass of the solute particle. The solute trajectory
is then determined using MD, and the interaction with the solvent is
performed every collision time $\Delta t$.

The random MPC solvent therefore has similar properties to
the MPC solvent, except that there are no HI. The relevant
parameters in both methods are the average number of particles per cell,
$M$, the rotation angle
$\alpha$, and the collision time $\Delta t$ which can be chosen to be the
same.  For small values of the density ($M< 5$), fluctuation
effects have been noticed \cite{kiku_03_tcm} and could also be included in
the random MPC solvent by a Poisson-distributed density.
The velocity autocorrelation functions \cite{ripo_05_drf} of a
random MPC solvent show a simple exponentially decay, which implies
some differences in the solvent
diffusion coefficients. Other transport coefficients such as the
viscosity depend on HI only weakly \cite{tuzel_06_dcs} and consequently are
expected to be essentially identical in both solvents.

\section{Applications to Colloid and Polymer Dynamics}
\label{sec:applications}

The relevance of hydrodynamic interactions for the dynamics of
complex fluids---such as dilute or semidilute polymer solutions,
colloid suspensions, and microemulsions---is well known
\cite{doi:86,dhon96}. From the simulation point of view,
however, these systems are difficult to study because of the large
gap in length- and time-scales between solute and solvent dynamics.
One possibility for investigating complex fluids is the straightforward
application of molecular dynamics simulations (MD), in which the
fluid is course-grained and represented by Lennard-Jones
particles. Such simulations provide valuable insight into polymer
dynamics \cite{pier:91,duen:91,pier:92,duen:93,aust:99}.
Similarly, mesoscale algorithms such as Lattice-Boltzmann and
dissipative particle dynamics have been widely used for modeling
of colloidal and polymeric systems
\cite{boek97,ahlr:98,ahlr:01,spen:00,lowe:04}.

Solute molecules, e.g., polymers, are typically composed of a
large number of individual particles, whose interactions are
described by a force-field. As discussed in Sec.~\ref{sec:BOUND_COND},
the particle-based character of the MPC solvent allows for
an easy and controlled coupling between the solvent and solute
particles. Hybrid simulations combining MPC and molecular
dynamics simulations are therefore easy to implement. Results of such
hybrid simulations are discussed in the following.

\subsection{Colloids}
\label{sec:colloids}

Many applications in chemical engineering, geology, and biology involve
systems of particles immersed in a liquid or gas flow.
Examples include sedimentation processes, liquid-solid fluidized beds,
and flocculation in suspensions.

Long-ranged solvent-mediated hydrodynamic interactions have a profound
effect on the non-equilibrium properties of colloidal suspensions, and
the many-body hydrodynamic backflow effect makes it difficult to answer even
relatively simple questions such what happens when a collection of particles
sediments through a viscous fluid. Batchelor \cite{batch_72_sed}
calculated the lowest-order volume fraction correction to the average
sedimentation
velocity, $v_s=v_s^0(1-6.55\,\phi)$, of hard spheres of hydrodynamic
radius $R_H$
where $v_s^0$ is the sedimentation velocity of a single sphere.
Due to the complicated interplay between short-ranged contact forces and
long-ranged HI, it is hard to extend this result to the high volume fraction
suspensions of interest for ceramics and soil mechanics. An additional
complication is that the Brownian motion of solute particles in water
cannot be neglected if they are smaller than $1\mu m$ in diameter.

The dimensionless Peclet number $Pe=v_s^0R_H/D$,
where $D$ is the self-diffusion coefficient of the suspended
particles,  measures the relative strength of HI and thermal motion.
Most studies of sedimentation have focused on the limit of infinite
Peclet number, where Brownian forces are negligible. For example, Ladd
\cite{ladd_97_lba} employed a Lattice-Boltzmann method (LB), and Hoefler
and Schwarzer \cite{hoef_00_nss} used a marker-and-cell Navier-Stokes
solver to simulate such non-Brownian suspensions. The main difficulty
with such algorithms is the solid-fluid coupling which can be very tricky:
in LB simulations, special ``boundary nodes'' were inserted on the colloid
surface, while in Ref.~\cite{hoef_00_nss}, the coupling was mediated by
inertia-less
markers which are connected to the colloid by stiff springs
and swim in the fluid, effectively dragging the colloid, but also exerting
a force on the fluid.  Several methods for coupling embedded particles
to an MPC solvent were discussed in Sec.~\ref{sec:BOUND_COND}.

Using the force-based solvent-colloid coupling described in
Sec.~\ref{sec:ADDIT_FORCE}, Padding and Louis \cite{padd_04_hbf}
investigated the importance of HI during sedimentation at small Peclet
numbers. Surprisingly, they found that the sedimentation velocity does
not change if the Peclet number is varied between 0.1 and 15 for a range of
volume fractions. For small volume fractions, the numerical results agree
with the Batchelor law; for intermediate $\phi$ they are consistent with
the semi-empirical Richardson-Zaki law, $v_s=v_s^0(1-\phi)^n$, $n=6.55$.
Even better agreement was found with theoretical predictions by
Hayakawa and Ichiki \cite{haya_95_sph,haya_97_shs}, who took higher order
HI into account. Purely hydrodynamic arguments are therefore still valid in an
average sense at low $Pe$, {\em i.e.}, for strong Brownian motion and
relatively weak HI.
This also means that pure Brownian simulations without HI, which lead to
$v_s=v_s^0(1-\phi)$, strongly underestimate the effect of backflow.

On the other hand, it is known that the velocity autocorrelation function
of a colloidal particle embedded in a fluctuating liquid at equilibrium
exhibits a hydrodynamic long-time tail,
$\langle v(t) v(0)\rangle\sim t^{-d/2}$,
where $d$ is the spatial dimension \cite{ernst_70_atb}. These tails have
been measured earlier for point-like SRD particles in two
\cite{ihle_01_srd,ihle_03_srd_b} and three \cite{ripo_05_drf} spatial
dimensions, and found to be in quantitative agreement with analytic
predictions, with no adjustable parameters.
It is therefore not surprising that good agreement was also obtained
for embedded colloids \cite{padd_04_hbf}.
MPC therefore correctly describes two of the most important effects in
colloidal suspensions, thermal fluctuations and hydrodynamic interactions.

In a series of papers, Hecht {\it et al.}
\cite{hecht_05_scc,hecht_06_she,hecht_07_sta} used hybrid SRD-molecular
dynamics simulations to investigate a technologically important colloidal
system---$Al_2O_3$-particles of diameter $0.5\mu m$ (which is often
used in ceramics) suspended in water---with additional colloid-colloid
interactions. These colloids usually carry a charge which, by
forming an electric double layer with ions in water, results in a screened
electrostatic repulsion. The interaction can be approximated
by the Derjaguin-Landau-Verwey-Overbeek (DLVO) theory \cite{DLVO_1,DLVO_2}.
The resulting potential
contains a repulsive Debye-H{\"u}ckel contributions,
$V_{EL}\sim {\rm exp}(-\kappa[r-d])/r$, where $d$ is the particle diameter,
$\kappa$ is the inverse screening length, and $r$ is the distance of the
particle centers. The second part of the DLVO-potential is a short-range
van der Waals attraction,
\begin{equation}
\label{VANDERWAALS}
V_{vdW}=-{A_H\over 12}\left[ {d^2\over r^2-d^2}+{d^2\over r^2}+2
{\rm ln}\left( {r^2-d^2\over r^2}\right) \right] ,
\end{equation}
which turns out to be important at the high volume fractions
($\phi > 20 \%$) and high salt concentrations of interest.
$A_H$ is the Hamaker
constant which involves the polarizability of the particles and the solvent.
DLVO theory makes the assumption of linear polarizability and is
valid only at larger distances. It therefore does not include
the so-called primary potential minimum at particle contact, which is
observed experimentally and is about $30 k_B T$ deep. Because of this
potential minimum, colloids which come in contact rarely become free again.
In order to ensure numerical stability for reasonable values of the
time step, this minimum was modeled by an additional parabolic potential
with depth of order $6 k_B T$.
The particle Reynolds number of the real system is very small, of order
$10^{-6}$ to $10^{-7}$. Since it would be too time-consuming to model this
Reynolds number, the simulations were performed at $Re\approx 0.02$,
which still ensures that the contribution of momentum convection is
negligible compared to that of momentum diffusion. However,
due to the remaining inertial effects and the non-zero time step,
it was still possible that particles partially overlapped in the simulation.
This overlap was penalized by an additional potential, frequently
used in simulations of granular matter, given by a Hertz-law,
\begin{equation}
V_{Hertz}\sim(d-r)^{5/2}\;\;\;\;{\rm if\;}r<d\,.
\end{equation}

\begin{figure}
\begin{center}
\vspace{2cm}
\includegraphics*[width=4.0in,angle=0]{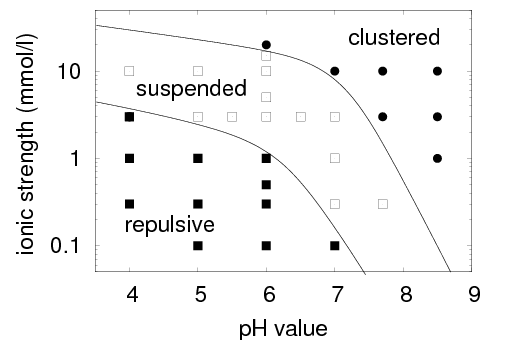}
\caption{
Phase diagram of a colloidal suspension (plotted at zero shear and
volume fraction $\phi=35\%$) in the ionic-strength-pH plane
depicting three regions: a clustered region, a suspended regime,
and a repulsive structure.  From Ref.~\cite{hecht_07_sta}.
}
\label{fig_hecht}
\end{center}
\end{figure}

SRD correctly describes long-range HI, but it can only resolve hydrodynamic
interactions on scales larger than both the mean free path $\lambda$ and
the cell size $a$. In a typical simulation with about 1000 colloid particles,
a relatively small colloid diameter of about four lattice units was chosen
for computational efficiency. This means that HI are not fully resolved
at interparticle distances comparable to the colloid diameter, and
lubrication forces have to be inserted by hand. Only the most
divergent mode, the so-called squeezing mode, was used,
$F_{lub}\sim v_{rel}/r_{rel}$,  where $r_{rel}$ and $v_{rel}$ are the
relative distance and velocity of two colloids, respectively.
This system of interacting $Al_2O_3$-particles was simulated in order to
study the dependence of the suspension's viscosity and structure
on shear rate, pH, ionic strength and volume fraction.
The resulting stability diagram of the suspension as a function
of ionic strength and pH value is shown in Fig.~\ref{fig_hecht}
(plotted at zero shear) \cite{hecht_07_sta}. The pH controls the surface
charge density which, in turn, affects the electrostatic interactions
between the colloids. Increasing the ionic strength, experimentally
achieved by ``adding salt'', decreases the screening length $1/\kappa$,
so that the attractive forces become more important; the particles start
forming clusters. Three different states are observed: (i) a clustered regime,
where particles aggregate when van der Waals attractions dominate,
(ii) a suspended regime where particles are distributed homogeneously
and can move freely---corresponding to a stable suspension favored when
electrostatic repulsion prevents clustering but is not strong enough to
induce order. At very strong Coulomb repulsion the repulsive regime
(iii) occurs, where the mobility of the particles is restricted, and
particles arrange in local order which maximizes nearest neighbor distances.

The location of the phase boundaries in Fig. \ref{fig_hecht} depends on the
shear rate. In the clustered phase, shear leads to a breakup of clusters, and
for the shear rate $\dot\gamma=1000\,s^{-1}$, there are many small clusters
which behave like single particles. In the regime where the particles are
slightly clustered, or suspended, shear thinning is observed. Shear thinning
is more pronounced in the slightly
clustered state, because shear tends to reduce cluster size. Reasonable
agreement with experiments was achieved, and discrepancies
were attributed to polydispersity and the manner in which lubrication
forces were approximated, as well as uncertainties how
the pH and ionic strength enter the model force parameters.

In the simulations of Hecht {\it et al.} \cite{hecht_05_scc}, the
simple collisional coupling
procedure described in Sec.~\ref{sec:COLL_COUPL} was used.
This means that the colloids were treated as point particles, and
solvent particles could flow right through them. Hydrodynamic interactions
were therefore only resolved in an average sense, which is acceptable
for studies of the general properties of an ensemble of many colloids.
The heat from viscous heating was removed using the stochastic thermostat
described in Sec.~\ref{sec:THERMOSTAT}.

Various methods for modeling no-slip boundary conditions at colloid
surfaces---such as the thermal wall coupling described in
Sec.~\ref{sec:THERMAL_WALL}---were systematically investigated in
Ref.~\cite{padd_05_stick}. No-slip boundary conditions are
important, since colloids are typically not completely spherical or smooth, and
the solvent molecules also transfer angular momentum to the colloid. Using
the SRD algorithm without angular momentum conservation, it was found that
the rotational friction coefficient was larger than predicted by Enskog-theory
when the ghost-particle coupling was used \cite{padding_private}.
On the other hand, in a detailed study of the translational and
rotational velocity autocorrelation function of a sphere coupled to the
solvent by the thermal-wall boundary condition, quantitative agreement with
Enskog theory was observed at short times, and with mode-coupling theory
at long times. However, it was also noticed that for small
particles, the Enskog and hydrodynamic contributions to the friction
coefficients were not clearly separated. Specifically, mapping the system
to a density matched colloid in water, it appeared that the Enskog and the
hydrodynamic contributions are equal at a particle radius of $6\,nm$ for
translation and $35.4\,nm$ for rotation; even for a particle radius of
$100\,nm$, the Enskog contribution to the friction is still of order
$30\%$ and cannot be ignored.

In order to clarify the detailed character of the hydrodynamic interactions
between colloids in SRD, Lee and Kapral \cite{lee_05_tfm} numerically
evaluated the fixed-particle friction tensor for two nano-spheres
embedded in an
SRD solvent. They found that for intercolloidal spacings less than
$1.2\, d$, where $d$ is the colloid diameter, the measured friction
coefficients start to deviate from the expected theoretical curve.
The reader is referred to the review by Kapral \cite{kapr_08_mpc} for
more details.

\subsection{Polymer Dynamics}
\label{sec:polymer}

The dynamical behavior of macromolecules in solution is strongly
affected or even dominated by hydrodynamic interactions
\cite{kirk:48,erpe:58,doi:86}. From a theoretical point of
view, scaling relations predicted by the Zimm model for, e.g.,
the dependencies of dynamical quantities on the length of the
polymer are, in general, accepted and confirmed \cite{petr:06}.
Recent advances in experimental
single-molecule techniques provide insight into the dynamics of
individual polymers, and raise the need for a quantitative
theoretical description in order to determine molecular parameters
such as diffusion coefficients and relaxation times.
Mesoscale hydrodynamic simulations can be used to verify the
validity of theoretical models. Even more, such simulations are
especially valuable when analytical methods fail, as for more
complicated molecules such as polymer brushes, stars, ultrasoft
colloids, or semidilute solutions, where hydrodynamic interactions
are screened to a certain degree. Here, mesoscale simulations
still provide a full characterization of the polymer dynamics.

We will focus on the dynamics of polymer chains in dilute
solution. In order to compare simulation results with theory---in
particular the Zimm approach \cite{zimm:56,doi:86}---and scaling
predictions, we address the dynamics of Gaussian as well as
self-avoiding polymers.

\subsubsection{Simulation Method and Model}
\label{subsubsec:model}

Polymer molecules are composed of a large number of equal repeat
units called monomers. To account for the generic features of
polymers, such as their conformational freedom, no detailed
modeling of the basic units is necessary. A coarse-grained
description often suffices, where several monomers are comprised
in an effective particle. Adopting such an approach, a polymer
chain is introduced into the MPC solvent by adding $N_m$ point
particles, each of mass $m_m$, which are connected linearly by
bonds. Two different models are considered, a Gaussian polymer and
a polymer with excluded-volume (EV) interactions. Correspondingly,
the following potentials are applied:\\
(i) {\em Gaussian chain}: The monomers, with the positions ${\vec
r}_i$ ($i=1,\ldots,N_m$), are connected by the harmonic potential
\begin{equation}\label{pot_gauss}
U_G = \frac{3 k_B T}{2 b^2} \sum_{i=1}^{N_m-1} \left({\vec r}_{i+1}
- {\vec r }_i \right)^2 ,
\end{equation}
with zero mean bond length, and $b$ the root-mean-square bond length. Here, the
various monomers freely penetrated each other. This simplification allows for
an analytical treatment of the chain dynamics as in the Zimm
model \cite{zimm:56,doi:86}.\\
(ii) {\em Excluded-volume chain}: The monomers are connected by the harmonic
potential
\begin{equation}\label{pot_ev}
U_{B} = \frac{\kappa}{2} \sum_{i=1}^{N_m-1} \left(|{\vec r}_{i+1} -
{\vec r }_i| -b \right)^2 ,
\end{equation}
with mean bond length $b$. The force constant $\kappa$ is chosen
such that the fluctuations of the bond lengths are on the order of
a percent of the mean bond length. In addition, non-bonded
monomers interact via the repulsive, truncated Lennard-Jones
potential
\begin{eqnarray}\label{pot_lj}
U_{LJ} = \left\{
\begin{array}{cc}
4 \epsilon \left[\left(\frac{\displaystyle  \sigma}{\displaystyle r}
\right)^{12} - \left(\frac{\displaystyle \sigma}{\displaystyle r}
\right)^{6} \right] + \epsilon,  & r < 2^{1/6} \sigma \\
0, & \mbox{otherwise}
\end{array}.
\right.
\end{eqnarray}
The excluded volume leads to swelling of the polymer structure
compared to a Gaussian chain, which is difficult to fully account
for in analytical calculations \cite{muss:05}.

The dynamics of the chain monomers is determined by Newtons'
equations of motion between the collisions with the solvent. These
equations are integrated using the velocity Verlet algorithm with the
time step $\Delta t_p$. The latter is typically smaller than the
collision time $\Delta t$. The monomer-solvent interaction is
taken into account by inclusion of the monomer of mass $m_m=\rho
m$ in the collision step \cite{male_00_dsp,muss:05}, compare
Sec.~\ref{sec:COLL_COUPL}.
Alternatively, a Lennard-Jones potential can be used to account
for the monomer-MPC particle interaction, where a MPC particle is
of zero interaction range \cite{male_00_smd,lee:06}.

We scale length and time according to $\hat x = x/a$ and $\hat t = t
\sqrt{k_BT/ma^2}$, which corresponds to the choice $k_BT =1$, $m=1$,
and $a=1$. The mean free path of a fluid particle   $\Delta t
\sqrt{k_BT/m}$ is then given by $\lambda =  \Delta \hat t$. In
addition, we set $b = a$, $\sigma = a$, and $\epsilon/k_BT =1$.

The equilibrium properties of a polymer are not affected by
hydrodynamic interactions. Indeed, the results for various
equilibrium quantities---such as the radius of gyration---of
MPC simulation are in excellent agreement with the results of
molecular dynamics of Monte Carlo simulations without explicit
solvent \cite{muss:05}.

Simulations of Gaussian chains, i.e., polymers with the bond
potential (\ref{pot_gauss}), can be compared with analytical
calculations based on the Zimm approach \cite{zimm:56,doi:86}.
Note, however, that the simulations are {\em not} performed in the Zimm
model. The Zimm approach relies on the preaveraging approximation
of hydrodynamic interactions, whereas the simulations take into
account the configurational dependence of the hydrodynamic
interactions, and therefore hydrodynamic fluctuations. Hence, the
comparison can serve as a test of the validity of the
approximations employed in the Zimm approach.

\begin{figure}[t]
\begin{center}
\includegraphics*[width=8.5cm,clip]{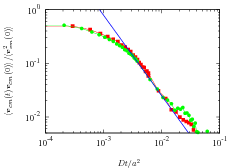}
\end{center}
\caption{\label{fig_1} Center-of-mass velocity autocorrelation
functions for Gaussian polymers of length $N_m =20$, $N_m=40$, and
$\lambda = 0.1$ as a function of $Dt$. The solid line is
proportional to $(Dt)^{-3/2}$. From Ref.~\cite{muss:05}.}
\end{figure}

The Zimm model rests upon the Langevin equation for over-damped
motion of the monomers, i.e., it applies for times larger than the
Brownian time scale $\tau_B \gg m_m/\zeta$, where $\zeta$ is
Stokes' friction coefficient \cite{dhon96}. On such time
scales, velocity correlation functions have decayed to zero and
the monomer momenta are in equilibrium with the solvent. Moreover,
hydrodynamic interactions between the various parts of the polymer
are assumed to propagate instantaneously. This is not the case in
our simulations. First of all, the monomer inertia term is taken
into account, which implies non-zero velocity autocorrelation
functions. Secondly, the hydrodynamic interactions build up
gradually. The center-of-mass velocity autocorrelation function
displayed in Fig.~\ref{fig_1} reflects these aspects. The
correlation function exhibits a long-time tail, which decays as
$\lla {\vec v}_{cm}(t) {\vec v}_{cm}(0) \rra \sim t^{-3/2}$ on
larger time scales. The algebraic decay is associated with a
coupling between the motion of the polymer and the hydrodynamic
modes of the fluid
\cite{ernst_70_atb,ernst_71_atbc,hansen_86_tsl}. A scaling of
time with the diffusion coefficient $D$ shows that the
correlation function is a universal function of $D t$. This is in
agreement with results of DPD simulations of dilute polymer
systems \cite{lowe:04}.

The polymer center-of-mass diffusion coefficient follows either
via the Green-Kubo relation from the velocity autocorrelation
function or by the Einstein relation from the  center-of-mass mean
square displacement. According to the Kirkwood formula
\cite{kirk:48,erpe:58,liu:03}
\begin{equation}  \label{kirkwood}
D^{(K)} = \frac{D_0}{N_m} + \frac{k_BT}{6 \pi \eta } \frac{1}{R_H} ,
\end{equation}
where the hydrodynamic radius $R_H$ is defined as
\begin{equation}  \label{hydro_dyn_rad}
\frac{1}{R_H} = \frac{1}{N_m^2}\lla \sum_{i=1}^{N_m} \sum_{j=1}^{N_m}{'}
\frac{1}{|{\vec r}_i - {\vec r}_j|} \rra
\end{equation}
and the prime indicates that the term with $j=i$ has to be left
out in the summation. The diffusion coefficient is composed of the
local friction contribution $D_0/N_m$, where $D_0$ is the
diffusion coefficient of a single monomer in the same solvent, and
the hydrodynamic contribution.

\begin{figure}[t]
\begin{center}
\includegraphics*[width=8.5cm,clip]{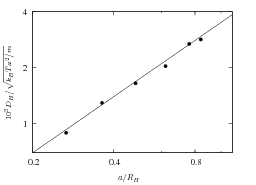}
\end{center}
\caption{\label{fig_3} Dependence of the hydrodynamic part of the
diffusion coefficient, $D_H = D- D_0/N_m$, on the hydrodynamic
radius for Gaussian chains of lengths $N_m =5$, $10$, $20$, $40$,
$80$, and $160$ (right to left). The mean free path is $\lambda = 0.1$.
From Ref.~\cite{muss:05}.}
\end{figure}

Simulation results for the hydrodynamic contribution, $D_H = D -D_0/N_m$,
to the diffusion coefficient are plotted in Fig.~\ref{fig_3} as a function
of the hydrodynamic radius (\ref{hydro_dyn_rad}). In the limit $N_m \gg
1$, the diffusion coefficient $D$ is dominated by the hydrodynamic
contribution $D_H$, since $D_H \sim N_m^{-1/2}$. For shorter
chains, $D_0/N_m$ cannot be neglected, and therefore has to be
subtracted in order to extract the scaling behavior of $D_H$. The
hydrodynamic part of the diffusion coefficient $D_H$ exhibits the
dependence predicted by the Kirkwood formula and the Zimm theory,
i.e., $D_H \sim 1/R_H$. The finite-size corrections
to $D$ show a dependence  $D= D_{\infty}- {\rm const}./L$ on
the size $L$ of a periodic system, in agreement with previous studies
\cite{male_00_dsp,duen:99,spen:00}. Simulations for various system
sizes for polymers of lengths $N_m=10$, $20$, and $40$ allow an
extrapolation to
infinite system size, which yields $D_0/\sqrt{k_BTa^2/m} \approx
1.7\times 10^{-2} $, in good agreement with the diffusion
coefficient of a monomer in the same solvent. The values of
$D_\infty$ are about $30\%$ larger than the finite-system-size
values presented in Fig.~\ref{fig_3}.
Similarly the diffusion coefficient for a polymer chain with
excluded volume interactions displays the dependence $D_H \sim
1/R_H$ \cite{muss:05}.

The Kirkwood formula neglects hydrodynamic fluctuations and is thus identical
with the preaveraging result of the Zimm approach. When only the hydrodynamic
part is considered, the Zimm model yields the diffusion coefficient
\begin{equation} \label{diff_zimm}
D_Z = 0.192 \frac{k_BT}{b \eta \sqrt{N_m}} .
\end{equation}
MPC simulations for polymers of length $N_m=40$ yield $D_Z/\sqrt{k_BTa^2/m}=
0.003$. This value agrees with the numerical value for an infinite
system, $D_H/\sqrt{k_BTa^2/m}= 0.0027$, within $10\%$. The MPC
simulations yield a diffusion coefficient smaller than $D^{(K)}$,
in agreement with previous studies presented in
Refs.~\cite{doi:86,fixm:83,liu:03}.
Note that the experimental values are also smaller by about $15
\%$ than those predicted by the Zimm approach
\cite{doi:86,schm:81,stoc:84}.

To further characterize the internal dynamics of the molecular
chain, a mode analysis in terms of the eigenfunctions
of the discrete Rouse model \cite{rous:53,doi:86} has been performed.
The mode amplitudes ${\vec \chi}_p$ are calculated according to
\begin{equation} \label{rouse_mode}
{\vec \chi}_p = \sqrt{\frac{2}{N_m}} \sum_{i=1}^{N_m} {\vec r}_i \cos \left[
\frac{p\pi}{N_m} \left(i-\frac{1}{2} \right)\right], \ p = 1 \ldots N_m .
\end{equation}
Due to hydrodynamic interactions, Rouse modes are no longer
eigenfunctions of the chain molecule. However, within the Zimm
theory, they are reasonable approximations and the autocorrelation
functions of the mode amplitudes decay exponentially, i.e.,
\begin{equation}
\left\langle {\vec \chi}_p (t) {\vec \chi}_p (0) \right\rangle =
\left\langle {\vec \chi}_p^2 \right\rangle \exp \left( - t/\tau_p
\right).
\end{equation}
For the Rouse model, the relaxation times $\tau_p$
depend on chain length and mode number according to $\tau_p \sim
1/\sin^2\left(p \pi / 2 N_m \right)$, whereas for the Zimm model
the dependence
\begin{equation}  \label{zimm_relax_time}
\tau_p \sim (p/N_m)^{1/2}/\sin^2\left(p \pi / 2N_m \right)
\end{equation}
is obtained. The extra contribution $\sqrt{p/N_m}$ follows from the
eigenfunction representation of the preaveraged hydrodynamic tensor, under the
assumption that its off-diagonal elements do not significantly contribute to
the relaxation behavior.

\begin{figure}[t]
\begin{center}
\includegraphics*[width=8.5cm,clip]{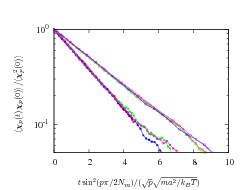}
\end{center}
\caption{\label{fig_modes} Correlation functions of the Rouse-mode
amplitudes for the modes $p= 1-4$ of Gaussian polymers. The chain
lengths are $N_m=20$ (right) and $N_m=40$ (left). From Ref.~\cite{muss:05}. }
\end{figure}

\begin{figure}[t]
\begin{center}
\includegraphics*[width=8.5cm,clip]{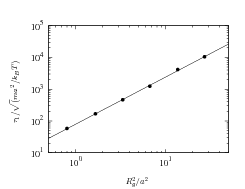}
\end{center}
\caption{\label{fig_tau} Dependence of the longest relaxation time
$\tau_1$ on the radius of gyration for Gaussian chains of the
lengths given in Fig.~\ref{fig_3}. From Ref.~\cite{muss:05}.}
\end{figure}

In Fig.~\ref{fig_modes}, the autocorrelation functions for the
mode amplitudes are shown for the mean free path $\lambda =0.1$.
Within the accuracy of the simulations, the correlation functions
decay exponentially and exhibit the scaling behavior predicted by
the Zimm model. Hence, for the small mean free path, hydrodynamic
interactions are taken into account correctly. This is no longer
true for the large mean free path, $\lambda = 2$. In this case,
a scaling behavior between that predicted by the Rouse and Zimm
models is observed. This implies that hydrodynamic interactions are
present, but are not fully developed or are small compared to the
local friction of the monomers. We obtain pure Rouse behavior for
a system without solvent by simply rotating the velocities of the
individual monomers \cite{muss:05}.

The dependence of the longest relaxation time on the radius of
gyration is displayed in Fig.~\ref{fig_tau} for $\lambda =0.1$.
The scaling behavior $\tau_1 \sim R^3_g$ is in very good agreement
with the predictions of the Zimm theory. We even find almost quantitative
agreement; the relaxation time of the $p=1$ mode of our
simulations is approximately $30$~\% larger than the Zimm value
\cite{doi:86}.

The scaling behavior of {\em equilibrium} properties of single
polymers with excluded-volume interactions has been studied
extensively \cite{doi:86,dege:79,decl:90,cepe:81,krem:88}. It has
been found that even very short chains  already follow the scaling
behavior expected for much longer chains. In particular, the
radius of gyration increases like $R_G \sim N_m^{\nu}$ with the
number of monomers, and the static structure factor $S({\vec q})$
exhibits a scaling regime for $2 \pi/R_G \ll q \ll 2 \pi/\sigma$,
with a $q^{-1/\nu}$ decay as a function of the scattering vector
$q$ and the exponent $\nu \approx 0.6$. For the interaction
potentials (\ref{pot_ev}), (\ref{pot_lj}) with the parameters
$b=\sigma=a$, $\epsilon/k_BT=1$, the exponent $\nu \approx 0.62$ is obtained
from the chain-length dependence of the radius of gyration, the
mean square end-to-end distance, as well as the $q-$dependence of
the static structure factor \cite{muss:05}.

\begin{figure}[t]
\begin{center}
\includegraphics*[width=8.5cm,clip]{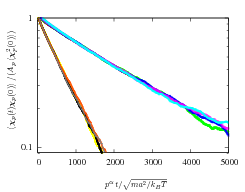}
\end{center}
\caption{\label{fig_corr_modes_ev} Correlation functions of the
Rouse-mode amplitudes for various modes as a function of the
scaled time $tp^{\alpha}$ for polymers with excluded volume
interactions. The chain lengths are $N_m=20$ (left) and $N_m=40$
(right). The calculated correlations where fitted by $A_p \exp
(-t/\tau_p)$ and have been divided by $A_p$. The scaling exponents
of the mode numbers are $\alpha= 1.93$ ($N_m=20$) and $\alpha=
1.85$ ($N_m=40$), respectively. From Ref.~\cite{muss:05}. }
\end{figure}

An analysis of the intramolecular dynamics in terms of the Rouse
modes yields non-exponentially decaying autocorrelation functions
of the mode amplitudes. At very short times, a fast decay is found,
which turns into a slower exponential decay which is well fitted
by $A_p \exp (-t/\tau_p)$, see Fig.~\ref{fig_corr_modes_ev}.
Within the accuracy of
these calculations, the correlation functions exhibit universal
behavior. Zimm theory predicts the dependence $\tau_p \sim
p^{-3\nu}$ for the relaxation times on the {\em mode number} for
polymers with excluded-volume interactions \cite{doi:86}. With
$\nu = 0.62$, the exponent $\alpha$ for the polymer of length
$N_m=40$ is found to be in excellent agreement with the theoretical
prediction. The exponent for the polymers with $N_m=20$ is slightly
larger.

\begin{figure}[t]
\begin{center}
\includegraphics*[width=8.5cm,clip]{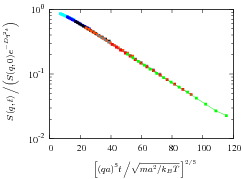}
\end{center}
\caption{\label{fig_dsf_ev_n40} Normalized dynamic structure
factor $S({\bf q},t)/ (S({\bf q},0) \exp \left(-D q^2 t \right))$
of polymers with excluded volume interactions for $N_m=40$ and
various $q$-values in the range $0.7 < qa < 2$ as a function of
$q^{2}t^{2/3}$. From Ref.~\cite{muss:05}. }
\end{figure}

Zimm theory predicts that the dynamic structure factor, which is
defined by
\begin{equation} \label{dyn_struc_def}
S({\vec q},t) = \frac{1}{N_m}\sum_{i=1}^{N_m} \sum_{j=1}^{N_m} \lla \exp
\left(i {\vec q} [{\vec r}_i(t) - {\vec r}_j(0) ] \right) \rra,
\end{equation}
scales as \cite{doi:86}
\begin{equation} \label{dyn_struc_scal}
S({\vec q},t) = S({\vec q},0) f(q^{\alpha} t)
\end{equation}
with $\alpha =3$ for $q R_G \gg 1$, independent of the solvent conditions
($\Theta$ or good solvent). To extract the scaling relation for
the intramolecular dynamics, which corresponds to the
prediction (\ref{dyn_struc_scal}), we resort to the following
considerations. As is well known, the dynamic structure factor for
a Gaussian distribution of the differences ${\vec r}_i(t) - {\vec
r}_j(0)$ and a linear equation of motion is given by
\cite{doi:86,wink:97}
\begin{eqnarray} \label{dyn_struc_gauss}
S({\vec q},t)  & = & S({\vec q},0) e^{-D q^2 t}
\frac{1}{N_m}\sum_{i=1}^{N_m} \sum_{j=1}^{N_m} \exp \left(- q^2 \lla ({\vec
r}'_i(t) -{\vec r}'_j(0))^2 \rra /6\right),
\end{eqnarray}
where $D q^2 t$ accounts for the center-of-mass dynamics and
${\vec r}'_i$ denotes the position of monomer $i$ in the center-of-mass
reference frame. Therefore, in order to obtain the dynamics in the
center-of-mass reference frame, we plot $S({\vec q},t) / (S({\vec
q},0) \exp \left(-D q^2 t \right))$. The simulation results for
the polymer of length $N_m=40$, shown in
Fig.~\ref{fig_dsf_ev_n40}, confirm the predicted scaling behavior.
Thus, MPC-MD hybrid simulations are very well suited to study
the dynamics of even short polymers in dilute solution.

As mentioned above, the structure of a polymer depends on the
nature of the solvent. In good solvent excluded volume
interactions lead to expanded conformations and under bad solvent
conditions the polymer forms a dense coil. In a number of
simulations the influence of hydrodynamic interactions on the
transition from an extended to a collapsed state has been studied,
when the solvent quality is abruptly changed. Both, molecular
dynamics simulations with an explicit solvent \cite{chan:01} as
well as MPC simulations \cite{yeo02,yeo05b,lee:06} yield a
significant different dynamics in the presence of hydrodynamic
interactions. Specifically, the collapse is faster, with a much
weaker dependence of the characteristic time on polymer length
\cite{yeo05b,lee:06}, and the folding path is altered.

Similarly, a strong influence of hydrodynamic interactions has
been found on the polymer translocation dynamics through a small
hole in a wall \cite{ali:05} or in polymer packing in a virus
capsid \cite{ali:04,ali:06}. Cooperative backflow effects lead to
a rather sharp distribution of translocation times with a peak at
relatively short times. The fluid flow field, which is created as
a monomer moves through the hole, guides following monomers to
move in the same direction.

\subsection{Polymers in Flow Fields}

Simulations of a MPC fluid confined between surfaces and exposed
to a constant external force yield the expected parabolic velocity
profile for appropriate boundary conditions
\cite{lamu_01_mcd,alla_02_mss,wata:07}. The ability of MPC to account for the
flow behavior of mesoscale objects, such as polymers, under
non-equilibrium conditions has been demonstrate for a number of
systems. Rod-like colloids in shear flow exhibit flow induced
alignment \cite{wink:04}. The various diagonal components of the
radius of gyration tensor exhibit a qualitative and quantitative
different behavior. Due to the orientation, the component in the
flow direction increases with increasing Peclet number larger than
united and saturates at large shear rates because of finite size
effects. The transverse components decrease with shear rate, where
the component in the gradient direction is reduced to a greater
extent. The rod rotational velocity in the shear plane shows two
distinct regimes. For Peclet numbers much smaller than unity, the
rotational velocity increases linearly with the shear rate,
because the system is isotropic. At Peclet numbers much larger
than unity, the shear induced anisotropies lead to a slower
increase of the rotational velocity with the shear rate
\cite{wink:04}.

The simulations of a tethered polymer in a Poiseuille flow
\cite{webs_05_mtp} yield a series of morphological transitions from
sphere to deformed sphere to trumpet to stem and flower to rod,
similar to theoretically predicted structures
\cite{broch:93,broch:94,broch:95}. The crossovers between the
various regimes occur at flow rates close to the theoretical
estimates for a similar system. Moreover, the simulations in
Ref.~\cite{webs_05_mtp} show that backflow effects lead to an effective
increase in viscosity, which is attributed to the fluctuations of
the free polymer end rather than its shape.

The conformational, structural, and transport properties of free
flexible polymers in microchannel flow have been studied in
Refs.~\cite{wata:07,cann:07} by hybrid MPC-molecular dynamics
simulations. These simulations confirm the cross streamline
migration of the molecules as previously observed in
Refs.~\cite{agar:94,jend:04,usta:06,khar:06,stei:06,usta:07}. In
addition, various other polymer properties are addressed in
Ref.~\cite{cann:07}.

All these hybrid simulations confirm that MPC is an excellent
method to study the non-equilibrium behavior of polymers in flow
fields. In the next section, we will provide a more detailed
example for a more complicated object, namely an ultrasoft colloid
in shear flow.

\begin{figure}[t]
\begin{center}
\includegraphics[height=8.5cm,angle=-90]{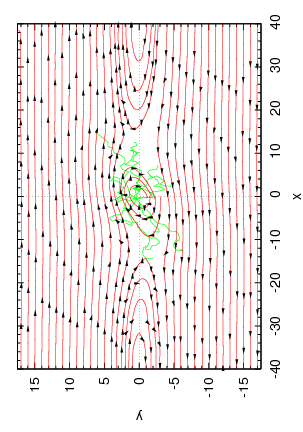}
\includegraphics[height=8.5cm,angle=-90]{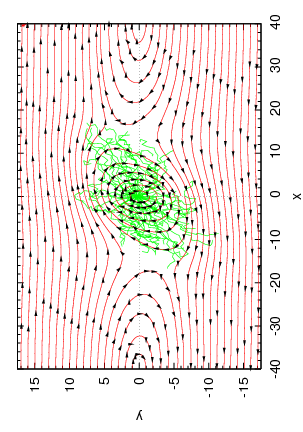}
\end{center}
\caption{Fluid flow lines in the flow-gradient plane of the star
polymer's center of mass reference frame for  $f=10$ (top) and
$f=50$ (bottom) arms, both with $L_f=20$ monomers per arm and an
applied shear field with $ {\rm Wi}=\dot{\gamma}\tau=22$.
From Ref.~\cite{ripo_07_hss}.}
\label{fig:fs.f}
\end{figure}

\subsection{Ultra-soft Colloids in Shear Flow}
\label{sec:ultrasoft_shear}

Star polymers present a special macromolecular architecture, in
which several linear polymers of identical length are linked
together by one of their ends at a common center. This structure
is particularly interesting because it allows for an almost continuous
change of properties from that of a flexible linear polymer to a
spherical colloidal particle with very soft interactions. Star
polymers are therefore also often called ultrasoft colloids. The
properties of star polymers in and close to equilibrium have been
studied intensively, both theoretically \cite{gres:87,liko:01} and
experimentally \cite{vlas:01}. A star polymer is a ultrasoft
colloid, where the core extension is very small compared to the
length of an arm. By anchoring polymers on the surface of a hard
colloid, the softness can continuously be changed from ultrasoft
to hard by increasing the ratio between the core and shell radius
at the expense of the thickness of the soft polymer corona.
Moreover, star polymers have certain features in common with
vesicles and droplets. Although their shell can be softer than
that of the other objects, the dense packing of the monomers will
lead to a cooperative dynamical behavior resembling that of
vesicles or droplets \cite{ripo:06}.

Vesicles and droplets encompass fluid which is not exchanged with
the surrounding. In contrast, for star-like molecules fluid is
free to penetrate into the molecule and internal fluid is
exchanged with the surrounding in the course of time. This
intimate coupling of the star-polymer dynamics and the fluid flow
leads to a strong modification of the flow behavior at and next to
the ultrasoft colloid particular in non-equilibrium systems.

\begin{figure}[t]
\begin{center}
\includegraphics[height=8.5cm,angle=-90]{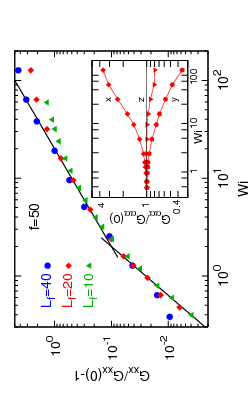}
\end{center}
\caption{Normalized component $G_{xx}$ of the average gyration
tensor as a function of the Weissenberg number ${\rm Wi}$, for
star polymers of $f=50$ arms and the arm lengths $L_f = 10$, $20$,
and $40$ monomers. Power-law behaviors with quadratic and linear
dependencies on $\rm Wi$ are indicated by lines. The inset shows
all three diagonal components of the gyration tensor (for $L_f =
20$). From Ref.~\cite{ripo:06}.}
\label{fig:Gaa.f}
\end{figure}

In the following, we will discuss a few aspects in the behavior of
star polymers in shear flow as a function of arm number $f$, arm
length $L_f$, and shear rate $\dot \gamma$. The polymer model
is the same as described in Sec.~\ref{subsubsec:model}, where the
chain connectivity is determined by
the bond potential (\ref{pot_ev}) and the excluded-volume
interaction is described by the Lennard-Jones potential
(\ref{pot_lj}). A star polymer of functionality $f$ is modeled as
$f$ linear polymer chains of $L_f$ monomers each, with one of
their ends linked to a central particle. Linear polymer molecules
are a special case of star polymers with functionality $f=2$.
Lees-Edwards boundary conditions \cite{alle:87} are employed in
order to impose a linear velocity profile $(v_x,v_y,v_z) =
(\dot{\gamma} r_y,0,0)$ in the fluid in the absence of a polymer.
For small shear rates, the conformations of star polymers remain
essentially unchanged compared to the equilibrium state. Only when
the shear rate exceeds a characteristic value, a structural
anisotropy as well as an alignment is induced by the flow (cf.
Fig.~\ref{fig:fs.f}). The shear rate dependent quantities are
typically presented in terms of the Weissenberg number $\rm Wi=
\dot \gamma \tau$ rather than the shear rate itself, where $\tau$
is the longest characteristic relaxation time of the considered
system. For the star polymers, the best data collapse for
stars of various arm lengths is found when the relaxation time
$\tau = \eta b^3 L^{2}_f/k_BT$ is used \cite{ripo:06}. Remarkably,
there is essentially
no dependence on the functionality. Within the range of investigated
star sizes, this relaxation time has to be considered as
consistent with the prediction for the blob model of Ref.~\cite{gres:89},
where $\tau \sim L^{1.8}_f f^{0.1}$ for the Flory exponent $\nu=0.6$.

In Fig.~\ref{fig:fs.f}, typical star conformations are shown
which indicate the alignment and induced anisotropy in the flow.
Moreover, the figure reveals the intimate coupling of the polymer
dynamics and the emerging fluid flow field. In the region, where
the fluid coexists with the star polymer, the externally imposed
flow field is strongly screened and the fluid velocity is no
longer aligned with the shear flow direction, but rotates around
the polymer center of mass. The fluid stream lines are calculated
by integration of the coarse-grained fluid velocity field. Outside
the region covered by the star polymer, the fluid adapts to the
central rotation by generating two counter-rotating vortices, and
correspondingly two {\em stagnation points} of vanishing fluid
velocity \cite{ripo_07_hss}.

The fluid flow in the vicinity of the star polymer is
distinctively different from that of a sphere but resembles the
flow around an ellipsoid \cite{miku:04}. In contrast to the
latter, the fluid penetrates into the area covered by the star
polymer. While the fluid in the core of the star rotates together
with the polymer, the fluid in the corona follows the external
flow to a certain extent.

A convenient quantity to characterize the structural properties
and alignment of polymers in flow is the average gyration tensor,
which is defined as
\begin{equation}
G_{\alpha\beta}({\dot\gamma}) = \frac{1}{N_m}
       \sum_{i=1}^{N_m} \langle r'_{i,\alpha} r'_{i,\beta} \rangle \ ,
\end{equation}
where $N_m = f L_f +1$ is the total number of monomers,
$r'_{i,\alpha}$ is the position of monomer $i$ relative to the
polymer center of mass, and $\alpha \in\{x,y,z\}$. The average
gyration tensor is directly accessible in scattering experiments.
Its diagonal components $G_{\alpha\alpha}(\dot{\gamma})$ are the
squared radii of gyration of the star polymer along the axes of
the reference frame. In the absence of flow, scaling
considerations predict \cite{gres:87}
$G_{xx}(0)=G_{yy}(0)=G_{zz}(0)=R_g^2(0)/3 \sim
L_f^{2\nu}f^{1-\nu}$.

The diagonal components $G_{\alpha\alpha}$ of the average gyration
tensor are shown in Fig.~\ref{fig:Gaa.f} as a function of the
Weissenberg number for various functionalities and arm lengths. We
find that the extension of a star increases with increasing shear
rate in the shear direction ($x$), decreases in the gradient
direction ($y$), and is almost independent of $\rm Wi$ in the
vorticity direction ($z$). The deviation from spherical symmetry
exhibits a ${\rm Wi}^2$ power-law dependence for small shear rates
for all functionalities. A similar behavior has been found for
rod-like colloids \cite{wink:04} (due to the increasing alignment
with the flow direction) and for linear polymers \cite{doi:86}.
However, for stars of not too small functionality, a new scaling
regime appears, where the deformation seems to scales {\em
linearly} with the Weissenberg number. For large shear rates,
finite-size effects appear due to the finite monomer number. These
effects emerge when the arms are nearly stretched, and therefore
occur at higher Weissenberg numbers for larger arm lengths.

\begin{figure}[t]
\begin{center}
\includegraphics[height=8.5cm,angle=-90]{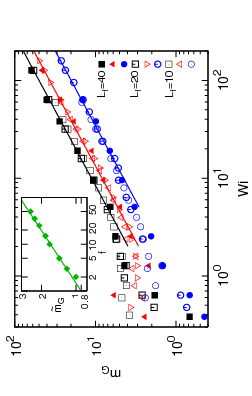}
\end{center}
\caption{Orientational resistance $m_G$ as a function of the
Weissenberg number $\rm Wi$ for star polymers with functionalities
$f=2$ (circles), $f=15$ (triangles), and $f=50$ (squares), and
different arm lengths indicated in the figure. Lines correspond to
the power law $m_G = {\tilde m}_G(f) {\rm Wi}^{0.65}$. The inset
shows that the amplitude also follows a power-law behavior with
$\tilde{m}_G(f) \sim f^{0.27}$. From Ref.~\cite{ripo:06}. }
\label{chi.f}
\end{figure}

The average flow alignment of a (star) polymer can be
characterized by the orientation angle $\chi_G$, which is the
angle between the eigenvector of the gyration tensor with the
largest eigenvalue and the flow direction.  It follows
straightforwardly \cite{aust:99} from the simulation data via
\begin{equation}
\tan(2 \chi_G) = 2G_{xy}/(G_{xx} - G_{yy}) \equiv m_G/{\rm Wi},
\end{equation}
where the right-hand-side of the equation defines the orientation
resistance parameter $m_G$ \cite{link:93}. It has been shown for
several systems including rod-like colloids and linear polymers
without self-avoidance \cite{doi:86} that close to equilibrium
$G_{xy}\sim {\dot\gamma}$ and $(G_{xx} - G_{yy}) \sim
{\dot\gamma}^2$, so that $m_G$ is independent of $\rm Wi$. Our
results for the orientation resistance are presented in
Fig.~\ref{chi.f} for various functionalities $f$ and arm lengths
$L_f$.  Data for different $L_f$ collapse onto universal curves,
which approach a plateau for small shear rates, as expected. For
larger shear rates, ${\rm Wi} \gg 1$, a power-law behavior
\cite{ripo:06}
\begin{equation}
m_G({\rm Wi}) \sim f^\alpha \ {\rm Wi}^\mu,
\end{equation}
is obtained with respect to the Weissenberg number and the
functionality, where $\alpha = 0.27 \pm 0.02$ and $\mu = 0.65 \pm 0.05$.
For self-avoiding linear polymers, a somewhat smaller exponent
$\mu=0.54 \pm 0.03$ was obtained in Refs.~\cite{aust:99,teix:05},
whereas theoretical calculations predict $\sim {\rm Wi}^{2/3}$ in
the limit of large Weissenberg numbers \cite{wink:06}.

\begin{figure}[t]
\begin{center}
\includegraphics[height=8.5cm,angle=-90]{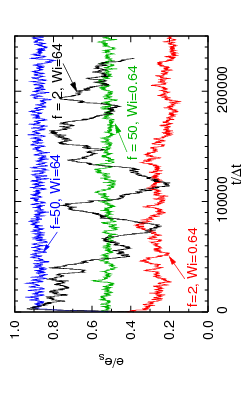}
\end{center}
\caption{Temporal evolution of the largest intramolecular distance
$e=\max_{ij} |{\bf r}_i-{\bf r}_j|$ of a linear polymer and a star
polymer with $50$ arms, for the Weissenberg numbers $\rm Wi=0.64$
and $\rm Wi=64$. In both cases, $L_f=40$. The time $t$ is measured
in units of the collision time $\delta t$. $e_s$ corresponds to
the fully stretched arms. From Ref.~\cite{ripo:06}.}
\label{fig:extens}
\end{figure}

\begin{figure}[h]
\begin{center}
\includegraphics[height=8.5cm,angle=-90]{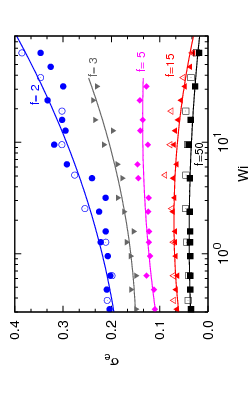}
\end{center}
\caption{Widths of the distribution functions of the largest
intramolecular distances, $\sigma_e = (\langle e^2 \rangle -
\langle e \rangle^2)/ \langle e \rangle^2$, of a linear polymer
and star polymers with up to $50$ arms as a function of the
Weissenberg number. From Ref.~\cite{ripo:06}.}
\label{fig:sigma_ext}
\end{figure}

\begin{figure}[t]
\begin{center}
\includegraphics*[width=12cm]{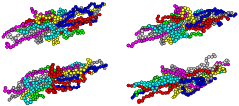}
\end{center}
\caption{\label{star_tumbling} Successive snapshots of a star
polymer of functionality $f=25$ and arm length $L_f=20$, which
illustrate the tank-treating motion.}
\end{figure}

The data for the average orientation and deformation of a star
polymer described so far seem to indicate that the properties vary
smoothly and monotonically from linear polymers to star polymers
of high functionality. However, this picture changes when the {\em
dynamical behavior} is considered. It is well known by now that
linear polymers show a tumbling motion in flow, with alternating
collapsed and stretched configurations during each cycle
\cite{dege:74,ledu:99,smit:99,teix:05}. For large Weissenberg
numbers, this leads to very large fluctuations of the largest
intramolecular distance of a linear polymer with time, as
demonstrated experimentally in Refs.~\cite{smit:99,teix:05}, and
reproduced in the MPC simulations \cite{ripo:06}, see Fig.~\ref{fig:extens}.
A similar behavior is found for $f=3$. However, for $f > 5$, a
quantitatively different behavior is observed as displayed in
Fig.~\ref{fig:sigma_ext}. Now, the fluctuations of the largest
intramolecular distance are much smaller and {\em decrease} with
increasing Weissenberg number as shown in Fig.~\ref{fig:extens},
and the dynamics resembles much more the continuous tank-treading
motion of fluid droplets and capsules.  The shape and orientation
of such stars depends very little on time, while the whole object
is rotating.  On the other hand, a single, selected arm resembles
qualitatively the behavior of a linear polymer---it also
collapses and stretches during the tank-treading motion. The
successive snapshots of Fig.~\ref{star_tumbling} illustrate the
tank-treating motion. Following the top left red polymer in the
top left image, we see that the extended polymer collapses in the
course of time, moves to the right and stretches again. In
parallel, other polymers exhibit a similar behavior on the bottom
side. Moreover, the images show that the orientation of a star
hardly changes in the course of time.

\begin{figure}[t]
\begin{center}
\includegraphics[height=8.5cm,angle=-90]{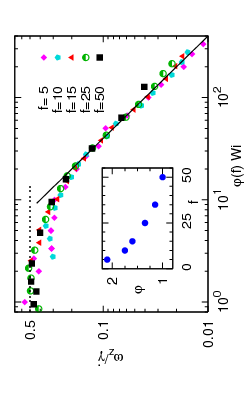}
\end{center}
\caption{Scaled rotation frequencies as function of a rescaled
Weissenberg number for various functionalities. Dashed and full
lines correspond to $\omega/{\dot\gamma}=1/2$ for small $\rm Wi$,
and $\omega/{\dot\gamma}\sim 1/{\rm Wi}$ for large $\rm Wi$,
respectively. The inset shows the dependence of the rescaling
factor $\varphi$ on the functionality. From Ref.~\cite{ripo:06}.}
\label{fig:frequency}
\end{figure}

The rotational dynamics of a star polymer can be characterized
quantitatively by calculating the rotation frequency
\begin{equation}\label{rot_freq}
\omega_{\alpha} = \sum_{\beta=x}^{z}
     \langle \Theta_{\alpha\beta}^{-1} L_\beta \rangle
\end{equation}
of a star, where
\begin{equation}\label{inertia}
\Theta_{\alpha\beta} = \sum_{i=1}^{N_m} [{\vec r'}_i^2
\delta_{\alpha\beta} - r'_{i,\alpha}r'_{i,\beta}]
\end{equation}
is the instantaneous moment-of-inertia tensor and $L_\beta$ is the
instantaneous angular momentum. Since the rotation frequency for
all kinds of soft objects---such as rods, linear polymers, droplets
and capsules---depends linearly on $\dot\gamma$ for small shear
rates, the reduced rotation frequency $\omega/{\dot\gamma}$ is shown
in Fig.~\ref{fig:frequency} as a function of the Weissenberg
number.  The data approach $\omega/{\dot\gamma}=1/2$ for small
$\rm Wi$, as expected \cite{aust:99,aust:02}. For larger shear
rates, the reduced frequency decreases due to the deformation and
alignment of the polymers in the flow field. With increasing arm
number, the decrease of $\omega/{\dot\gamma}$ at a given
Weissenberg number becomes smaller, since the deviation from the
spherical shape decreases. Remarkably, the frequency curves for
all stars with $f>5$ are found to collapse onto a universal scaling
function when $\omega/{\dot\gamma}$ is plotted as a function of a
rescaled Weissenberg number, see Fig.~\ref{fig:frequency}.
For high shear rates, $\omega/{\dot\gamma}$ decays as ${\rm
Wi}^{-1}$, which implies that the rotation frequency becomes {\em
independent} of $\dot\gamma$. A similar behavior has been observed
for capsules at high shear rates \cite{navo:98}.

The presented results show that star polymers in shear flow show a
very rich structural and dynamical behavior. With increasing
functionality, stars in flow change from linear-polymer-like to
capsule-like behavior. These macromolecules are therefore
interesting candidates to tune the viscoelastic properties of
complex fluids.


\section{Vesicles and Cells in Hydrodynamic Flows}
\label{sec:vesicles}

\subsection{Introduction}
\label{sec:ves_intro}

The flow behavior of fluid droplets, capsules, vesicles and cells
is of enormous importance in science and technology. For example,
the coalescence and break-up of fluid droplets is essential for
emulsion formation and stability. Capsules and vesicles are discussed
and used as drug carriers. Red blood cells (RBC) flow in the blood stream,
and may coagulate or be torn apart under unfavorable flow conditions.
Red blood cells also have to squeeze through narrow capillaries to
deliver their oxygen cargo.
White blood cells in capillary flow adhere to, roll along and
detach again from the walls of blood vessels under normal physiological
conditions; in inflamed tissue, leukocyte rolling leads to firm adhesion
and induces an immunological response.

These and many other applications have induced an intensive theoretical
and simulation activity to understand and predict the behavior of such
soft, deformable objects in flow.  In fact, there are some general,
qualitative properties in simple shear flow, which are shared by droplets,
capsules, vesicles and cells. When the internal viscosity is low and
when they are highly deformable, a tank-treading motion is observed
(in the case of droplets for not too high shear rates),
where the shape and orientation are
stationary, but particles localized at the interface or attached to the
membrane orbit around the center of mass with a rotation axis in the
vorticity direction. On the other hand, for high internal viscosity or
small deformability, the whole object performs a tumbling motion, very
much like a colloidal rod in shear flow. However, if we take a more
careful look, then the behavior of droplets, capsules, vesicles and cells
is quite different. For example, droplets can break up easily at higher
shear rates, because their shape is determined by the interfacial tension;
fluid vesicles can deform much more easily then capsules, since their
membrane has no shear elasticity; etc.
We focus here on the behavior of fluid vesicles and red blood cells.

\subsection{Modeling Membranes}
\label{sec:memb_modeling}

\subsubsection{Modeling Lipid-Bilayer Membranes}
\label{sec:bilayer_models}

The modeling of lipid bilayer membranes depends very much on the
length scale of interest. The structure of the bilayer itself or
the embedding of membrane proteins in a bilayer are best studied
with {\em atomistic
models} of both lipid and water molecules. Molecular dynamics
simulations of such models are restricted to about $10^3$ lipid
molecules. For larger system sizes, {\em coarse-grained models} are
required \cite{goet98,gg:gomp99a,otte03}. Here, the hydrocarbon
chains of lipid molecules are
described by short polymer chains of Lennard-Jones particles,
which have a repulsive interaction with the lipid head groups as well
as with the water molecules, which are also modeled as single
Lennard-Jones spheres. Very similar models, with Lennard-Jones
interactions replaced by linear ``soft'' dissipative-particle dynamics
(DPD) potentials, have also
been employed intensively \cite{shil02,rekv04,lara04,orti05,vent06}.
For the investigation of shapes and thermal fluctuations of single-
or multi-component membranes, the hydrodynamics of the solvent is
irrelevant. In this case, it can be advantageous to use a {\em
solvent-free membrane model}, in which the hydrophobic effect of
the water molecules is replaced by an effective attraction among the
hydrocarbon chains \cite{nogu01b,nogu02c,fara03,cook05a}. This approach
is advantageous in the case of membranes in dilute solution, because
it reduces the number of molecules---and thus the degrees of freedom
to be simulated---by orders of magnitude. However,
it should be noticed that the basic length
scale of atomistic and coarse-grained or solvent-free models
is still on the same order of magnitude.

In order to simulate larger systems, such as giant unilamellar
vesicles (GUV) or red blood cells (RBC), which have a radius on the
order of several micrometers, a different approach is required.
It has been shown that in this limit
the properties of lipid bilayer membranes are described very well
by modeling the membrane as a two-dimensional manifold embedded in
three-dimensional space, with the shape and fluctuations controlled
by the curvature elasticity \cite{helf73},
\begin{equation} \label{eq:helf}
{\cal H} = \int dS \ 2\kappa H^2 \ ,
\end{equation}
where $H=(c_1+c_2)/2$ is the mean curvature, with the local
principal curvatures $c_1$ and $c_2$, and the integral is over the
whole membrane area.
To make the curvature elasticity amenable to computer simulations,
it has to be discretized. This can be done either by using
triangulated surfaces \cite{gg:gomp97f,gg:gomp04c}, or by employing
particles with properly designed interactions which favor the
formation of self-assembled, nearly planar sheets \cite{drou91,nogu06}.
In the latter case, both scalar particles with isotropic multi-particle
interactions (and a curvature energy obtained from a moving least-squares
method) \cite{nogu06} as well as particles with an internal spin variable
and anisotropic, multi-body forces \cite{drou91} have been employed
and investigated.

\subsubsection{Dynamically Triangulated Surfaces}
\label{sec:triang_surf}

\begin{figure}
\begin{center}
\includegraphics*[width=10.0cm]{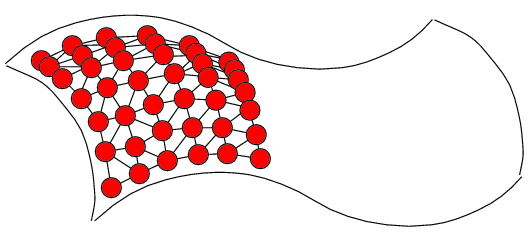}
\end{center}
\caption{ \label{fig:dyn_triang}
Triangulated-network model of a fluctuating membrane. All vertices have
short-range repulsive interactions symbolized by hard spheres. Bonds
represent attractive interactions with imply a maximum separation
$\ell_0$ of connected vertices.
From Ref.~\cite{gomp02_FS}.
}
\end{figure}

In a dynamically-triangulated surface model
\cite{ho90,gg:gomp92a,boal92a,gg:gomp97f,gg:gomp04c} of vesicles and
cells, the membrane is described by $N_{mb}$ vertices which are
connected by tethers to form a triangular network of spherical topology,
see Fig.~\ref{fig:dyn_triang}.
The vertices have excluded volume and mass $m_{mb}$.
Two vertices connected by a bond have an attractive interaction, which
keeps their distance below a maximum separation $\ell_0$. A short-range
repulsive interaction among all vertices makes the network self-avoiding
and prevents very short bond lengths.
The curvature energy can be discretized in different ways
\cite{gg:gomp96b,gg:gomp97f}. In particular, the discretization
\cite{itzy86,gg:gomp96b}
\begin{equation}
U_{cv} = \frac{\kappa}{2} \sum_i \frac{1}{\sigma_i}
  \left\{ \sum_{j(i)} \frac{\sigma_{i,j}{\bf r}_{i,j}}{r_{i,j}} \right\}^2
\end{equation}
has been found to give reliable results in comparison with the
continuum expression (\ref{eq:helf}). Here,
the sum over $j(i)$ is over the neighbors of a vertex $i$ which are
connected by tethers. The bond vector between the vertices $i$ and $j$ is
${\bf r}_{i,j}$, and $r_{i,j}=|{\bf r}_{i,j}|$.
The length of a bond in the dual lattice is
$\sigma_{i,j}=r_{i,j}[\cot(\theta_1)+\cot(\theta_2)]/2$,
where the angles $\theta_1$ and $\theta_2$ are opposite to bond $ij$ in
the two triangles sharing this bond.  Finally,
$\sigma_i=0.25\sum_{j(i)} \sigma_{i,j}r_{i,j}$ is the area of the dual
cell of vertex $i$.

To model the fluidity of the membrane, tethers can be flipped between
the two possible diagonals of two adjacent triangles.
A number $\psi N_{b}$ of bond-flip attempts is performed with
the Metropolis Monte Carlo method \cite{gg:gomp96b} at time intervals
$\Delta t_{BF}$,
where $N_{b}=3(N_{mb}-2)$ is the number of bonds in the
network, and $0<\psi<1$ is a parameter of the model. Simulation results
show that the vertices of a dynamically triangulated membrane show
diffusion, i.e., the squared distance of two initially neighboring
vertices increases linearly in time.

\subsubsection{Vesicle Shapes}
\label{sec:ves_shapes}

\begin{figure}[ht]
\begin{center}
\includegraphics[width=11.0cm]{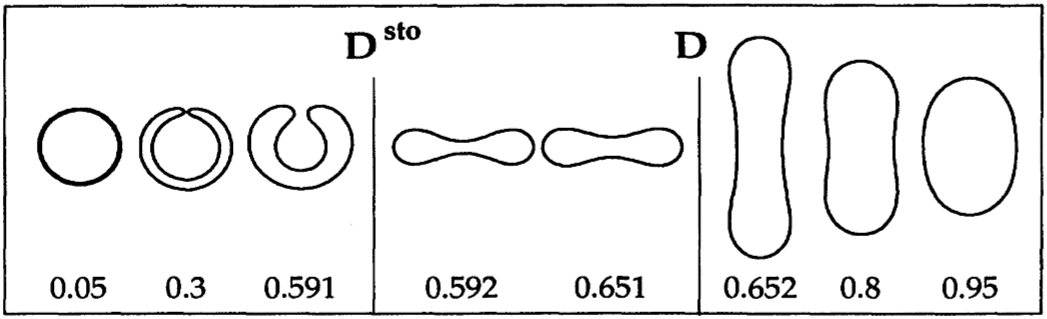}
\end{center}
\caption{ \label{fig:ves_shapes}
Shapes of fluid vesicles as a function of the reduced volume $V^*$.
$D$ and $D^{sto}$ denote the discontinuous prolate-oblate and
oblate-stomatocyte transitions, respectively. All shapes display
rotational symmetry with respect to the vertical axis.
From Ref.~\cite{seif91b}.
}
\end{figure}

Since the solubility of lipids in water is very low, the number of lipid
molecules in a membrane is essentially constant over typical experimental
time scales. Also, the osmotic pressure generated by a small number of
ions or macromolecules in solution, which cannot penetrate the lipid
bilayer, keeps the internal volume essentially constant.
The shape of fluid vesicles \cite{seif91b} is therefore determined by
the competition
of the curvature elasticity of the membrane, and the constraints of
constant volume $V$ and constant surface area $S$. In the simplest case
of vanishing spontaneous curvature, the curvature elasticity is given
by Eq.~(\ref{eq:helf}). In this case, the vesicle shape in the absence
of thermal fluctuations depends on a single dimensionless parameter, the
reduced volume $V^*= V/V_0$, where $V_0 = (4\pi/3) R_0^3$ and
$R_0=(S/4\pi)^{1/2}$ are the volume and radius of a sphere of the same
surface area $S$, respectively. The calculated vesicle shapes are shown
in Fig.~\ref{fig:ves_shapes}. There are three phases. For reduced volumes
not too far from the sphere, elongated prolate shapes are stable. In a
small range of reduced volumes of $V^* \in [0.592,0.651]$, oblate discocyte
shapes have the lowest curvature energy. Finally, at very low reduced
volumes, cup-like stomatocyte shapes are found.

These shapes are very well reproduced in simulations with dynamically
triangulated surfaces \cite{gg:gomp94k,gg:gomp95a,gg:gomp03c,nogu_05_dfv}.

\subsubsection{Modeling Red Blood Cells}
\label{sec:RBC_models}

Red blood cells have a biconcave disc shape, which can hardly be
distinguished from the discocyte shape of fluid vesicles with
reduced volume $V^*\simeq 0.6$, compare Fig.~\ref{fig:ves_shapes}.
However, the membrane of red blood cells is more complex, since a
spectrin network is attached to the plasma
membrane \cite{albe07}, which helps to retain the integrity of the
cell in strong shear gradients or capillary flow.  Due to the
spectrin network, the red blood cell membrane has a non-zero shear
modulus $\mu$.

The bending rigidity $\kappa$ of RBCs has been measured by micropipette
aspiration \cite{moha94} and atomic force microscopy \cite{sche01} to
be approximately $\kappa=50 k_BT$.  The shear modulus
of the composite membrane, which is induced by the spectrin
network, has been determined by several techniques; it is found
to be $\mu=2\times 10^{-6}$ N/m from optical tweezers manipulation
\cite{leno01}, while the value $\mu=6\times 10^{-6}$ N/m is obtained
from micropipette aspiration \cite{moha94}. Thus, the
dimensionless ratio $\mu R_0^2/\kappa \simeq 100$, which implies
that bending and stretching energies are roughly of equal importance.

Theoretically, the shapes of RBCs in the absence of flow have been
calculated very successfully on the basis of a mechanical model of
membranes, which includes both curvature and shear elasticity
\cite{disc98,lim02}. In particular, it has been shown recently
that the full stomatocyte-discocyte-echinocyte sequence
of RBCs can be reproduced by this model \cite{lim02}.

The composite membrane of a red blood cell, consisting of the lipid
bilayer and the spectrin network, can be modeled as a composite network,
which consists
of a dynamically-triangulated surface as in the case of fluid vesicles,
coupled to an additional network of harmonic springs with fixed
connectivity (no bond-flip) \cite{disc98,gg:gomp05g}. Ideally, the
bond length of the elastic network is larger than of the fluid mesh
\cite{disc98}---in order to mimic the situation of the red blood cell
membrane, where the average distance of anchoring points is about 70 nm,
much larger than the size of a lipid molecule---and thereby allow, for
example, for thermal fluctuations of the distances between neighboring
anchoring points. On the other hand, to investigate the behavior of cells
on length scales much larger the mesh size of the spectrin network,
it is more efficient to use the same number of the bonds for
both the fluid and the tethered networks \cite{gg:gomp05g}.
The simplest case is a harmonic tethering potential,
$(1/2)k_{\rm {el}}({\bf r}_{i}-{\bf r}_{j})^2$.
This tether network generates a shear modulus $\mu =\sqrt{3}k_{\rm {el}}$.

\subsection{Modeling Membrane Hydrodynamics}
\label{sec:memb_hydro}

Solvent-free models, triangulated surfaces and other discretized curvature
models have the disadvantage that they do not contain a solvent,
and therefore do not describe the hydrodynamic behavior correctly.
However, this apparent disadvantage can be turned into an advantage by
combining these models with a mesoscopic hydrodynamics technique.
This approach has been employed for dynamically triangulated surfaces
\cite{nogu_04_fvv,nogu_05_dfv} and for meshless
membrane models in combination with MPC \cite{gg:gomp06h}, as well as for
fixed membrane triangulations in combination with both MPC \cite{gg:gomp05g}
and the Lattice Boltzmann method (LBM) \cite{dupi07}.

The solvent particles of the MPC fluid interact with the membrane in two
ways to obtain an impermeable membrane with no-slip boundary conditions.
First, the membrane vertices are included in the MPC collision procedure,
as suggested for polymers in Ref.~\cite{male_00_dsp}, compare
Sec.~\ref{sec:COLL_COUPL}.
Second, the solvent particles are scattered with a bounce-back rule
from the membrane surface.
Here, solvent particles inside ($1\le i\le N_{in}$)
and outside ($N_{in}<i\le N_{s}$) of the vesicle have to be distinguished.
The membrane triangles are assumed to have a finite but very small
thickness $\delta = 2l_{bs}$.
The scattering process is then performed at discrete time steps
$\Delta t_{BS}$, so that scattering does not occur exactly on the membrane
surface, but the solvent particles can penetrate slightly into the
membrane film \cite{nogu_05_dfv}.
A similar procedure has bee suggested in Ref.~\cite{kiku_03_tcm} for spherical
colloidal particles embedded in a MPC solvent.
Particles which enter the membrane film, {\em i.e.}, which have a distance
to the triangulated surface smaller than $l_{bs}$, or interior
particles which reach
the exterior volume and vice versa, are scattered at the membrane
triangle with the closest center of mass. Explicitly \cite{nogu_05_dfv},
\begin{eqnarray}
{\bf v}^{(new)}_{s}(t) &=& {\bf v}_{s}(t)
- \frac{6m_{mb}}{m_{s}+3m_{mb}} ({\bf v}_{s}(t)-{\bf v}_{tri}(t))\\
{\bf v}^{(new)}_{tri}(t) &=& {\bf v}_{tri}(t)
+ \frac{2m_{s}}{m_{s}+3m_{mb}} ({\bf v}_{s}(t)-{\bf v}_{tri}(t)),
\end{eqnarray}
when $({\bf v}_{s}(t)-{\bf v}_{tri}(t))\cdot{\bf n}_{tri} <0$,
where ${\bf v}_{s}(t)$ and ${\bf v}_{tri}(t)$ are the
velocities of the solvent particle and of
the center of mass of the membrane triangle, respectively, and
${\bf n}_{tri}$ is the normal vector of the triangle, which is
oriented towards the outside (inside) for external (internal) particles.

\begin{figure}
\begin{center}
\includegraphics[width=8.0cm]{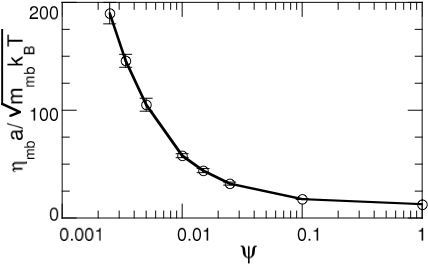}
\end{center}
\caption{ \label{fig:memb_visco}
Dependence of the membrane viscosity $\eta_{mb}$ on the probability
$\psi$ for the selection of a bond for a bond-flip attempt,
for a membrane with $N_{mb}=1860$ vertices.
From Ref.~\cite{nogu_05_dfv}.
}
\end{figure}

The bond flips provide a very convenient way to vary
the membrane viscosity $\eta_{mb}$, which increases
with decreasing probability $\psi$ for the selection of a bond
for a bond-flip attempt. The membrane viscosity has been determined
quantitatively from a simulation of a flat membrane in
two-dimensional Poiseuille flow.
The triangulated membrane is put in a rectangular box of size
$L_x\times L_y$
with periodic boundary conditions in the $x$-direction.
The edge vertices at the lower and upper boundary ($y=\pm L_y/2$) are
fixed at their positions.
A gravitational force $(m_{mb}g,0)$ is applied to all membrane
vertices to induce a flow. Rescaling of relative velocities is
employed as a thermostat.
Then, the membrane viscosity is calculated from
$\eta_{mb}= \rho_{mb}gL_y/8v_{max}$,
where $\rho_{mb}$ is average mass density of the membrane
particles, and $v_{max}$ the maximum velocity of the parabolic
flow profile.  The membrane viscosity $\eta_{mb}$,
which is obtained in this way, is shown in Fig.~\ref{fig:memb_visco}.
As $\psi$ decreases, it takes longer and longer for a membrane particle
to escape from the cage of its neighbors, and $\eta_{mb}$ increases.
This is very similar to the behavior of a hard-sphere fluid with increasing
density.  Finally, for $\psi=0$, the membrane becomes solid.

\subsection{Fluid Vesicles in Shear Flow}
\label{sec:ves_shear}

The dynamical behavior of {\em fluid vesicles} in simple shear flow has
been studied experimentally \cite{haas97,kant05,kant06,made06},
theoretically~\cite{kell82,seco82,seco83,tran84,seif99b,misb06,dank07b,lebe07},
numerically with the boundary-integral technique \cite{krau96,beau04b}
or the phase-field method \cite{beau04b,bibe05}, and
with mesoscale solvents \cite{nogu_04_fvv,nogu_05_dfv,gg:gomp07c}.
The vesicle shape is now determined by the competition of the curvature
elasticity of the membrane, the constraints of constant volume $V$ and
constant surface area $S$, and the external hydrodynamic forces.

Shear flow is characterized (in the absence of
vesicles or cells) by the flow field ${\bf v} = \dot\gamma y {\bf e}_x$,
where ${\bf e}_x$ is a unit vector, compare Sec.~\ref{sec:ultrasoft_shear}.
The control parameter of shear flow is the shear rate $\dot \gamma$, which
has the dimension of an inverse time.
Thus, a dimensionless, scaled shear rate $\dot\gamma^* = \dot\gamma \tau$
can be defined, where $\tau$ is the longest relaxation time of a vesicle.
It is given by $\tau = \eta_0 R_0^3/\kappa$,
where $\eta_0$ is the solvent viscosity, $R_0$ the average radius, and
$\kappa$ the bending rigidity \cite{broc75}.
For $\dot\gamma^* < 1$,
the internal vesicle dynamics is fast compared to the external
perturbation, so that the vesicle shape is hardly affected by the flow
field, whereas for $\dot\gamma^* > 1$, the flow forces
dominate and the vesicle is in a non-equilibrium steady state.

One of the difficulties in theoretical studies of the hydrodynamic effects
on vesicle dynamics is the no-slip boundary condition for the embedding
fluid on the vesicle surface, which changes its shape dynamically under
the effect of flow and curvature forces.
In early studies, a fluid vesicle was therefore modeled as an
ellipsoid with fixed shape \cite{kell82}. This simplified model is still
very useful as a reference for the interpretation of simulation results.

\subsubsection{Generalized Keller-Skalak Theory}
\label{sec:Keller-Skalak}

The theory of Keller and Skalak \cite{kell82} describes the hydrodynamic
behavior of vesicles of fixed ellipsoidal shape in shear flow, with
the viscosities $\eta_{in}$ and $\eta_0$ of the internal and external
fluids, respectively. Despite
of the approximations needed to derive the equation of motion for the
inclination angle $\theta$, which measures the deviation of the symmetry
axis of the ellipsoid with the flow direction, this theory describes
vesicles in flow surprisingly well. It has been generalized later
\cite{tran84} to describe the effects of a membrane viscosity $\eta_{mb}$.

The main result of the theory of Keller and Skalak is the equation of
motion for the inclination angle \cite{kell82},
\begin{equation} \label{eq:KS}
\frac{d}{dt}{\theta} = \frac{1}{2} \dot\gamma\{-1+B \cos(2\theta)\},
\end{equation}
where $B$ is a constant, which depends on the geometrical parameters
of the ellipsoid, on the viscosity contrast $\eta_{in}^*=\eta_{in}/\eta_{0}$,
and the scaled membrane viscosity $\eta_{mb}^*=\eta_{mb}/(\eta_0R_0)$
\cite{kell82,tran84,nogu_05_dfv},
\begin{equation}
\label{eq:KS_B}
B = f_0\left\{f_1+ \frac{f_1^{-1}}
      {1+f_2(\eta_{in}^*-1) + f_2f_3 \eta_{mb}^*}\right\},
\end{equation}
where $f_0$, $f_1$, $f_2$, and $f_3$ are geometry-dependent parameters.
In the spherical limit, $B \to \infty$.
Eq.~(\ref{eq:KS}) implies the following behavior:
\begin{itemize}
\item For $B>1$, there is a stationary solution, with $\cos(2\theta) =1/B$.
This corresponds to a {\em tank-treading} motion, in which the orientation of
the vesicle axis is time independent, but the membrane itself rotates
around the vorticity axis.
\item For $B<1$, no stationary solution exists, and the vesicle shows a
{\em tumbling} motion, very similar to a solid rod-like colloidal particle
in shear flow.
\item The shear rate $\dot\gamma$ only determines the time scale, but
does not affect the tank-treading or tumbling behavior. Therefore, a
transition between these two types of motion can only be induced by
a variation of the vesicle shape or the viscosities.
\end{itemize}

However, the vesicle shape in shear flow is often not as constant as
assumed by Keller and Skalak. In these situations, it is very helpful
to compare simulation results with a generalized Keller-Skalak theory,
in which shape deformation and thermal fluctuations are taken into
account.  Therefore, a phenomenological model has been suggested in
Ref.~\cite{nogu_05_dfv}, in which in addition to the inclination angle
$\theta$ a second parameter is introduced to characterize the vesicle
shape and deformation, the asphericity \cite{rudn86}
\begin{equation}
\alpha = \frac{({\lambda_1}-{\lambda_2})^2+ ({\lambda_2}-{\lambda_3})^2+
({\lambda_3}-{\lambda_1})^2}{2 R_{g}^4},
\end{equation}
where ${\lambda_1} \leq {\lambda_2} \leq {\lambda_3}$ are the
eigenvalues of the moment-of-inertia tensor and the squared
radius of gyration $R_{g}^2=\lambda_1+\lambda_2+\lambda_3$. This implies
$\alpha=0$ for spheres (with $\lambda_1 = {\lambda_2} = {\lambda_3}$),
$\alpha=1$ for long rods (with $\lambda_1 = {\lambda_2} \ll {\lambda_3}$),
and $\alpha=1/4$ for flat disks (with
$\lambda_1 \ll  {\lambda_2} = {\lambda_3}$).
The generalized Keller-Skalak model is then defined by the stochastic
equations
\begin{eqnarray}
\zeta_{\alpha} \frac{d}{dt}{\alpha}=
-\partial F/\partial \alpha + {A}\dot\gamma \sin(2\theta)
                + \zeta_{\alpha} g_{\alpha}(t) \label{eq:al}\\
\frac{d}{dt}{\theta}=
\frac{1}{2} \dot\gamma\{-1+B(\alpha) \cos(2\theta)\}
                + g_{\theta}(t) \label{eq:thet},
\end{eqnarray}
with Gaussian white noises $g_{\alpha}$ and $g_{\theta}$, which
are determined by
\begin{eqnarray}
\langle g_{\alpha}(t)\rangle = \langle g_{\theta}(t)\rangle &=&
      \langle g_{\alpha}(t) g_{\theta}(t')\rangle=0, \nonumber \\
\langle g_{\alpha}(t) g_{\alpha}(t')\rangle &=&
      2 D_\alpha \delta(t-t'), \\
\langle g_{\theta}(t) g_{\theta}(t')\rangle &=&
      2 D_\theta \delta(t-t'), \nonumber
\end{eqnarray}
friction coefficients $\zeta_{\alpha}$ and $\zeta_{\theta}$,
and diffusion constants $D_\alpha = k_{\rm B}T/\zeta_{\alpha}$ and
$D_\theta = k_{\rm B}T/\zeta_{\theta}$.
Note that $\zeta_{\theta}$ does not appear in Eq.~(\ref{eq:thet});
it drops out because the shear force is also caused by friction.

The form of the stochastic equations (\ref{eq:al}) and (\ref{eq:thet})
is motivated by the following considerations.
The first term in Eq.~(\ref{eq:al}), $\partial F/\partial \alpha$, is the
thermodynamic force due to bending energy and volume constraints; it is
calculated from the free energy $F(\alpha)$.
The second term of Eq.~(\ref{eq:al}) is the deformation force due to the
shear flow. Since the hydrodynamic forces elongate the vesicle for
$0 < \theta < \pi/2$ but push to reduce the elongation for
$-\pi/2 < \theta < 0$, the flow forces should be
proportional to $\sin(2\theta)$ to leading order.
The amplitude $A$ is assumed to be independent of the asphericity $\alpha$.
$\zeta_\alpha$ and $A$ can be estimated \cite{gg:gomp07c} from the
results of a perturbation theory \cite{misb06} in the quasi-spherical limit.
Eq.~(\ref{eq:thet}) is adapted from Keller-Skalak
theory.  While $B$ is a constant in Keller-Skalak theory, it is now
a function of the (time-dependent) asphericity $\alpha$ in
Eq.~(\ref{eq:thet}).

\subsubsection{Effects of Membrane Viscosity: Tank-Treading and Tumbling}
\label{sec:memb_visco}

The theory of Keller and Skalak \cite{kell82} predicts for fluid
vesicles a transition from tank-treading to tumbling with increasing
viscosity contrast $\eta_{in}/\eta_{0}$. This has been confirmed in
recent simulations based on a phase-field model \cite{beau04b}.

\begin{figure}[t]
\begin{center}
\includegraphics[width=8.0cm]{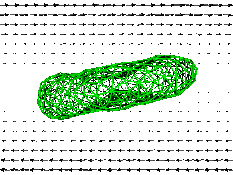}
\end{center}
\caption{\label{fig:discocyte_tt}
Snapshot of a discocyte vesicle in shear flow
with reduced shear rate $\dot\gamma^*=0.92$, reduced volume $V^*=0.59$,
membrane viscosity
$\eta_{mb}^*=0$ and viscosity contrast $\eta_{in}/\eta_{0}=1$.
The arrows represent the velocity field in
the $xz$-plane. From Ref.~\cite{nogu_05_dfv}.}
\end{figure}

The membrane viscosity $\eta_{mb}$ is also an important factor for
the vesicle dynamics in shear flow. For example,
the membrane of red blood cells becomes more viscous on aging
\cite{tran84,nash83} or in diabetes mellitus \cite{tsuk01}.
Experiments indicate that the energy dissipation in the membrane is
larger than that inside a red blood cell \cite{seco83,tran84}.
Furthermore, it has been shown recently that vesicles can not
only be made from lipid bilayers, but also from bilayers
of block copolymers \cite{disc99}. The membrane viscosity of these
``polymersomes'' is several orders of magnitude larger than
for liposomes, and can be changed over a wide range by varying the
polymer chain length \cite{dimo02}.

\begin{figure}[t]
\begin{center}
\includegraphics[width=8.0cm]{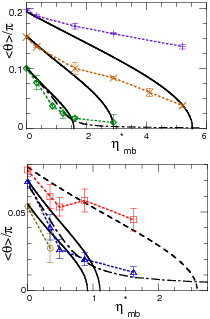}
\end{center}
\caption{\label{fig:theta_membvisco}
Average inclination angle $\langle\theta\rangle$ as a function of
reduced membrane viscosity $\eta_{mb}^*$, for the shear rate
$\dot\gamma^*=0.92$ and various reduced volumes $V^*$.
Results are presented for prolate (circles) and discocyte (squares) vesicles
with $V^*=0.59$, as well as prolate vesicles with $V^*=0.66$ (triangles),
$0.78$ (diamonds), $0.91$ (crosses), and $V^*=0.96$ (pluses).
The solid and dashed lines are calculated by K-S theory,
Eqs.~(\ref{eq:KS}) and (\ref{eq:KS_B}), for
prolate ($V^*=0.59$, $0.66$, $0.78$, $0.91$, and $0.96$) and
oblate ellipsoids ($V^*=0.59$), respectively.
The dashed-dotted lines are calculated from
Eq.~(\ref{eq:thet}) with thermal fluctuations, for
$V^*=0.66$, $V^*=0.78$, and
the rotational Peclet number $\dot\gamma/D_\theta=600$
(where $D_\theta$ is the rotational diffusion constant).
From Ref.~\cite{nogu_05_dfv}.}
\end{figure}

A variation of the membrane viscosity can be implemented easily in
dynamically triangulated surface models of membranes, as explained in
Sec.~\ref{sec:triang_surf}.  An example of a discocyte in tank-treading
motion, which is obtained by such a membrane model \cite{nogu_05_dfv},
is shown in Fig.~\ref{fig:discocyte_tt}.
Simulation results for the inclination angle as a function of the
reduced membrane viscosity $\eta^*_{mb}= \eta_{mb}/(\eta_0 R_0)$ are
shown in Fig.~\ref{fig:theta_membvisco}. This demonstrates the
tank-treading to tumbling transition of fluid vesicles with increasing
membrane viscosity. The threshold shear rate decreases with decreasing reduced
volume $V^*$, since with increasing deviation from the spherical shape,
the energy dissipation within the membrane increases. Interestingly, the
discocyte shape is less affected by the membrane viscosity than the
prolate shape for $V^*=0.59$, since it is more compact---in contrast
to a vesicle with viscosity contrast $\eta_{in}/\eta_{0} > 1$, where
the prolate shape is affected less \cite{nogu_05_dfv}.

Figure~\ref{fig:theta_membvisco} also shows a comparison of the simulation
data with results of K-S theory for fixed shape,
both without and with thermal fluctuations.
Note that there are no adjustable parameters.
The agreement of the results of theory and simulations is excellent
in the case of vanishing membrane viscosity, $\eta_{mb}=0$. For small
reduced volumes, $V^*\simeq 0.6$, the tank-treading to tumbling transition
is smeared out by thermal fluctuations, with an intermittent tumbling
motion occurring in the crossover region. This behavior is captured very
well by the generalized K-S model with thermal fluctuations. For larger
reduced volumes and non-zero membrane viscosity, significant deviations
of theory and simulations become visible. The inclination angle $\theta$
is found to decrease much more slowly with increasing membrane viscosity
than expected theoretically. This is most likely due to thermal
membrane undulations, which are not taken into account in K-S theory.

\subsubsection{Swinging of Fluid Vesicles}
\label{sec:swinging}

\begin{figure}[t]
\begin{center}
\includegraphics[width=8.0cm]{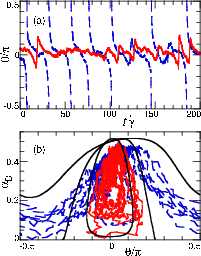}
\end{center}
\caption{\label{fig:swing_ext}
Temporal evolution of vesicle deformation $\alpha_{D}$ and
inclination angle $\theta$, for $V^*=0.78$ and $\eta_{mb}^*=2.9$.
Here, $\alpha_D = (L_1-L_2)/(L_1+L_2)$, where $L_1$ and $L_2$ are
the maximum lengths in the direction of the eigenvectors of the
gyration tensor in the vorticity plane.
The solid (red) and dashed (blue) lines represent simulation data
for $\dot\gamma^*=3.68$ and $0.92$ ($\kappa/k_BT=10$ and $40$, with
$\dot\gamma \eta_{0} R_0^3/k_BT=36.8$), respectively.
The solid lines in (b) are obtained from Eqs.~(\ref{eq:al}),
(\ref{eq:thet}) without thermal noise for $\dot\gamma^*=1.8$, $3.0$, and
$10$ (from top to bottom).
From Ref.~\cite{gg:gomp07c}.}
\end{figure}

Recently, of a new type of vesicle dynamics in shear flow has been observed
experimentally \cite{kant06}, which is characterized by
oscillations of the inclination angle $\theta$
with $\theta(t) \in [-\theta_0,\theta_0$] and $\theta_0 < \pi/2$.
The vesicles were found to transit from tumbling to this oscillatory motion
with increasing shear rate $\dot\gamma$.
Simultaneously with the experiment, a ``vacillating-breathing" mode for
quasi-spherical fluid vesicles was predicted theoretically, based on a
spherical-harmonics expansion of the equations of motion to leading order
(without thermal fluctuations) \cite{misb06}.
This mode exhibits similar dynamical behavior as seen experimentally;
however, it ``coexists" with the tumbling mode, and its orbit depends on
the initial deformation, i.e., it is not a limit cycle. Furthermore, the shear
rate appears only as the basic time scale, and therefore cannot
induce any shape transitions.
Hence it does not explain the tumbling-to-oscillatory transition seen in
the experiments \cite{kant06}.

\begin{figure}[t]
\begin{center}
\includegraphics*[width=8.0cm]{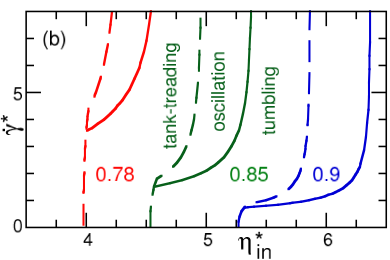}
\end{center}
\caption{\label{fig:swing_phase}
Dynamical phase diagram as a function of viscosity contrast
$\eta_{in}^*=\eta_{in}/\eta_{0}$, for $\eta_{\rm {mb}}^*=0$ and
various reduced volumes $V^*$, calculated from Eqs.~(\ref{eq:al}),
(\ref{eq:thet}) without thermal noise.
The tank-treading phase is located on the left-hand-side of the dashed
lines. The solid lines represent the tumbling-to-swinging transitions.
From Ref.~\cite{gg:gomp07c}.}
\end{figure}

Simulation data for the oscillatory mode---which has also been denoted
``trembling'' \cite{kant06} or ``swinging'' \cite{gg:gomp07c} mode---are
shown in Fig.~\ref{fig:swing_ext}. The simulation results demonstrate
that the transition can indeed be induced by increasing shear rate, and
that it is robust to thermal fluctuations. Figure~\ref{fig:swing_ext} also
shows that the simulation data are well captured by the generalized
K-S model, Eqs.~(\ref{eq:al}) and (\ref{eq:thet}), which takes into
account higher-order contributions in the
curvature energy of a vesicle. The theoretical model can therefore be
used to predict the full dynamic phase diagram of prolate vesicles
as a function of shear rate and membrane viscosity or viscosity
contrast, compare Fig.~\ref{fig:swing_phase}.
The swinging phase appears at the boundary between the tank-treading
and the tumbling phase for sufficiently large shear rates. The
phase diagram explains under which conditions the swinging phase
can be reached from the tumbling phase with increasing shear
rate---as observed experimentally \cite{kant06}.

The generalized K-S model is designed to capture the vesicle flow
behavior for non-spherical shapes sufficiently far from a sphere.
For quasi-spherical vesicles, a derivation of the
equations of motion by a systematic expansion in the undulation
amplitudes gives quantitatively more reliable results.
An expansion to third order results in a phase diagram \cite{dank07b,lebe07},
which agrees very well with Fig.~\ref{fig:swing_phase}.

\subsubsection{Flow-Induced Shape Transformations}

\begin{figure}[t]
\begin{center}
\includegraphics[width=8.0cm]{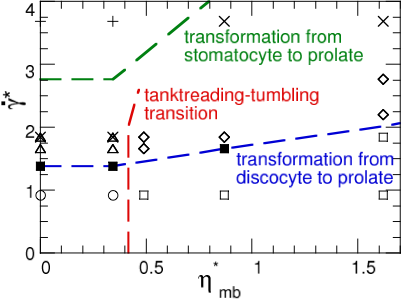}
\end{center}
\caption{\label{fig:shear_diag}
Dynamical phase diagram of a vesicle in shear flow for reduced volume
$V^*=0.59$. Symbols show simulated parameter values, and indicate
tank-treading discocyte and tank-treading prolate ($\circ$),
tank-treading prolate and unstable discocyte ($\triangle$),
tank-treading discocyte and tumbling (transient) prolate ($\square$),
tumbling with shape oscillation ($\diamond$),
unstable stomatocyte ($+$), stable stomatocyte ($\times$), and
near transition ($\blacksquare$).
The dashed lines are guides to the eye.
From Ref.~\cite{nogu_05_dfv}.}
\end{figure}

Shear flow does not only induce different dynamical modes of prolate
and oblate fluid vesicles, it can also induce phase transformations.
The simplest case is a oblate fluid vesicle with $\eta_{mb}=0$ and
viscosity contrast $\eta_{in}/\eta_{0}=1$. When the reduced shear
rate reaches $\dot\gamma^*\simeq 1$, the discocyte vesicles are
stretched by the
flow forces into a prolate shape \cite{krau96,nogu_04_fvv,nogu_05_dfv}.
A similar transition is found for stomatocyte vesicles, except that
in this case a larger shear rate $\dot\gamma^*\simeq 3$ is required.
In the case of non-zero membrane viscosity, a rich phase
behavior appears, see Fig.~\ref{fig:shear_diag}.

\begin{figure}[ht]
\begin{center}
\includegraphics[width=8.0cm]{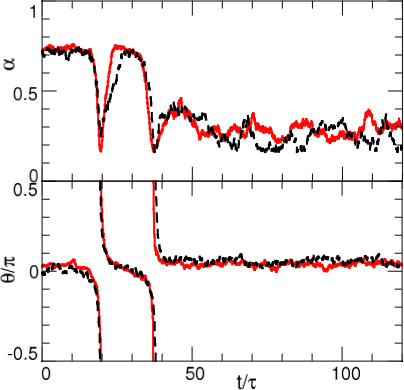}
\end{center}
\caption{\label{fig:pro2dis_dyn}
Time dependence of asphericity $\alpha$ and inclination angle
$\theta$, for $\dot\gamma^*=1.84$,
$\eta_{\rm {mb}}^*=1.62$, and $V^*=0.59$.
The dashed lines are obtained from Eqs.~(\ref{eq:al}) and (\ref{eq:thet}),
with $\zeta_{\alpha}=100$, $A=12$, and $B(\alpha)=1.1-0.17\alpha$.
From Ref.~\cite{gg:gomp05i}.}
\end{figure}

Surprisingly, flow forces
can not only stretch vesicles into a more elongated shape, but can also
induce a transition from an elongated prolate shape into a more
compact discocyte shape \cite{nogu_05_dfv}. Simulation results for
the latter transition
are shown in Fig.~\ref{fig:pro2dis_dyn}. This transition is possible,
because in a range of membrane viscosities, the prolate shape is in
the tumbling phase, while the oblate shape is tank-treading,
compare Fig.~\ref{fig:theta_membvisco}. Of course, this requires
that the free energies of the two shapes are nearly equal, which
implies a reduced volume of $V^*\simeq 0.6$.  Thus,
a prolate vesicle in this regime starts tumbling; as the inclination
angle becomes negative, shear forces push to shrink the long axis of
the vesicle; when this force is strong enough to overcome the
free-energy barrier between the prolate and the oblate phase, a
shape transformation can be induced, compare Fig.~\ref{fig:pro2dis_dyn}.
The vesicle then remains in the stable tank-treading state.

At higher shear rates, on intermittent behavior has been observed, in which
the vesicle motion changes in irregular intervals between tumbling and
tank-treading \cite{nogu_05_dfv}.

\subsubsection{Vesicle Fluctuations under Flow in Two Dimensions}

At finite temperature, stochastic fluctuations of the membrane due
to thermal motion affect the dynamics of vesicles.
Since the calculation of thermal fluctuations under flow conditions
requires long times and large membrane sizes (in order to
have a sufficient range of undulation wave vectors), simulations
have been performed for a two-dimensional system in the stationary
tank-treading state \cite{gg:gomp08xxc}. For comparison, in the limit
of small deformations from a circle, Langevin-type
equations of motion have been derived, which are highly nonlinear due
to the constraint of constant perimeter length \cite{gg:gomp08xxc}.

\begin{figure}[ht]
\begin{center}
\includegraphics*[angle=0,width=9.0cm]{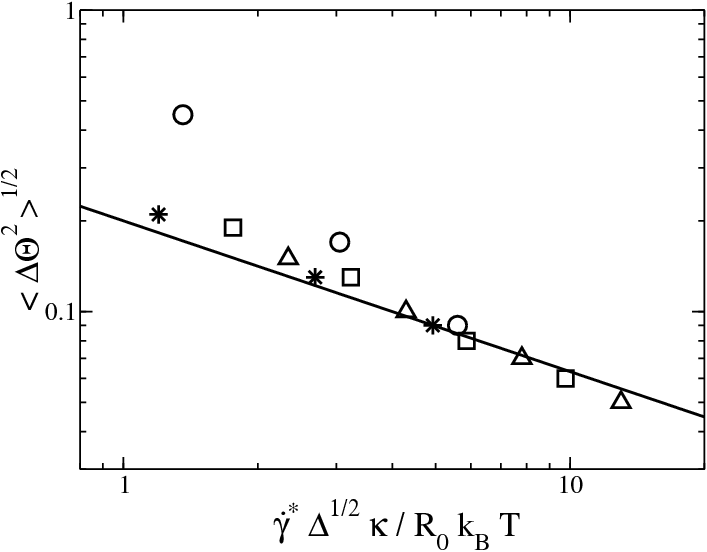}
\end{center}
\caption{ \label{fig:ves_2dim}
Fluctuations of the inclination angle $\langle \Delta
\theta^{2}\rangle^{1/2}$ of a two-dimensional fluctuating vesicle in
shear flow, as a function of scaled shear rate
$\dot\gamma^* \Delta^{1/2} \kappa / (R_{0} k_{B}T)$, where
$\Delta$ is the dimensionless excess membrane length.
Symbols indicate simulation data for different internal vesicle areas $A$
for fixed membrane length, with $A^{*} \equiv A/\pi R_0^2= 0.95$ (squares),
$A^{*} = 0.90$ (triangles), $A^{*}=0.85$ (stars), and $A^{*} = 0.7$
(circles). The solid line is the theoretical result in the
quasi-circular limit. From Ref.~\protect{\cite{gg:gomp08xxc}}.
}
\end{figure}

The effect of the shear flow is to induce a tension in the membrane,
which reduces the amplitude of thermal membrane undulations. This
tension can be extracted directly from simulation data for the
undulation spectrum. The reduction of the undulation amplitudes also
implies that the fluctuations of the inclination angle $\theta$ get
reduced with increasing shear rate. The theory for quasi-circular
shapes predicts a universal behavior as a function of the scaled
shear rate $\dot\gamma^* \Delta^{1/2} \kappa / (R_{0} k_{B}T)$, where
$\dot\gamma^*=\dot{\gamma}\eta_0 R_{0}^{3}/\kappa$ is the reduced
shear rate in two dimensions, and $\Delta$ is the dimensionless
excess membrane length.
Theory and simulation results for the inclination angle as a function
of the reduced shear rate are shown in Fig.~\ref{fig:ves_2dim}.
There are no adjustable parameters.
The agreement is excellent as long as the deviations from the circular
shape are not too large \cite{gg:gomp08xxc}.

\subsection{Fluid Vesicles and Red Blood Cells in Capillary Flow}
\label{sec:cap_flow}

\subsubsection{RBC Deformation in Narrow Capillaries}

The deformation of single RBCs and single fluid vesicles
in capillary flows were studied theoretically by lubrication theories
\cite{seco86,skal90,brui96} and boundary-integral methods
\cite{queg97,pozr05a,pozr05b}. In most of these studies,
axisymmetric shapes which are coaxial with the center of the capillary
were assumed and cylindrical coordinates were employed.
In order to investigate non-axisymmetric shapes as well as flow-induced
shape transformations, a fully three-dimensional simulation approach
is required.

\begin{figure}[ht]
\begin{center}
\includegraphics[width=11.8cm]{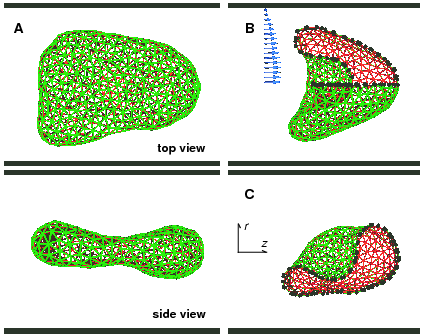}
\end{center}
\caption{\label{fig:RBC_capillary}
Snapshots of vesicles in capillary flow, with bending rigidity
$\kappa/k_BT=20$ and capillary radius $R_{cap} = 1.4 R_0$.
(A) Fluid vesicle with discoidal shape at the mean fluid
velocity $v_{m}\tau/R_{cap}=41$, both in side and top views.
(B) Elastic vesicle (RBC model) with parachute shape at
$v_{m}\tau/R_{cap}=218$ (with shear modulus $\mu R_0^2/k_BT=110$).
The blue arrows represent the velocity field of the solvent.
(C) Elastic vesicle with slipper-like shape at
$v_{m}\tau/R_{cap}=80$ (with $\mu R_0^2/k_BT=110$).
The inside and outside of the membrane are depicted in red and green,
respectively. The upper front quarter of the vesicle in (B) and the
front half of the vesicle in (C) are removed to allow for a look into the
interior; the black circles indicate the lines where the membrane has
been cut in this procedure.
Thick black lines indicate the walls the cylindrical capillary.
From Ref.~\cite{gg:gomp05g}.}
\end{figure}

We focus here on the behavior of single red blood cells in capillary
flow \cite{gg:gomp05g}, as described by a triangulated surface model
for the membrane (compare Sec.~\ref{sec:RBC_models})
immersed in a MPC solvent (see Sec.~\ref{sec:RBC_models}). The radius
of the capillary, $R_{cap}$, is taken to be slightly larger than the
mean vesicle or RBC radius, $R_0=\sqrt{S/4\pi}$, where $S$ is the
membrane area. Snapshots of vesicle and RBC shapes in flow are shown
in Fig.~\ref{fig:RBC_capillary} for a reduced volume of $V^*=0.59$,
where the vesicle shape at rest is a discocyte. For sufficiently small
flow velocities, the discocyte shape is retained. However, the
discocyte is found {\em not} in a coaxial orientation; instead
the shortest eigenvalue of the gyration tensor is oriented perpendicular
to the cylinder axis \cite{gg:gomp05g}. Since two opposite sides of the
rim of the discocyte are closer to wall where the flow velocity is
small, the rotational symmetry is slightly disturbed and the top view
looks somewhat triangular, see Fig.~\ref{fig:RBC_capillary}A. With increasing
flow velocity, a shape transition to an axisymmetric shape occurs.
In the case of fluid vesicles this is a prolate shape, while in the
case of RBCs a parachute shape is found. Such parachute shapes of
red blood cells have previously been observed experimentally
\cite{skal69,tsuk01}.

The fundamental difference between the flow behaviors of fluid vesicles
and red blood cells at high flow velocities is due to the shear
elasticity of the spectrin network. Its main
effect for $\mu R_0^2/\kappa \gtrsim 1$ is to suppress
the discocyte-to-prolate transition, because the
prolate shape would acquire an elastic energy of order $\mu R_0^2$.
In comparison, the shear stress in the parachute shape is much
smaller.

\begin{figure}[ht]
\begin{center}
\includegraphics[width=5.9cm]{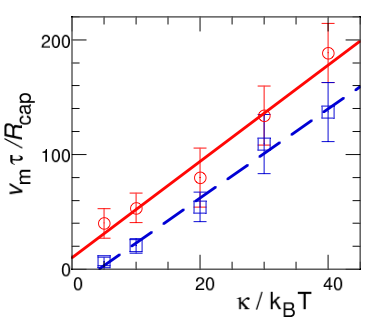}
\includegraphics[width=5.8cm]{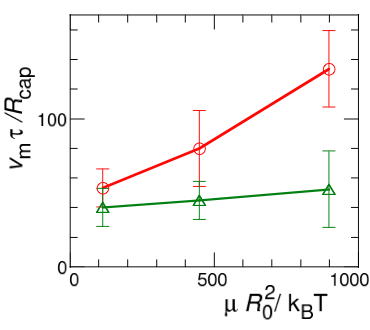}
\end{center}
\caption{\label{fig:cap_flow_velocity}
Critical flow velocity $v_m$ of the discocyte-to-parachute transition
of elastic vesicles and of the discocyte-to-prolate transition of fluid
vesicles, as a function of the bending rigidity $\kappa$.
From Ref.~\cite{gg:gomp05g}.}
\end{figure}

Some diseases, such as diabetes mellitus and sickle cell anemia,
change the mechanical properties of RBCs; a reduction of RBC
deformability was found to be responsible for an enhanced flow
resistance of blood \cite{chie87}. Therefore, it is very important to
understand the relation of RBC elasticity and flow properties in
capillaries. The flow velocity at the discocyte-to-prolate transition
of fluid vesicles and at the discocyte-to-parachute transition is
therefore shown in Fig.~\ref{fig:cap_flow_velocity} as a function of
the bending rigidity and the shear modulus, respectively. In both
cases, an approximately linear dependence is obtained \cite{gg:gomp05g},
\begin{equation}
v_{m}^{c}\frac{\tau}{R_{cap}} =
0.1\frac{\mu R_0^2}{k_BT}  + 4.0\frac{\kappa}{k_BT}.
\end{equation}
This result suggests that parachute shapes of RBCs should appear for
flow velocities larger than $v_m^{c} = 800
R_{cap}/\tau$ $\simeq 0.2$mm/s under physiological conditions.
This is consistent with the experimental results of
Ref.~\cite{suzu96}, and is in the range of micro-circulation in
the human body.

Figure~\ref{fig:cap_flow_velocity} (right) also shows that there is a
metastable region, where discocytes are seen for
increasing flow velocity, but parachute shapes for decreasing flow
velocity. This hysteresis becomes more pronounced with increasing
shear modulus. It is believed to be
due to a suppression of thermal fluctuations with increasing
$\mu R_0^2/k_BT$.

\subsubsection{Flow in Wider Capillaries}

The flow of many red blood cells in wider capillaries has also been
investigated by several simulation techniques. Discrete
fluid-particle simulations---an extension of dissipative-particle
dynamics (DPD)---in combination with bulk-elastic discocyte cells
(in contrast to the membrane elasticity of real red blood cells)
have been employed to investigate the dynamical clustering of red blood
cells in capillary vessels \cite{bory03,dzwi03}.
An immersed finite-element model---a combination of the immersed
boundary method for the solvent hydrodynamics \cite{eggl98} and a
finite-element method to describe the membrane elasticity---has been
developed to study red blood cell aggregation \cite{liu06}.
Finally, it has been demonstrated that the Lattice-Boltzmann method for
the solvent in combination with a triangulated mesh model with curvature
and shear elasticity for the membrane can be used efficiently to
simulate RBC suspensions in wider capillaries \cite{dupi07}.


\section{Viscoelastic Fluids}
\label{sec:viscoelastic}

One of the unique properties of soft matter is its viscoelastic
behavior \cite{lars_99_src}. Due to the long structural relaxation times,
the internal
degrees of freedom cannot relax sufficiently fast in an oscillatory
shear flow already at moderate frequencies, so that there is an
elastic restoring force which
pushes the system back to its previous state. Well studied
examples of viscoelastic fluids are polymer solutions and polymer
melts \cite{doi:86,lars_99_src}.

The viscoelastic behavior of polymer solutions leads to many unusual
flow phenomena, such as viscoelastic phase separation \cite{tana_00_vps}.
There is also a second level of complexity in soft matter systems,
in which a colloidal component is dispersed in a solvent, which is
itself a complex fluid. Examples are spherical or rod-like colloids
dispersed in polymer solutions. Shear flow can induce
particle aggregation and alignment in these systems \cite{verm_05_fis}.

It is therefore desirable to generalize the MPC technique to model
viscoelastic fluids, while retaining as much as possible of the computational
simplicity of standard MPC for Newtonian fluids. This can be done
by replacing the point particles of standard MPC by harmonic
dumbbells with spring constant $K$ \cite{tao_08}.

As for point particles, the MPC algorithm consists of two steps, streaming and
collisions.  In the streaming step, within a time interval $\Delta t$,
the motion of all dumbbells is governed by Newton's equations of motion.
The center-of-mass coordinate of each dumbbell follows a simple
ballistic trajectory.
The evolution of the relative coordinates of dumbbell $i$, which
consists of two monomers at positions ${\bf r}_{i1}(t)$ and ${\bf r}_{i2} (t)$
with velocities ${\bf v}_{i1}(t)$ and ${\bf v}_{i2} (t)$, respectively,
is determined by the harmonic interaction potential, so that
\begin{eqnarray}
 \label{rr}
 {\bf r}_{i1}(t+\Delta t) - {\bf r}_{i2} (t+\Delta t)
 & = & {\bf A}_{i}(t) \cos(\omega_0 \Delta t)
            + {\bf B}_{i}(t) \sin(\omega_0 \Delta t)\;; \\
 \label{rv}
 {\bf v}_{i1}(t+\Delta t) - {\bf v}_{i2} (t+\Delta t)
 & = & - \omega_0 {\bf A}_{i}(t) \sin(\omega_0 \Delta t) \nonumber \\
 & & \hspace*{2cm}  + \omega_0 {\bf B}_{i}(t) \cos(\omega_0 \Delta t) \;,
\end{eqnarray}
with angular frequency $\omega_0 = \sqrt{2K/m}$. The amplitudes
${\bf A}_{i}(t)$ and ${\bf B}_{i}(t)$ are determined by the initial positions
and velocities at time $t$.
The collision step is performed for the two point particles constituting
a dumbbell in exactly the same way as for MPC point-particle fluids.
This implies, in particular, that the various collision rules of MPC,
such as SRD, AT-a or AT+a, can all be employed also for simulations of
viscoelastic solvents, depending on the requirements of the system under
consideration. Since the streaming step is only a little more time
consuming and the collision step is identical,
simulations of the viscoelastic MPC fluid can be performed
with essentially the same efficiency as for the standard point-particle
fluid.

\begin{figure}[ht]
  \begin{center}
    \includegraphics*[width=9cm]{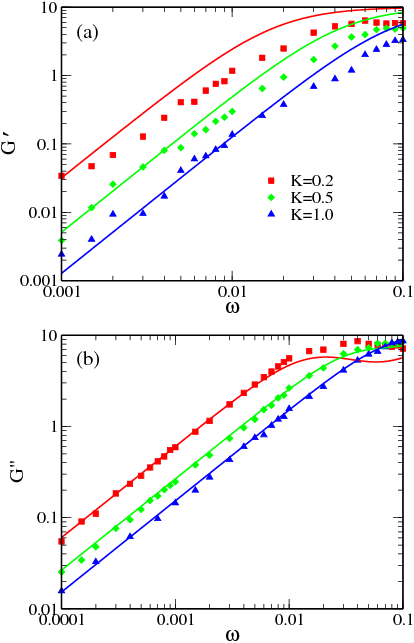}
  \end{center}
  \caption{
    \label{k-g}
     (a) Storage $G'$ and (b) loss moduli $G''$,
     as function of oscillation
     frequency $\omega$ on a double-logarithmic scale, for systems of
     dumbbells with various spring constants ranging from
     $K=0.2$ to $K=1.0$. Simulations are performed in two dimensions
     with the SRD collision rule.
     The wall separation and the collision time are $L_y=10$ and
     $\Delta t=0.02$, respectively.  From Ref.~\cite{tao_08}.
   }
\end{figure}

The behavior of harmonic dumbbells in dilute solution has been studied
in detail analytically \cite{bird_87_dpl}. These results can be used to
predict the zero-shear viscosity $\eta$ and the storage and
loss moduli, $G'(\omega)$ and $G''(\omega)$ in oscillatory shear with
frequency $\omega$, of the MPC dumbbell fluid. This requires the
solvent viscosity and diffusion constant of monomers in the solvent.
Since the viscoelastic MPC fluid consists of dumbbells only, the
natural assumption is to employ the viscosity $\eta_{MPC}$ and diffusion
constant $D$ of an MPC point-particle fluid of the same density.
The zero-shear viscosity is then found to be \cite{tao_08}
\begin{equation}
\eta = \eta_{MPC} \, + \frac{\rho}{2} \, \frac{k_BT}{\omega_H},
\label{eq:visc_theo_MPC}
\end{equation}
where
\begin{equation}\label{eq:omega_H}
\omega_H=\frac{4K}{\zeta}=\frac{4DK}{k_BT}.
\end{equation}
Similarly, the storage and loss
modulus, and the average dumbbell extension, are predicted to
be \cite{tao_08}
\begin{eqnarray}
\label{eq:storage}
G^\prime &=& \frac{\rho k_BT}{2} \,
           \frac{(\omega / \omega_H)^2}{1+(\omega / \omega_H)^2}, \\
\label{eq:loss}
G^{\prime\prime} &=& \eta_{MPC}\,\omega  +
  \frac{\rho k_BT}{2} \, \frac{\omega / \omega_H}{1+(\omega / \omega_H)^2},
\end{eqnarray}
and
\begin{equation}
\frac{\langle r^2 \rangle}{\langle r^2 \rangle_{\rm eq}} =
                1+\frac{2}{3}(\dot{\gamma} / \omega_H)^2.
\label{eq:R2}
\end{equation}
Simulation data are shown in Fig.~\ref{k-g}, together with the
theoretical predications (\ref{eq:storage}) and (\ref{eq:loss}).
The comparison shows a very good agreement. This includes not only the
linear and quadratic frequency dependence of $G''$ and $G'$ for small
$\omega$, respectively,
but also the leveling off when $\omega$ reaches $\omega_H$. In case
of $G''$, there is quantitative agreement without any adjustable
parameters, whereas $G'$ is somewhat overestimated by Eq.~(\ref{eq:storage})
for small spring constants $K$.
The good agreement of theory and simulations implies that the
characteristic frequency
decreases linearly with decreasing spring constant $K$ and mean free
path $\lambda$ (since $D\sim \lambda$).
A comparison of other quantities, such as the zero-shear viscosity,
shows a similar quantitative agreement \cite{tao_08}.


\section{Conclusions and Outlook}
\label{sec:conclusions}

In the short time since Malevanets and Kapral introduced
multi-particle collision (MPC) dynamics as a particle-based
mesoscale simulation technique for studying the hydrodynamics of
complex fluids, there has been enormous progress. It has been
shown that kinetic theory can be generalized to calculate transport
coefficients, several collision algorithms have been proposed and
employed, and the method has been generalized to describe multi-phase
flows and viscoelastic fluids. The primary applications to
date---which include studies of the equilibrium dynamics and flow
properties of colloids, polymers, and vesicles in
solution---have dealt with mesoscopic particles embedded in a
single-component Newtonian solvent. An important advantage of
this algorithm is that it is very straightforward to model
the dynamics for the embedded particles using a hybrid MPC-molecular
dynamics simulations approach. The results of these studies are in excellent
quantitative agreement with both theoretical predictions and results
obtained using other simulation techniques.

How will the method develop in the future? This is of course
difficult to predict. However, it seems clear that there will be
two main directions, a further development of the method itself,
and its application to new problems in Soft Matter hydrodynamics.
On the methodological front, there are several very recent
developments, like angular-momentum conservation, multi-phase
flows and viscoelastic fluids, which have to be explored in more
detail. It will also be interesting to combine them to study, {\em e.g.},
multi-phase flows of viscoelastic fluids. On the application side,
the trend will undoubtedly be towards more complex systems, in which
thermal fluctuations are important. In such systems, the
method can play out its strengths, because the interactions of colloids,
polymers, and membranes with the mesoscale solvent can all be treated
on the same basis.

\section*{Acknowledgments}
Financial support from the Donors of the American Chemical Society
Petroleum Research Fund, the National Science Foundation under grant No.
DMR-0513393, the German Research Foundation (DFG) within the SFB TR6
``Physics of Colloidal Dispersion in External Fields'', and the
Priority Program ``Nano- and Microfluidics'' are gratefully acknowledged.
We thank Elshad Allahyarov, Luigi Cannavacciuolo, Jens Elgeti,
Reimar Finken, Ingo G\"otze, Jens Harting, Martin Hecht, Hans Herrmann,
Antonio Lamura, Kiaresch Mussawisade,
Hiroshi Noguchi, Guoai Pan, Marisol Ripoll, Udo Seifert, Yu-Guo Tao,
and Erkan T\"uzel for many stimulating discussions and enjoyable
collaborations.

\bibliographystyle{apsrev}
\bibliography{gompper,amphiphile,Library,rev_article,comments}

\end{document}